\documentclass[12pt]{article}
\usepackage{epsfig}
\usepackage{amsmath}
\usepackage{hhline}
\usepackage{amssymb}
\usepackage{times}
\usepackage{cite}
\usepackage{rotating}
\usepackage{dcolumn}
\usepackage{booktabs}
\usepackage{multicol}
\usepackage{tabularx}

\newlength{\dinwidth}
\newlength{\dinmargin}
\begin{document}  
\newcommand{\pom}{{I\!\!P}}
\newcommand{\reg}{{I\!\!R}}
\newcommand{\pis}{\pi_{\rm{s}}}
\newcommand{\mm}{\mathrm}
\newcommand {\gapprox}
   {\raisebox{-0.7ex}{$\stackrel {\textstyle>}{\sim}$}}
\newcommand {\lapprox}
   {\raisebox{-0.7ex}{$\stackrel {\textstyle<}{\sim}$}}
\def\gsim{\,\lower.25ex\hbox{$\scriptstyle\sim$}\kern-1.30ex%
\raise 0.55ex\hbox{$\scriptstyle >$}\,}
\def\lsim{\,\lower.25ex\hbox{$\scriptstyle\sim$}\kern-1.30ex%
\raise 0.55ex\hbox{$\scriptstyle <$}\,}
\newcommand{\gap}{\stackrel{>}{\sim}}
\newcommand{\lap}{\stackrel{<}{\sim}}
\newcommand{\fem}{$F_2^{em}$}
\newcommand{\tsnmp}{$\tilde{\sigma}_{NC}(e^{\mp})$}
\newcommand{\tsnm}{$\tilde{\sigma}_{NC}(e^-)$}
\newcommand{\tsnp}{$\tilde{\sigma}_{NC}(e^+)$}
\newcommand{\st}{$\star$}
\newcommand{\sst}{$\star \star$}
\newcommand{\ssst}{$\star \star \star$}
\newcommand{\sssst}{$\star \star \star \star$}
\newcommand{\tw}{\theta_W}
\newcommand{\sw}{\sin{\theta_W}}
\newcommand{\cw}{\cos{\theta_W}}
\newcommand{\sww}{\sin^2{\theta_W}}
\newcommand{\cww}{\cos^2{\theta_W}}
\newcommand{\trm}{m_{\perp}}
\newcommand{\trp}{p_{\perp}}
\newcommand{\trmm}{m_{\perp}^2}
\newcommand{\trpp}{p_{\perp}^2}
\newcommand{\alp}{\alpha_s}

\newcommand{\alps}{\alpha_s}
\newcommand{\sqrts}{$\sqrt{s}$}
\newcommand{\LO}{$O(\alpha_s^0)$}
\newcommand{\Oa}{$O(\alpha_s)$}
\newcommand{\Oaa}{$O(\alpha_s^2)$}
\newcommand{\PT}{p_{\perp}}
\newcommand{\pt}{p_{T}}
\newcommand{\JPSI}{J/\psi}
\newcommand{\sh}{\hat{s}}
\newcommand{\uh}{\hat{u}}
\newcommand{\MP}{m_{J/\psi}}
\newcommand{\PO}{I\!\!P}
\newcommand{\xbj}{x}
\newcommand{\xpom}{x_{\PO}}
\newcommand{\ttbs}{\char'134}
\newcommand{\xpomlo}{3\times10^{-4}}  
\newcommand{\xpomup}{0.05}  
\newcommand{\dgr}{^\circ}
\newcommand{\pbarnt}{\,\mbox{{\rm pb$^{-1}$}}}
\newcommand{\gev}{\,\mbox{GeV}}
\newcommand{\WBoson}{\mbox{$W$}}
\newcommand{\fbarn}{\,\mbox{{\rm fb}}}
\newcommand{\fbarnt}{\,\mbox{{\rm fb$^{-1}$}}}
\newcommand{\dsdx}[1]{$d\sigma\!/\!d #1\,$}
\newcommand{\eV}{\mbox{e\hspace{-0.08em}V}}
\newcommand{\shat}{\ensuremath{\hat{s}}}
\newcommand{\rb}[1]{\raisebox{1.5ex}[-1.5ex]{#1}}
%
%
\newcommand{\qsq}{\ensuremath{Q^2} }
\newcommand{\gevsq}{\ensuremath{{\rm GeV}^2} }
\newcommand{\et}{\ensuremath{E_t^*} }
\newcommand{\rap}{\ensuremath{\eta^*} }
\newcommand{\gp}{\ensuremath{\gamma^*}p }
\newcommand{\dsiget}{\ensuremath{{\rm d}\sigma_{ep}/{\rm d}E_t^*} }
\newcommand{\dsigrap}{\ensuremath{{\rm d}\sigma_{ep}/{\rm d}\eta^*} }

\newcommand{\dstar}{\ensuremath{D^*}}
\newcommand{\dstarp}{\ensuremath{D^{*+}}}
\newcommand{\dstarm}{\ensuremath{D^{*-}}}
\newcommand{\dstarpm}{\ensuremath{D^{*\pm}}}
\newcommand{\zDs}{\ensuremath{z(\dstar )}}
\newcommand{\Wgp}{\ensuremath{W_{\gamma p}}}
\newcommand{\ptds}{\ensuremath{p_t(\dstar )}}
\newcommand{\etads}{\ensuremath{\eta(\dstar )}}
\newcommand{\ptj}{\ensuremath{p_t(\mbox{jet})}}
\newcommand{\ptjn}[1]{\ensuremath{p_t(\mbox{jet$_{#1}$})}}
\newcommand{\etaj}{\ensuremath{\eta(\mbox{jet})}}
\newcommand{\DM}{\ensuremath{\Delta m}}
\newcommand{\detadsj}{\ensuremath{\eta(\dstar )\, \mbox{-}\, \etaj}}
\newcommand{\ftwocc}{\ensuremath{F_2^{c\bar{c}}}}
\newcommand{\rnorm}{\ensuremath{R^{\rm norm}}}
\newcommand{\lesim}{\mbox{$\;\raisebox{-1mm}{$\stackrel{\scriptstyle<}
{\scriptstyle\sim}$}\;$}}
\newcommand{\gesim}{\mbox{$\;\raisebox{-1mm}{$\stackrel{\scriptstyle>}
{\scriptstyle\sim}$}\;$}}
\newcommand{\matrixA}{\ensuremath{\textbf{A}}}
\newcolumntype{d}{D{.}{.}{-1}}
\newcolumntype{-}{D{d}{d}{1}}
\newcolumntype{e}{D{e}{10^}{3}}

\def\Journal#1#2#3#4{{#1} {\bf #2} (#3) #4}
\def\NCA{Nuovo Cimento}
\def\NIM{Nucl. Instrum. Methods}
\def\NIMA{{Nucl. Instrum. Methods} {\bf A}}
\def\NPB{{Nucl. Phys.}   {\bf B}}
\def\PLB{{Phys. Lett.}   {\bf B}}
\def\PRL{Phys. Rev. Lett.~}
\def\PRD{{Phys. Rev.}    {\bf D}}
\def\ZPC{{Z. Phys.}      {\bf C}}
\def\EJC{{Eur. Phys. J.} {\bf C}}
\def\EPJC{{Eur. Phys. J.} {\bf C}}
\def\CPC{Comp. Phys. Commun.~}


\begin{titlepage}

\noindent
\begin{flushleft}
{\tt DESY 11-066    \hfill    ISSN 0418-9833} \\
{\tt April 2011}                  \\
\end{flushleft}

\noindent

\vspace{2cm}
\begin{center}
\begin{Large}

{\bf   Measurement of \boldmath ${ D^{*\pm}}$ Meson Production and 
Determination of \ftwocc\ at low ${Q^2}$ in 
Deep-Inelastic Scattering at HERA  }

\vspace{2cm}

H1 Collaboration

\end{Large}
\end{center}

\vspace{2cm}

\begin{abstract}
Inclusive production of \dstar~mesons in deep-inelastic $ep$ scattering at HERA 
is studied in the range $5<Q^2<100 \ {\rm GeV}^2$ of the photon virtuality 
and $0.02<y<0.7$ of the inelasticity of the scattering process. The observed 
phase space for the \dstar~meson is $p_T (D^*) > 1.25\ {\rm GeV}$ and 
$|\eta (D^*)| < 1.8$. The data sample corresponds to an integrated 
luminosity of $348\ {\rm  pb}^{-1}$ collected with the H1 detector.
Single and double differential cross sections are measured and the charm 
contribution \ftwocc\ to the proton structure
function $F_2$ is determined. The results are compared to perturbative
QCD predictions at next-to-leading order implementing different schemes 
for the charm mass treatment and with Monte Carlo models based 
on leading order matrix elements with parton showers.
\end{abstract}

\vspace{1.5cm}

\begin{center}
Accepted by \EJC
\end{center}

\end{titlepage}

%
%
%
\begin{flushleft}

F.D.~Aaron$^{5,48}$,           
C.~Alexa$^{5}$,                
V.~Andreev$^{25}$,             
S.~Backovic$^{30}$,            
A.~Baghdasaryan$^{38}$,        
S.~Baghdasaryan$^{38}$,        
E.~Barrelet$^{29}$,            
W.~Bartel$^{11}$,              
K.~Begzsuren$^{35}$,           
A.~Belousov$^{25}$
% ...

\bigskip{\it
% ...
$ ^{5}$ National Institute for Physics and Nuclear Engineering (NIPNE) ,
          Bucharest, Romania$^{ m}$ \\
% ...
$ ^{11}$ DESY, Hamburg, Germany \\
% ...
$ ^{25}$ Lebedev Physical Institute, Moscow, Russia$^{ e}$ \\
% ...
$ ^{29}$ LPNHE, Universit\'e Pierre et Marie Curie Paris 6,
          Universit\'e Denis Diderot Paris 7, CNRS/IN2P3, Paris, France \\
 $ ^{30}$ Faculty of Science, University of Montenegro,
          Podgorica, Montenegro$^{ n}$ \\
% ...
$ ^{35}$ Institute of Physics and Technology of the Mongolian
          Academy of Sciences, Ulaanbaatar, Mongolia \\
% ...
$ ^{38}$ Yerevan Physics Institute, Yerevan, Armenia \\
% ...
$ ^{48}$ Also at Faculty of Physics, University of Bucharest,
          Bucharest, Romania \\
% ...
$ ^e$ Supported by the Deutsche Forschungsgemeinschaft \\
% ...
$ ^m$ Supported by the Romanian National Authority for Scientific Research
      under the contract PN 09370101 \\
 $ ^n$ Partially Supported by Ministry of Science of Montenegro,
      no. 05-1/3-3352 \\
}

\end{flushleft}
%

\newpage
\section{Introduction}
The measurement of the charm production cross section and the derived structure 
function  \ftwocc\ in deep-inelastic electron\footnote{In this paper ``electron''
is used to denote both electron and positron.}-proton scattering (DIS) at 
HERA allows tests of the theory of the strong interaction, 
quantum chromodynamics (QCD). Previous
measurements~\cite{h1dnull,zeusdstar97,h1gluon,zeusdstar00,h1f2c,zeusdstar04,
h1dmesons,h1vertex05,h1vertex06,h1dstardis07,zeusdmesons,zeusdplus09,zeusmu,
h1vertex09,h1dstarhighQ2,zeusdpluslambda,h1cbjets} 
of charm production in DIS at HERA have demonstrated
that charm quarks are predominantly produced by the boson gluon fusion 
process, which is sensitive to the gluon density in the proton. 
The production of charm quarks contributes
up to $30\%$ to the inclusive $ep$ scattering cross section. 
The correct treatment of effects related to the charm quark 
contribution in perturbative QCD calculations, in particular the mass 
effects, is therefore important for the determination of parton 
distribution functions (PDFs). 

At HERA several different techniques have been used to determine 
the charm contribution \ftwocc\ to the proton structure function $F_2$. 
Besides the full reconstruction of a $D$ or 
\dstar\ meson~\cite{h1dnull,zeusdstar97,h1gluon,zeusdstar00,h1f2c,zeusdstar04,
h1dmesons,h1dstardis07,zeusdmesons,zeusdplus09,h1dstarhighQ2,zeusdpluslambda}, 
the lifetime of heavy flavoured
hadrons~\cite{h1dmesons,h1vertex05,h1vertex06,zeusdplus09,h1vertex09,h1cbjets} 
or the semi-leptonic
decay~\cite{zeusmu} are exploited.
Compared to the other methods, the measurement of \dstar\ mesons 
provides a charm sample with a large signal-to-background ratio.
The results presented here are based on a data sample 
collected by the H1 experiment, corresponding to an integrated
luminosity of $348\,{\rm pb}^{-1}$. Increased statistics, extended phase space, 
as well as reduced systematic uncertainties compared to the previous 
H1 analysis~\cite{h1dstardis07} make more detailed
tests of pQCD predictions possible. Compared to earlier H1 analyses 
the phase space for the 
\dstar~meson is extended in transverse momentum from 
$p_T (D^*) > 1.5\ {\rm GeV}$ to $p_T (D^*) > 1.25\ {\rm GeV}$ 
and in pseudo-rapidity from $|\eta (D^*)| < 1.5$ to 
$|\eta (D^*)| < 1.8$. This extension reduces the amount of 
extrapolation needed for the determination of \ftwocc.

\section{QCD Models and Monte Carlo Simulation}
\label{models}

The QCD models employed for data corrections and for comparison with measured
cross sections are introduced in the following. 
Different Monte Carlo (MC) generators based on leading order (LO) QCD 
calculations 
complemented with parton showers are used to simulate detector 
effects in order to determine the acceptance and the efficiency for selecting 
DIS events with a \dstar~meson and to estimate the systematic uncertainties 
associated with the measurements. The generated events are passed through a detailed 
simulation of the H1 detector response based on the GEANT program~\cite{geant} 
and through the same reconstruction and analysis algorithms as used for the 
data.

The MC program RAPGAP~\cite{rapgap} is used for the generation of 
the direct process of boson gluon fusion to a $c\bar{c}$
pair. It uses a LO matrix element with massive charm quarks. 
Parton showers~\cite{partonshower}
based on the DGLAP evolution equations~\cite{DGLAP} model higher order QCD 
effects. The hadronisation of partons is performed with 
PYTHIA~\cite{pythia} which implements the Lund String 
Model~\cite{lundstring}. For the fragmentation 
of the charm quark into the \dstar~meson the Bowler 
parameterisation~\cite{bowler} is chosen and the longitudinal part of the 
fragmentation function is reweighted to the Kartvelishvili
parameterisation~\cite{kartvelish}. The parameter $\alpha$ of the
Kartvelishvili parameterisation is set to the values measured by
H1~\cite{h1dstarfrag} which depend on the centre-of-mass energy squared
\shat\ of the hard subprocess ($\gamma g \rightarrow c \bar{c}$).
The threshold between the two regions in \shat\ is chosen such that 
the mean value of \shat\ in the lower region is in agreement 
with the mean \shat\ of the event sample without a jet associated with the 
\dstar~meson~\cite{h1dstarfrag}.
RAPGAP is interfaced to the HERACLES program~\cite{heracles} in order to 
simulate the 
radiation of a real photon from the incoming or outgoing lepton and virtual 
electro-weak effects. For the determination of the detector acceptance and 
efficiency, RAPGAP is used with the PDF set CTEQ6.6M~\cite{Nadolsky:2008zw} 
which is derived at next-to-leading order (NLO), but gives a good
description of the data. 
Alternatively, RAPGAP is used with CTEQ6LL~\cite{Pumplin:2002vw} 
derived at LO.
The mass of the charm quark is set to $m_c = 1.5\ {\rm GeV}$. 
The renormalisation scale $\mu_r$ and the factorisation scale $\mu_f$ are 
set to $\mu_r = \mu_f = \sqrt{Q^2 + 4m_c^2 + (p^{*}_{T})^2}$, where 
$Q^2$ denotes the photon virtuality and $p^*_{T}$ the transverse momentum 
of the charm quark in the photon-gluon centre-of-mass frame.
The relevant parameter settings and their variations are
summarised in table~\ref{Tab_MC_parameters}. 

\begin{table}[htb]
\begin{center}
\begin{tabular}{|l|l|l|}
\hline
\multicolumn{3}{|c|}{ }\\[-0.3cm]
\multicolumn{3}{|c|}{\bf RAPGAP}\\[0.1cm]\hline
 & & \\[-0.35cm]
Parameter name & Central value & Variation \\[0.05cm]\hline
 & & \\[-0.35cm]
Charm mass & $m_c=1.5\ {\rm GeV}$ &  \\[0.2cm]
Renormalisation scale & $\mu_{r} = \sqrt{Q^2 + 4m_c^2 + (p^*_{T})^2}$ & \\[0.2cm]
Factorisation scale & $\mu_{f} = \sqrt{Q^2 + 4m_c^2 + (p^*_{T})^2}$ & \\[0.2cm]
 & $\alpha = 10.3$ for $\shat < \shat_{threshold}$  & $8.7 < \alpha < 12.2$ \\ 
Fragmentation & $\alpha = \ \ 4.4$ for $\shat > \shat_{threshold}$ & $3.9 < \alpha < 5.0$ \\
 & $\shat_{threshold} = 70 \ {\rm GeV}^2$ & 
   $50 < \shat_{threshold} < 90 \ {\rm GeV}^2$ \\[0.2cm]
PDF & CTEQ6.6M &  CTEQ6LL\\[0.05cm]
\hline
\multicolumn{3}{|c|}{ }\\[-0.3cm]
\multicolumn{3}{|c|}{\bf CASCADE}\\[0.1cm]\hline
 & & \\[-0.35cm]
Parameter name & Central value & Variation \\[0.05cm]\hline
 & & \\[-0.35cm]
Charm mass & $m_c=1.5\ {\rm GeV}$ &  \\[0.2cm]
Renormalisation scale & $\mu_{r,0} = \sqrt{Q^2 + 4m_c^2 + p_{T}^2}$ & 
   $1/2 < \mu_{r}/\mu_{r,0} <2$\\[0.2cm]
Factorisation scale & $\mu_{f,0} = \sqrt{\hat{s} + Q^2_{T}}$ & 
$1/2 < \mu_{f}/\mu_{f,0} <2$\\[0.2cm]
 & $\alpha=8.4$ for $\shat < \shat_{threshold}$ & $7.3 < \alpha < 9.7$ \\ 
Fragmentation & $\alpha=4.5$ for $\shat > \shat_{threshold}$ & $3.9 < \alpha < 5.1$ \\
 & $\shat_{threshold} = 70 \ {\rm GeV}^2$ & 
   $50 < \shat_{threshold} < 90 \ {\rm GeV}^2$ \\[0.2cm]
PDF & A0 & $\mu_{r}$ variation: A0-, A0+ \\[0.05cm]
\hline
\end{tabular}
\caption{\label{Tab_MC_parameters}
Parameters used in the MC simulations. The central choice of the
renormalisation (factorisation) scale is denoted by $\mu_{r,0}$ ($\mu_{f,0}$). 
The invariant mass squared and the transverse momentum squared of the 
$c\bar{c}$ pair are denoted by $\shat$ and $Q_{T}^2$, respectively, 
$m_c$ is the charm quark mass and $p^*_T$ and $p_T$ are the transverse momentum
of the charm quark in the photon-gluon centre-of-mass frame and in the
electron-proton centre-of-mass frame, respectively.
$\alpha$ is the fragmentation parameter 
in the Kartvelishvili parameterisation. Two values of $\alpha$ in 
two regions of $\shat$ with the boundary $\shat_{threshold}$ are 
used~\cite{h1dstarfrag}.
For CASCADE different PDF sets are available which were determined for a 
variation of the renormalisation scale by a factor of $1/2$ or $2$. These 
are used consistently for the $\mu_r$ variation here.
}
\end{center}
\end{table}

The CASCADE~\cite{cascade} program is based on 
$k_T$-factorisation and the CCFM evolution equations~\cite{CCFM}. In CASCADE 
the direct boson gluon fusion process is implemented using off-shell matrix 
elements convolved with a $k_T$-unintegrated gluon distribution of the 
proton. The PDF set A0~\cite{a0} is used. 
Time-like parton showers of the charm quark and anti-quark are implemented,
but those from initial state gluons are not. The hadronisation of partons 
is performed in the same way as for RAPGAP. 
When CASCADE is used for the extrapolation to \ftwocc, the renormalisation and 
factorisation scales are varied to estimate the theoretical uncertainty.
For the variation of the renormalisation scale,
the PDF sets A0- and A0+ are used,
which were extracted with the corresponding scale variation~\cite{a0}.
The parameter variations used in CASCADE 
are also listed in table~\ref{Tab_MC_parameters}.

In addition to RAPGAP and CASCADE, the data are also 
compared to predictions of an NLO calculation~\cite{riemersma,Harris} 
based on collinear factorisation and the DGLAP evolution equations. This 
calculation assumes three active flavours ($u, d, s$) 
in the proton (fixed-flavour-number scheme: FFNS) and massive charm quarks 
are produced dynamically via boson gluon fusion. The 
predictions are calculated using the 
program HVQDIS~\cite{Harris}. Corresponding fixed-flavour NLO parton density 
functions 
of the proton, CT10f3~\cite{Lai:2010vv} (with the strong coupling
set to $\alpha_S(M_Z)=0.106$)
and the NLO variant of MSTW2008f3~\cite{Martin:2010db}, are used. 
Charm quarks are fragmented in the $\gamma p$ centre-of-mass frame into 
\dstar~mesons using the Kartvelishvili~\cite{kartvelish} 
parameterisation for the fragmentation function with the value of the 
parameter $\alpha$ as
measured by H1~\cite{h1dstarfrag}. The renormalisation and 
factorisation scales are set to $\mu_r= \mu_f = \sqrt{Q^2 + 4m_c^2}$. The 
value used for the charm mass is $1.5\ {\rm GeV}$. To obtain the theoretical
systematic uncertainty for the extrapolation to \ftwocc\ the parameters 
are varied according 
to table \ref{Tab_hvqdis_variation}. Each of the variations is performed 
independently. The resulting
uncertainties are added in quadrature to obtain a conservative estimate of 
the total theoretical uncertainty.

The results are also compared with a NLO calculation~\cite{zmvfns} based on 
the zero-mass variable-flavour-number scheme (ZM-VFNS), 
where the charm quark is considered as a massless constituent of the proton.
This calculation is only valid for a sufficiently large transverse
momentum of the \dstar~meson $p_{T}^*(\dstar)$ in the $\gamma p$ 
centre-of-mass frame. For the comparison to this prediction
the analysis is therefore restricted to $p_{T}^*(\dstar) >2\ {\rm GeV}$. 
The ZM-VFNS uses the fragmentation function
determined in~\cite{KKKS}. The scales are chosen to be 
$\mu_r=\mu_f = \sqrt{(Q^2 +(p^*_{T})^2)/2}$.

\begin{table}[htd]
\begin{center}
\begin{tabular}{|l|l|l|}
\hline
\multicolumn{3}{|c|}{ }\\[-0.3cm]
\multicolumn{3}{|c|}{\bf HVQDIS}\\[0.1cm]\hline
 & & \\[-0.35cm]
Parameter name & Central value & Variation \\[0.05cm]\hline
 & & \\[-0.35cm]
Charm mass & $m_c=1.5\ {\rm GeV}$ & $1.3 < m_c < 1.7\ {\rm GeV}$ \\[0.2cm]
Renormalisation scale & $\mu_{r,0} = \sqrt{Q^2 + 4m_{c}^{2}}$ & 
       $1/2 < \mu_{r}/\mu_{r,0} <2$\\[0.2cm]
Factorisation scale & $\mu_{f,0} = \sqrt{Q^2 + 4m_{c}^{2}}$ & 
       $1/2 < \mu_{f}/\mu_{f,0} <2$\\[0.2cm]
 & $\alpha=6.1$ for $\shat < \shat_{threshold}$ & $5.3 < \alpha < 7.0$ \\ 
Fragmentation & $\alpha=3.3$ for $\shat > \shat_{threshold}$ & $2.9 < \alpha < 3.7$ \\
 & $\shat_{threshold} = 70 \ {\rm GeV}^2$ & 
   $50 < \shat_{threshold} < 90 \ {\rm GeV}^2$ \\[0.2cm]
PDF & CT10f3 & MSTW2008f3 \\[0.2cm]
Fragmentation fraction & $f(c\rightarrow\dstar) = 23.8\pm0.8\%
$~\cite{gladilin} & \\[0.05cm]
\hline
\multicolumn{3}{|c|}{ }\\[-0.3cm]
\multicolumn{3}{|c|}{\bf ZM-VFNS}\\[0.1cm]\hline
 & & \\[-0.35cm]
Parameter name & Central value & Variation \\[0.05cm]\hline
 & & \\[-0.35cm]
Charm mass & $m_c=1.5\ {\rm GeV}$ &  \\[0.2cm]
Renormalisation scale & $\mu_{r,0} = \sqrt{(Q^2 +(p^*_{T})^2)/2}$ & 
        $1/2 < \mu_{r}/\mu_{r,0} <2$\\[0.2cm]
Factorisation scale & $\mu_{f,0} = \sqrt{(Q^2 +(p^*_{T})^2)/2}$ & 
        $1/2 < \mu_{f}/\mu_{f,0} <2$\\[0.2cm]
Fragmentation & KKKS08 \cite{KKKS} &  \\[0.2cm]
PDF & CTEQ6.6M & \\[0.05cm]
\hline
\end{tabular}
\end{center}
\caption{Parameter variations used for the uncertainty estimation of the NLO 
calculations. The central choice of the
renormalisation (factorisation) scale is denoted by $\mu_{r,0}$ ($\mu_{f,0}$).
$m_c$ is the charm quark mass and $\alpha$ is the fragmentation 
parameter in the Kartvelishvili parameterisation. 
In the two regions of $\shat$, separated by the
boundary $\shat_{threshold}$, different values of $\alpha$ are 
used~\cite{h1dstarfrag}.
}
\label{Tab_hvqdis_variation}
\end{table}

\section{H1 Detector}
A detailed description of the H1 detector can be found 
elsewhere~\cite{h1detector,spacal}.
Only the components essential to the present analysis are described here.
The origin of the H1 coordinate system is the nominal $ep$ interaction point. 
The direction of the proton beam defines the positive $z$--axis (forward 
direction). Transverse momenta are measured in the $x$--$y$ plane. 
Polar~($\theta$) and~azimuthal~($\phi$) angles are measured with respect 
to this reference system. The pseudo-rapidity is defined as 
$\eta= -\ln{\tan (\theta/2)}$.

Charged particles are measured within the central tracking detector (CTD) 
in the pseudo-rapidity range $-1.85 < \eta < 1.85 $. The CTD consists of 
two large cylindrical jet chambers (CJCs), surrounding the silicon 
vertex detector CST~\cite{cst}. The CJCs are separated by a drift chamber which 
improves the $z$ coordinate reconstruction. A multiwire proportional 
chamber~\cite{mwpc}, which is mainly used in the trigger, is situated inside the inner CJC. 
These detectors are arranged concentrically around the interaction region in 
a magnetic field of \mbox{$1.16\ {\rm T}$}. The trajectories of charged
particles are measured with a transverse momentum resolution of
$\sigma(p_T)/p_T \approx 0.5\% \, p_T/{\rm GeV} \oplus 1.5\%$~\cite{ctdresolution}. The interaction
vertex is reconstructed from CTD tracks. The CTD also provides 
triggering information based on track segments measured in the 
CJCs~\cite{dcrphi,ftt}
and a measurement of the specific ionisation energy loss ${\rm d}E/{\rm d}x$
of charged particles.
The forward tracking detector measures tracks of charged particles at 
smaller polar angle \mbox{($1.5 < \eta <2.8$)} than the central tracker.
 
Charged and neutral particles are measured in the liquid argon (LAr) 
calorimeter, which surrounds the tracking chambers and covers the range 
$-1.5 < \eta < 3.4$ with full azimuthal acceptance. Electromagnetic shower 
energies are measured with a precision of $\sigma(E)/E = 12\% / \sqrt{E/{\rm GeV}} \oplus 1\%
$ and hadronic energies with $\sigma(E)/E = 50\% / \sqrt{E/{\rm GeV}} \oplus 2\%
$, as determined in test beam measurements~\cite{h1testbeam}.
A lead-scintillating fibre calorimeter 
(SpaCal)~\cite{spacal} covering the backward region $-4.0 < \eta < -1.4$ 
completes the measurement of charged and neutral particles. In this analysis 
the SpaCal is used in particular for the identification and reconstruction of the 
scattered electron. For electrons a relative energy resolution of 
$\sigma(E)/E = 7\% / \sqrt{E/{\rm GeV}} \oplus 1\%
$ is achieved, as determined in test beam measurements~\cite{spacaltestbeam}. 
The SpaCal provides energy and time-of-flight information 
used for triggering purposes. 
A Backward Proportional Chamber (BPC) in front of the SpaCal is used to
improve the angular measurement of the scattered electron. 

The hadronic final state is reconstructed using an energy flow algorithm 
which combines charged particles measured in the CTD and the forward
tracking detector with information from 
the SpaCal and LAr calorimeters~\cite{hadroo2}. 

The luminosity determination is based on the measurement of the 
Bethe-Heitler process ($ep \rightarrow ep \gamma$) where the photon is 
detected in a calorimeter located at $z=-103\ {\rm m}$ downstream of 
the interaction region in the electron beam direction.  

\section{Event Selection and Signal Extraction}
The data sample corresponds to an integrated luminosity
${\cal L}=348~{\rm pb}^{-1}$ and was recorded with the H1 detector 
in $e^+ p$ ($185~{\rm pb}^{-1}$) and $e^- p$ interactions
($163~{\rm pb}^{-1}$) in the years 2004 to 2007. 
During this period electrons at an energy of 
\mbox{$27.6\ {\rm GeV}$} were collided with protons at 
\mbox{$920\ {\rm GeV}$}. The events were triggered by a local energy 
deposit in the SpaCal in coincidence with at least one track in the CTD,
with an overall trigger efficiency of $98\%$. 

DIS events are selected by requiring a high energy electromagnetic 
cluster in the SpaCal which is consistent with resulting from the scattered 
electron. 
The event kinematics including the photon virtuality $Q^2$, the Bjorken 
scaling variable $x$ and the inelasticity variable $y$ are reconstructed
with the $e\Sigma$ method~\cite{Bassler:1994uq}, which uses information 
from the scattered electron and the hadronic final state and provides good 
resolution in the covered $y$ range.
The kinematic region for the photon virtuality is restricted to 
$5 < Q^2 < 100\ {\rm GeV}^2$ corresponding to the geometric acceptance of the SpaCal.
In order to ensure a high trigger efficiency, the energy of the electron 
candidate is required to fulfil $E_e^\prime>10\ {\rm GeV}$. The 
inelasticity is restricted to the range $0.02 < y < 0.7$.

Charm production is identified by the reconstruction
of \dstar\ mesons using the decay channel
$D^{*\pm} \rightarrow D^0 \pi_{\rm s}^\pm  \rightarrow K^\mp \pi^{\pm} 
\pi_{\rm s}^\pm$ which has a branching fraction ${\cal BR}=2.63 \pm 0.04\% 
$~\cite{pdg10}. The tracks of the decay particles are 
reconstructed in the CTD. The invariant mass of the $K^\mp \pi^\pm$ system 
is required to be consistent with the $D^0$ mass~\cite{pdg10} within 
$\pm 80\,{\rm MeV}$.
A loose particle identification criterion is applied to the kaon candidates 
using the measurement of the specific energy loss, ${\rm d}E/{\rm d}x$, in 
the CTD.
This improves the signal-to-background ratio, especially at low transverse 
momenta of the \dstar\ meson. The kinematic range of the measurement is
summarised in table~\ref{tab:range}. Details of the selection are described
in~\cite{thesis}.

\begin{figure}[htbp]
\begin{center}
\includegraphics[width=12cm, angle=0]{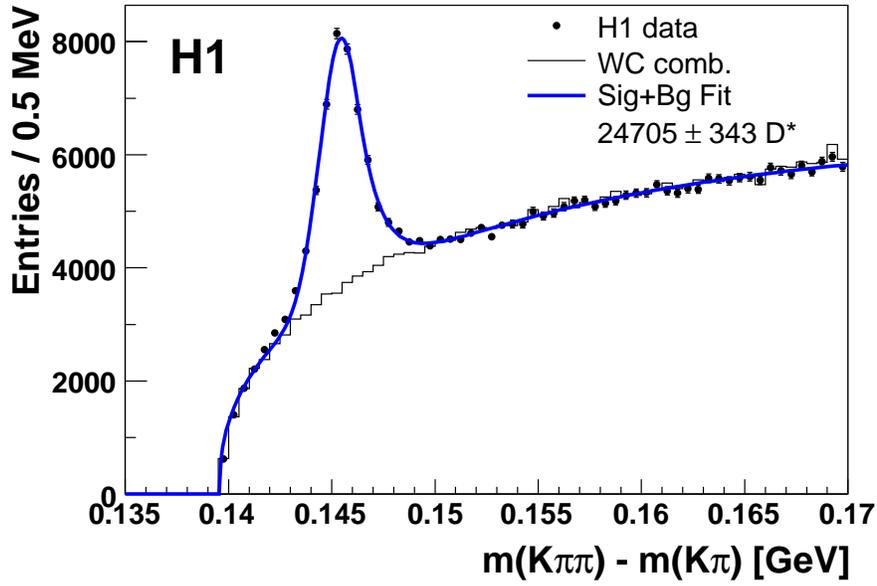}
\end{center} 
\caption{Distribution of the mass difference $\DM=m(K^\mp \pi^{\pm} 
\pi_{\rm s}^\pm)-m(K^\mp \pi^{\pm})$ for \dstar~candidates in the 
kinematic range of $5 < Q^2 < 100\ {\rm GeV}^2$,
$0.02 < y < 0.7$, $|\eta(\dstar)|<1.8$ and 
$\pt(\dstar) > 1.25\ {\rm GeV}$. The histogram shows the wrong charge 
combinations, $K^{\pm}\pi^\pm \pi_{\rm s}^{\mp}$.
The solid line represents the result of the fit described in the text.}
\label{SignalDist}
\end{figure}

\dstarpm\ candidates are selected using the mass difference 
method~\cite{deltammethod}. The distribution of the mass difference 
$\DM=m(K^\mp \pi^{\pm} \pi_{\rm s}^\pm)-m(K^\mp \pi^{\pm})$ 
is shown in figure~\ref{SignalDist}. 
A clear signal peak around the nominal mass difference of 
$145.4 \,{\rm MeV}$~\cite{pdg10} is observed. 

The wrong charge combinations, defined as $K^{\pm}\pi^\pm \pi_{\rm s}^{\mp}$ 
with $K^\pm \pi^\pm$ pairs in the accepted $D^0$ mass range, are used to 
constrain the shape of the combinatorial background in the signal region. 
The number of \dstar~mesons is determined in each analysis bin by a 
simultaneous fit to the right and the wrong charge \DM~distribution. 
As the signal has a tail towards larger \DM~values, the asymmetric 
Crystal Ball function~\cite{Gaiser} is used for the signal description.  
The shape of the background is parameterised with the Granet 
function~\cite{Granet}. 
An unbinned extended log likelihood 
fit~\cite{barlow} 
is performed using the RooFit~framework~\cite{Verkerke}. 

The fit to the complete data set yields $24705 \pm 343$ \dstar~mesons. This 
represents an increase in statistics of an order of magnitude compared to the previous 
analysis~\cite{h1dstardis07}. For each analysis bin the fit to the 
\DM~spectrum uses the two parameters describing the signal asymmetry 
obtained from the fit to the complete data set. The width of the peak varies in 
dependence of the \dstar\ kinematics and is therefore treated as a free 
parameter of the fit. 

\renewcommand{\arraystretch}{1.2}
\begin{table}[htb]
\begin{center}
\begin{tabular}[t]{|l|l|} \hline
    Photon virtuality $Q^2$ & $5<Q^2<100\ {\rm GeV}^2$  \\
    Inelasticity $y$ & $0.02<y<0.7$  \\
    Pseudo-rapidity of \dstarpm & $-1.8<\eta(D^*)<1.8$  \\
    Transverse momentum of \dstarpm  & $p_{T}(D^*)>1.25\ {\rm GeV}$  \\ \hline
    \end{tabular}
   \caption{Definition of the kinematic range of the present measurement.}
   \label{tab:range}
\end{center}
\end{table}
\renewcommand{\arraystretch}{1.0}


\section{Cross Section Determination and Systematic Errors}
\label{toc:sysError}
 
The following formula is used to calculate the inclusive \dstar~meson 
production cross section at the Born level in the visible kinematic range
defined in table~\ref{tab:range}: 
\begin{equation}
\sigma_{\rm vis}(ep \rightarrow e\dstarpm X)  = 
\frac{N(\dstarpm) 
\cdot (1-r)} { {\cal L}   \cdot {\cal BR} (D^* \rightarrow K \pi \pis) \cdot
 (1 +\delta_{\rm rad})   }   ~~. 
\label{eqn:totxsec}
\end{equation}
Here $N(\dstarpm)$ is the number of \dstar\ mesons obtained 
using an unfolding procedure defined below, $r$ is the contribution from 
reflections from other decay modes of the $D^0$ meson, ${\cal L}$ is the 
integrated luminosity, ${\cal BR}$ is the branching ratio and 
$\delta_{\rm rad}$ denotes the QED radiative corrections.
For the differential
measurements the cross section is also divided by the bin width. No
bin centre correction is applied.

To obtain $N(\dstar)$ in each measurement bin, the data are corrected for detector 
effects including the trigger efficiency by means of regularised 
matrix unfolding~\cite{unfolding1,unfolding2,unfolding3,TUnfold}. 
The response matrix \matrixA\ which reflects the acceptance 
and the resolution 
of the H1 
detector relates the distributions $\vec{y}_{\rm rec}$ of reconstructed 
variables to distributions $\vec{x}_{\rm true}$ of variables at the 
generator level, $\matrixA \vec{x}_{\rm true} =\vec{y}_{\rm rec}$. 
Each matrix element $A_{ij}$ is the probability for an event originating 
from bin $j$ of $\vec{x}_{\rm true}$ to be measured in 
bin $i$ of $\vec{y}_{\rm rec}$.
The response matrix is determined from simulation and has twice as many bins 
at the reconstruction level as at the generator level in order
to provide detailed information on the probability distribution and to
improve thereby the accuracy of the unfolding procedure. 
The procedure reduces statistical correlations between neighbouring bins
and the influence of model assumptions in the cross section determination. 
Additional bins outside of the kinematic range of this measurement are 
used to provide constraints on the migrations into the measured phase space.

The measured $D^*$ cross section includes decays from $B$ hadrons to 
$D^*$ mesons which are expected to contribute to less than $2\%
$. For the determination of \ftwocc\ the beauty contribution as calculated with
HVQDIS is subtracted.

For the present analysis the systematic error dominates over the statistical 
uncertainty for almost the 
whole phase space. The measurement is statistically limited 
only for large transverse momenta or large photon virtualities. 
The systematic uncertainties are determined in each bin separately 
and are summarised in table \ref{tab:sysError} for the total cross section.
They are 
divided into uncertainties which are bin-to-bin uncorrelated and uncertainties 
which are correlated between the bins. The uncertainties in the 
following are given in percent of the cross section values. 

The following sources for bin-to-bin uncorrelated systematic errors are 
considered:
\begin{description}
\item[Signal Extraction:] Using different parameterisations for the signal 
and background shapes~\cite{thesis} the systematic uncertainty on the signal 
extraction is estimated to be $2\%$. 

\item[Radiative Corrections:] For the correction of the measured cross section 
to the Born level, the HERACLES interface to RAPGAP is used. 
The corrections are of the order of $2.5\%$. An uncertainty of 
$2\%$ is assigned~\cite{h1cbjets}.

\item[Trigger efficiency:] The efficiency of the trigger conditions 
requiring an energy deposition in the SpaCal and a central track is at least
$95\%$. The combined uncertainty is estimated to be $1\%$.

\item[\boldmath $D^0$ mass cut:] The invariant mass resolution of the 
data is not fully reproduced by the MC simulation, leading 
to different efficiencies of the $D^0$ mass cut. 
The difference is evaluated by comparing the width of the $D^0$ signal 
in data and MC and extrapolating to the region outside of the mass cut 
assuming a Gaussian distribution~\cite{thesis}. The dependence on the 
\dstar\ kinematics is studied, and the maximum difference of $1.5\%$ is 
assigned as uncertainty.

\item[Reflections:] The amount of reflections $r$ from decay modes of the
$D^0$ meson other than $D^0 \rightarrow K^\mp \pi^{\pm}$ is determined using 
a RAPGAP MC sample of inclusive charm events and is found to amount to $3.8\%
$. The dependence of $r$ on kinematic quantities is studied in the simulation
and found to be constant within $1\%$, a value which is used as the 
systematic uncertainty. 

\item[Photoproduction background:]  The photoproduction background is 
estimated using a PYTHIA photoproduction MC sample to be less than $0.2\%$, 
which is used as systematic uncertainty. 

\item[\boldmath ${\rm d}E/{\rm d}x$ cut:] The loss of \dstar\ signal events due to the 
 ${\rm d}E/{\rm d}x$ requirement on the kaon track amounts to $3.4\%$ in data.
 The ${\rm d}E/{\rm d}x$ cut is not applied in the simulation, but corrected 
 for in the data by a global factor. 
 The dependence of the cut efficiency on kinematic variables is studied and 
 found to be within $2\%$, which is used as systematic uncertainty.
\end{description}

Where appropriate, the effect of the bin-to-bin correlated systematic 
uncertainties is evaluated by changing the response matrix and repeating 
the unfolding procedure. The following correlated error sources are considered:

\begin{description}

\item[Track finding efficiency:] The systematic error on the track efficiency
of $4.1\%$ per \dstar~meson is the dominant error of this measurement. It arises 
from two contributions: (i) The comparison of the track finding efficiency 
in data and simulation leads to an error of $2\%$ for the slow pion track 
and $1\%$ for the tracks of the $D^0$ daughter particles and is assumed to be 
correlated between the decay particles; (ii) the efficiency with which 
a track can be fitted to the event vertex
leads to a systematic error of $1\%$ per \dstar~meson.
The uncertainty on the track finding efficiency is considered to be half 
correlated between the bins of the measurement.

\item[Luminosity:] The systematic error on the luminosity measurement is 
estimated to be $3.2\%$.

\item[Branching ratio:] The uncertainty on the branching ratio of the 
\dstar\ meson is $1.5\%$ \cite{pdg10}.

\item[Model:] The parton shower model uncertainty is on average $2\%$, estimated by 
taking the difference in cross section obtained using RAPGAP or CASCADE
for the data correction.

\item[PDF:] Using different parton density functions in RAPGAP for the data 
correction leads to an uncertainty of below $1\%$.

\item[Fragmentation:] The parameter of the Kartvelishvili fragmentation 
function in RAPGAP is varied in the range given in table~\ref{Tab_MC_parameters}.
The resulting differences in the cross section are between $1\%$ and $5\%$.

\item[Electron energy:] The systematic uncertainty on the SpaCal energy 
scale is $0.5\%$ which results in a systematic error of typically 
below $1\%$, but up to $10\%$ at large \dstar\ inelasticity $z$ (see 
section~\ref{difwq}). 

\item[Electron angle:] The angular resolution of the SpaCal/BPC of 
$0.5\,{\rm mrad}$ leads to a systematic uncertainty of typically $2\%$. 

\item[Hadronic energy:] The systematic uncertainty on the energy scale of 
the hadronic final state is $2\%$. The influence in general 
is small (below $0.5\%$) but leads to larger uncertainties of up to 
$20\%$ at large \dstar\ inelasticity $z(\dstar)$ and small $y$. 

\end{description}

All sources of the systematic errors are assumed to be uncorrelated between 
the sources and added in quadrature. This results in an overall
systematic uncertainty of $7.6\%$. 

\newcolumntype{L}{l}
\begin{table}[htd]
\begin{center}
\begin{tabular}{|l|r|}
\hline
\multicolumn{2}{|c|}{ }\\[-0.3cm]
\multicolumn{1}{|l}{\bf Uncorrelated uncertainties} &   \\[0.1cm]  \hline
 & \\[-0.35cm]
Signal extraction &      	$2\%$ \\ 
Radiative correction &   	$2\%$ \\ 
Trigger efficiency &     	$1 \%$ \\ 
$D^0$ meson mass cut &   	$1.5\%$ \\ 
Reflections &            	$1.0\%$ \\ 
Photoproduction background &  	$<0.2\%$  \\
${\rm d}E/{\rm d}x$ cut &  	$2\%$ \\ [0.05cm]  \hline
\multicolumn{2}{|c|}{ }\\[-0.3cm]
\multicolumn{1}{|l}{\bf Correlated uncertainties} &   \\[0.1cm]  \hline
 & \\[-0.35cm]
Track efficiency &   		$4.1\%$ \\ 
Luminosity & 			$3.2\%$ \\
Branching ratio &  		$1.5\%$   \\ 
Model & 			$2.1\%$ \\ 
PDF &				$1\%$ \\ 
Fragmentation &  		$2.6\%$\\ 
Electron energy scale  & 	$1.3\%$ \\ 
Electron angle $\theta$ &    	$1.3\%$  \\ 
Hadronic energy scale &    	$0.3\%$  \\ [0.05cm]  \hline \hline
 & \\[-0.3cm]
{\bf Total systematic uncertainty} &  $7.6\%$\\ [0.1cm]  \hline
\end{tabular}
\end{center}
\caption{Summary of all sources of systematic uncertainties and their 
effect on the \dstar\ production cross section with the breakdown into 
bin-to-bin uncorrelated and bin-to-bin correlated sources.
}
\label{tab:sysError}
\end{table}

\section{Cross Sections}\label{difwq}

The cross section in the visible range defined in table~\ref{tab:range} 
is measured to be:
\begin{equation}
\sigma_{\rm vis} (e p \rightarrow e \dstarpm X  )=
6.44 \pm 0.09~({\rm stat.})  
\pm 0.49~({\rm syst.})~{\rm nb} ~~.
\end{equation}
The corresponding predictions from RAPGAP (CTEQ6LL), RAPGAP (CTEQ6.6M),
and CASCADE (A0) amount to $5.02~{\rm nb}$, $4.37~{\rm nb}$, 
and $5.09~{\rm nb}$, respectively.
The NLO calculation HVQDIS yields
$5.98 ^{+1.10}_{-0.88}~{\rm nb}$ 
with CT10f3 as the proton PDF and
$5.52 ^{+0.94}_{-0.82}~{\rm nb}$ 
with MSTW2008f3, where the uncertainty is determined by varying the 
parameters according to table \ref{Tab_hvqdis_variation} and adding the 
resulting uncertainties in quadrature.
The HVQDIS predictions are slightly below the measurement
but agree with the data within errors.

\begin{figure}[htbp]
\unitlength1.0cm
\begin{picture}(16,14)
\put(0,6){\includegraphics[width=8cm, angle=0]{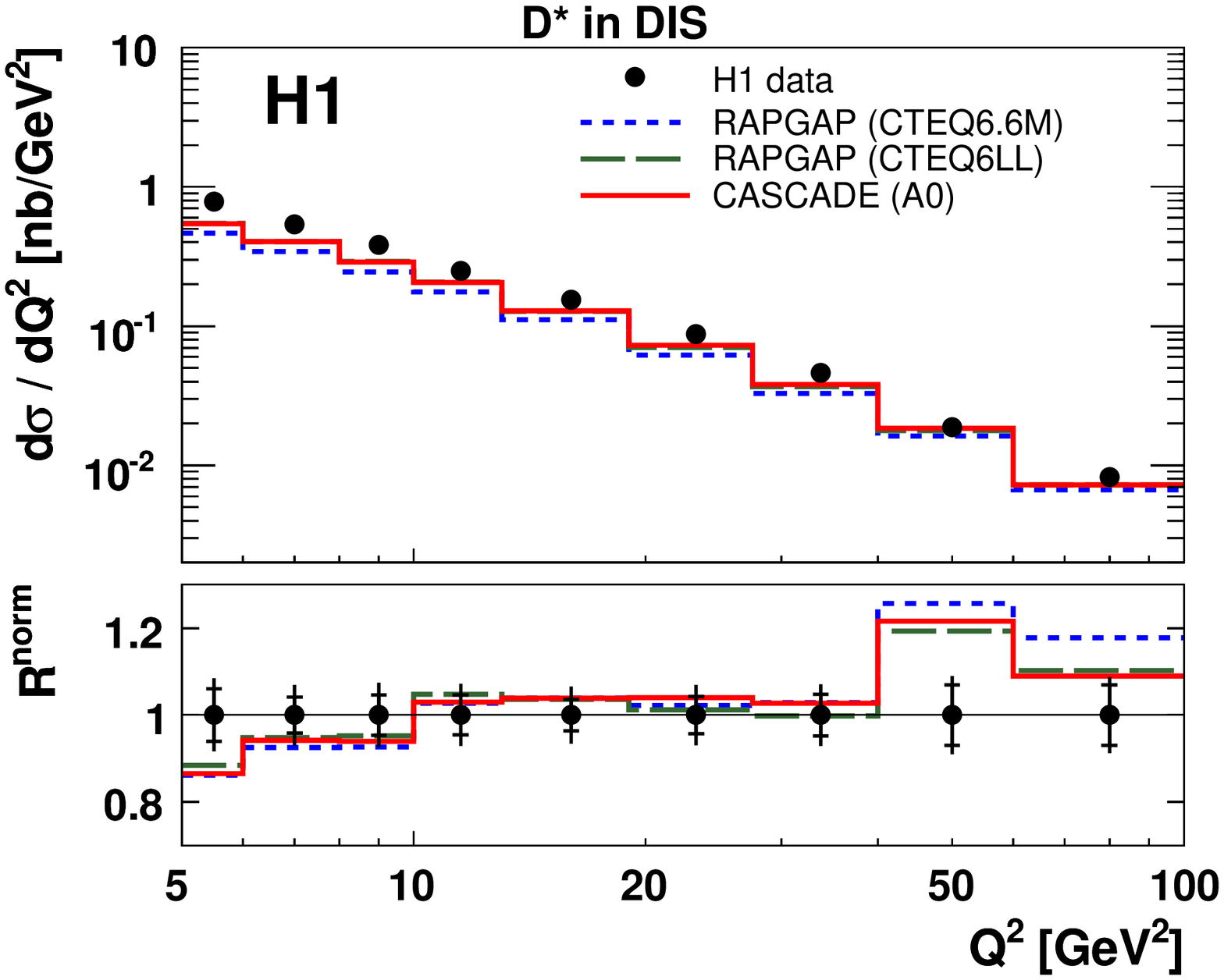}} 
\put(8,6){\includegraphics[width=8cm, angle=0]{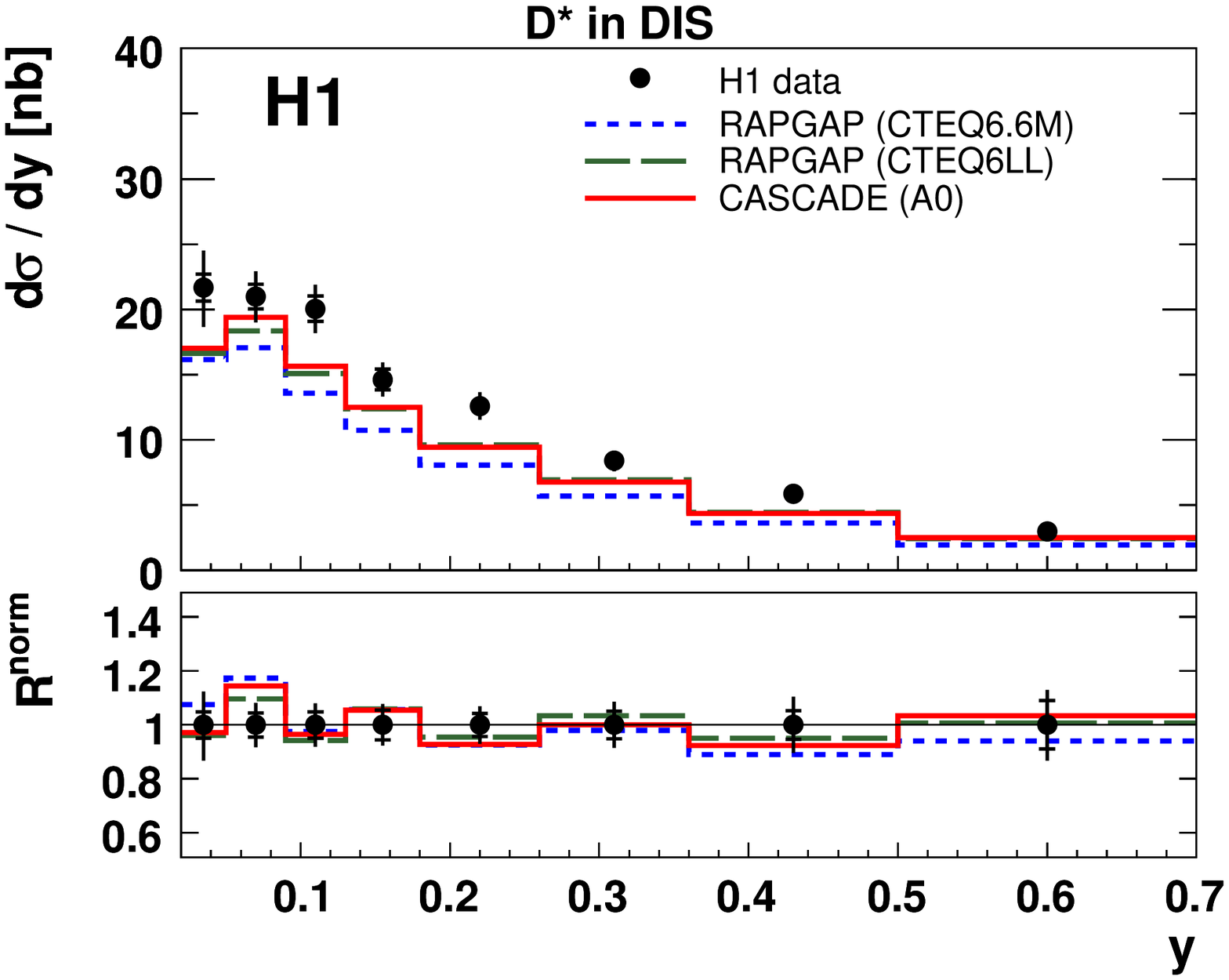}} 
\put(0,-1){\includegraphics[width=8cm, angle=0]{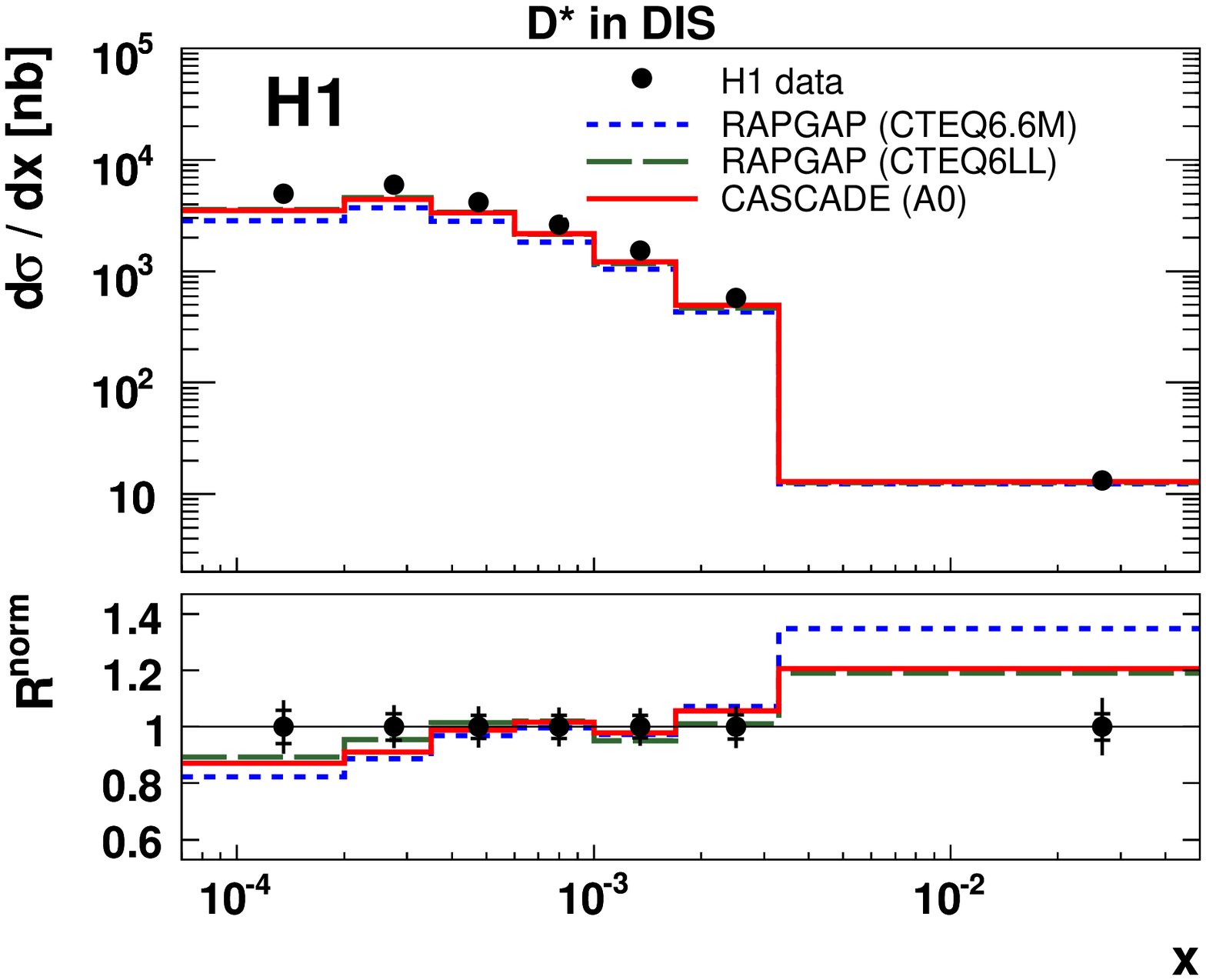}} 
\end{picture}
\caption{Differential \dstar\ cross section as a function of the photon
virtuality $Q^2$, the inelasticity $y$ and Bjorken $x$. 
The measurements correspond to the kinematic range of 
$5 < Q^2 < 100\ {\rm GeV}^2$, $0.02 < y < 0.7$, $|\eta(\dstar)|<1.8$ and
$\pt(\dstar) > 1.25\ {\rm GeV}$. 
The data are shown as points, the inner error bars show the statistical 
error, the outer error bars represent the statistical and systematic errors 
added in quadrature. The data are compared to predictions by the
MC program RAPGAP with two different proton parton densities and by the
MC program CASCADE.
In the lower part of the figures the normalised ratio $R^{\rm norm}$ of theory 
to data (equation \ref{eqn:normXsecs}) is shown, which has reduced
normalisation uncertainties.}
\label{fig:XSectionKinLO}
\end{figure}

\begin{figure}[htbp]
\unitlength1.0cm
\begin{picture}(16,14)
\put(0,6){\includegraphics[width=8cm, angle=0]{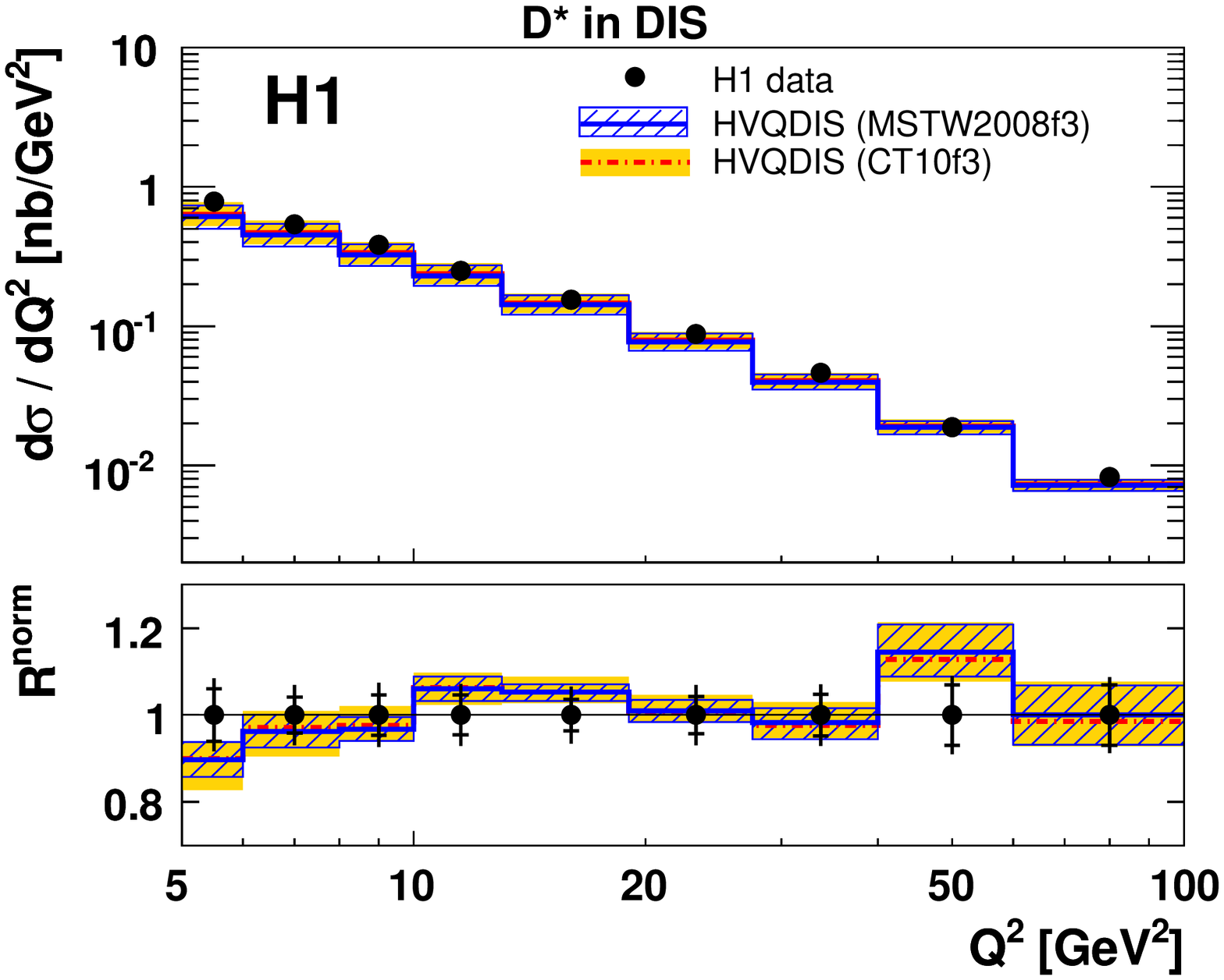}} 
\put(8,6){\includegraphics[width=8cm, angle=0]{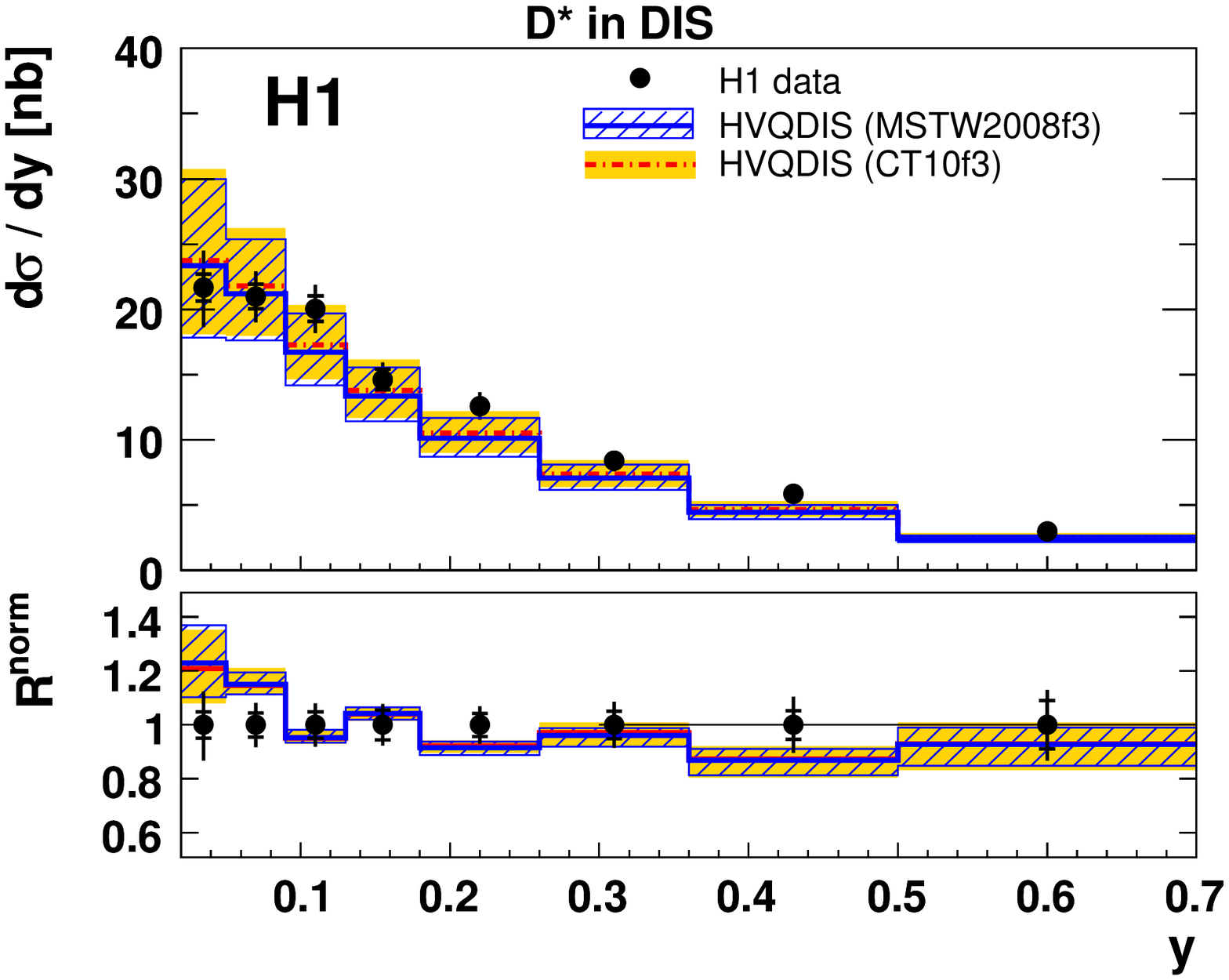}} 
\put(0,-1){\includegraphics[width=8cm, angle=0]{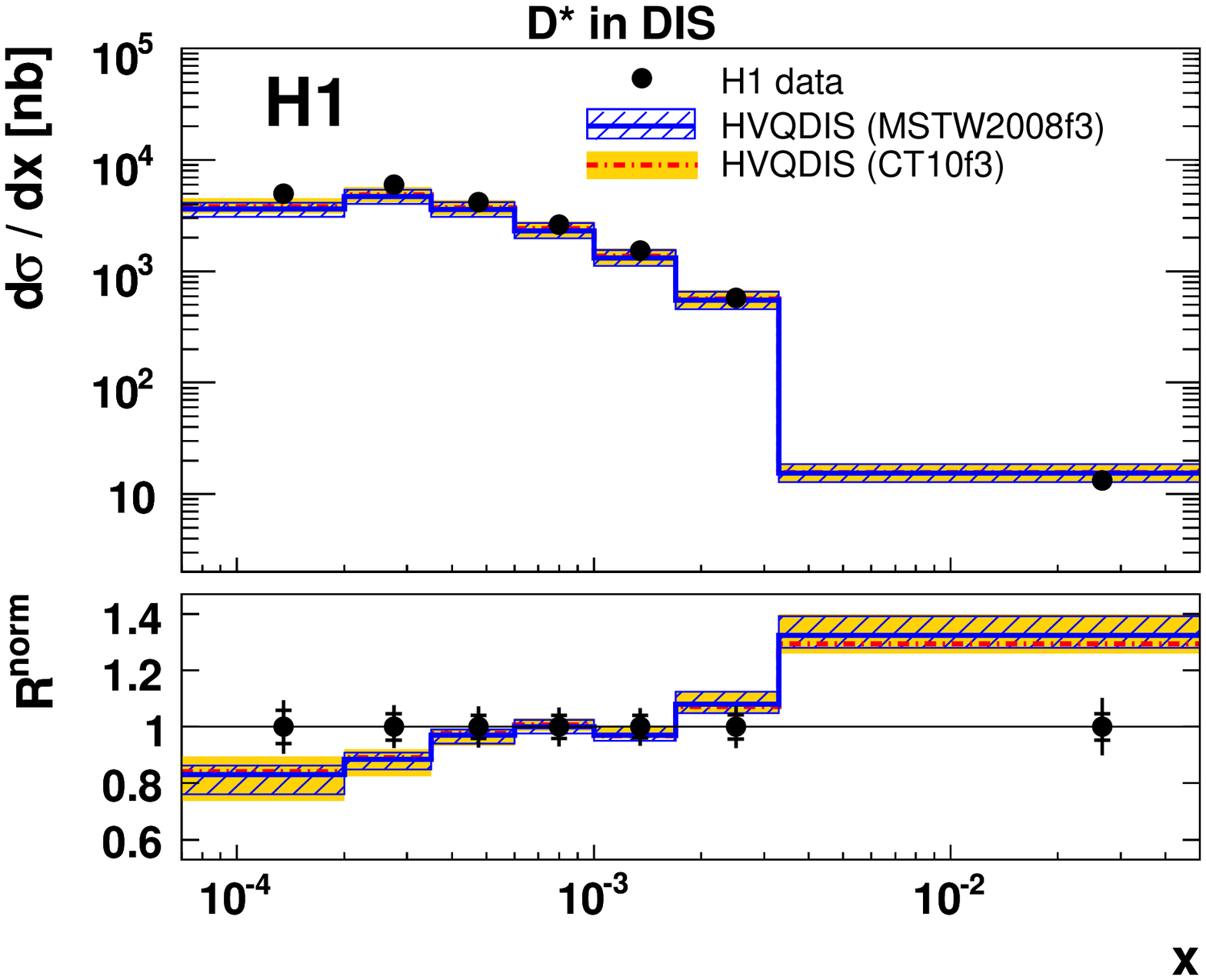}} 
\end{picture}
\caption{Differential \dstar\ cross section as a function of the photon
virtuality $Q^2$, the inelasticity $y$ and Bjorken $x$. 
The measurements correspond to the kinematic range of 
$5 < Q^2 < 100\ {\rm GeV}^2$, $0.02 < y < 0.7$, $|\eta(\dstar)|<1.8$ and
$\pt(\dstar) > 1.25\ {\rm GeV}$. The data are shown as points, the 
inner error bars show the statistical 
error, the outer error bars represent the statistical and systematic errors 
added in quadrature. The data are compared to predictions by the
next-to-leading order calculation HVQDIS with two different proton parton
densities. The bands indicate the theoretical uncertainties 
(table~\ref{Tab_hvqdis_variation}).
In the lower part of the figures the normalised ratio $R^{\rm norm}$ of theory 
to data (equation \ref{eqn:normXsecs}) is shown, which has reduced
normalisation uncertainties.}
\label{fig:XSectionKinNLO}
\end{figure}

In table~\ref{Tab_xsec_kin} and figures~\ref{fig:XSectionKinLO} 
and \ref{fig:XSectionKinNLO} the single 
differential cross sections are presented as a function of variables 
describing the event kinematics: the photon virtuality 
$Q^2$, the inelasticity $y$ and Bjorken $x$.
The measurements are compared to the predictions of the MC
programs RAPGAP and CASCADE (figure~\ref{fig:XSectionKinLO}) and of
the next-to-leading order calculation with the 
HVQDIS program (figure~\ref{fig:XSectionKinNLO}). 
Since the theoretical calculations predict smaller cross 
sections than the measurement, 
the normalised ratio \rnorm\ of theory to data is shown in the lower part 
of the figures, which facilitates the shape comparison between
the different theoretical predictions and the data. This ratio is defined as:
\begin{equation}
\rnorm = \frac{1/\sigma_{\rm vis}^{\rm theo} \cdot 
\frac{{\rm d}\sigma^{\rm theo}}{{\rm d}Y}}
{1/\sigma_{\rm vis}^{\rm data} \cdot 
\frac{{\rm d}\sigma^{\rm data}}{{\rm d}Y}}~~,
\label{eqn:normXsecs}
\end{equation}
where $\sigma_{\rm vis}^{\rm theo}$ and 
${{\rm d}\sigma^{\rm theo}}/{{\rm d}Y}$ are the total and differential cross
section of the model under consideration and $Y$ denotes any measured 
variable.
The normalisation uncertainties of the data (luminosity, branching ratio
and half of the tracking uncertainty) cancel in this ratio.
Similarly, uncertainty sources of the NLO predictions altering only the 
normalisation do not affect \rnorm\ since for each variation the 
total and the differential cross section are varied simultaneously. 
In all predictions the decrease with $Q^2$ is slightly less steep than in data.
The $y$ dependence is reasonably well described by all
predictions. The dependence on $x$ is slightly less steep in all predictions 
than in the data, an effect which is larger for the NLO calculations.

\begin{figure}[htbp]
\unitlength1.0cm
\begin{picture}(16,14)
\put(0,6){\includegraphics[width=8cm, angle=0]{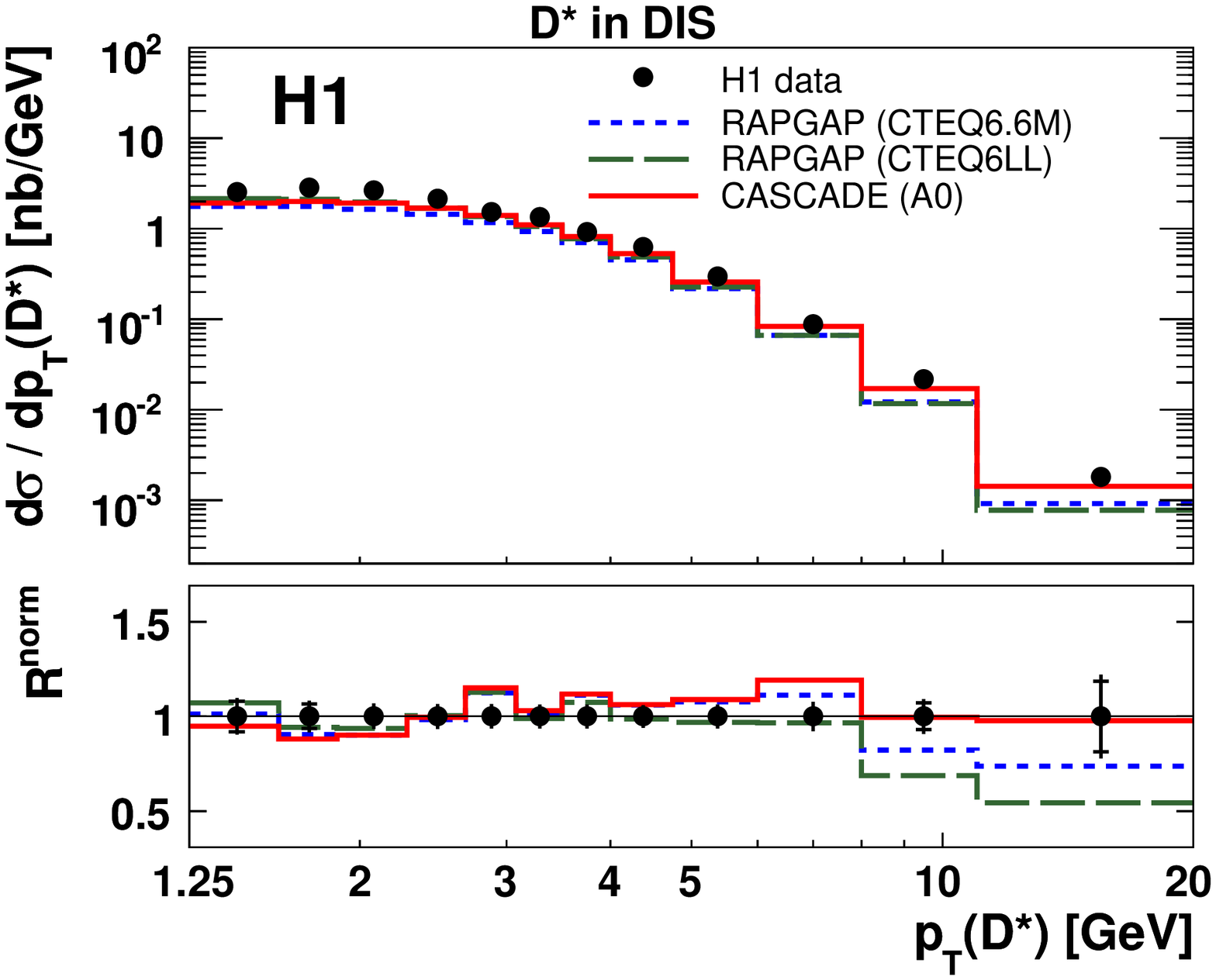}} 
\put(8,6){\includegraphics[width=8cm, angle=0]{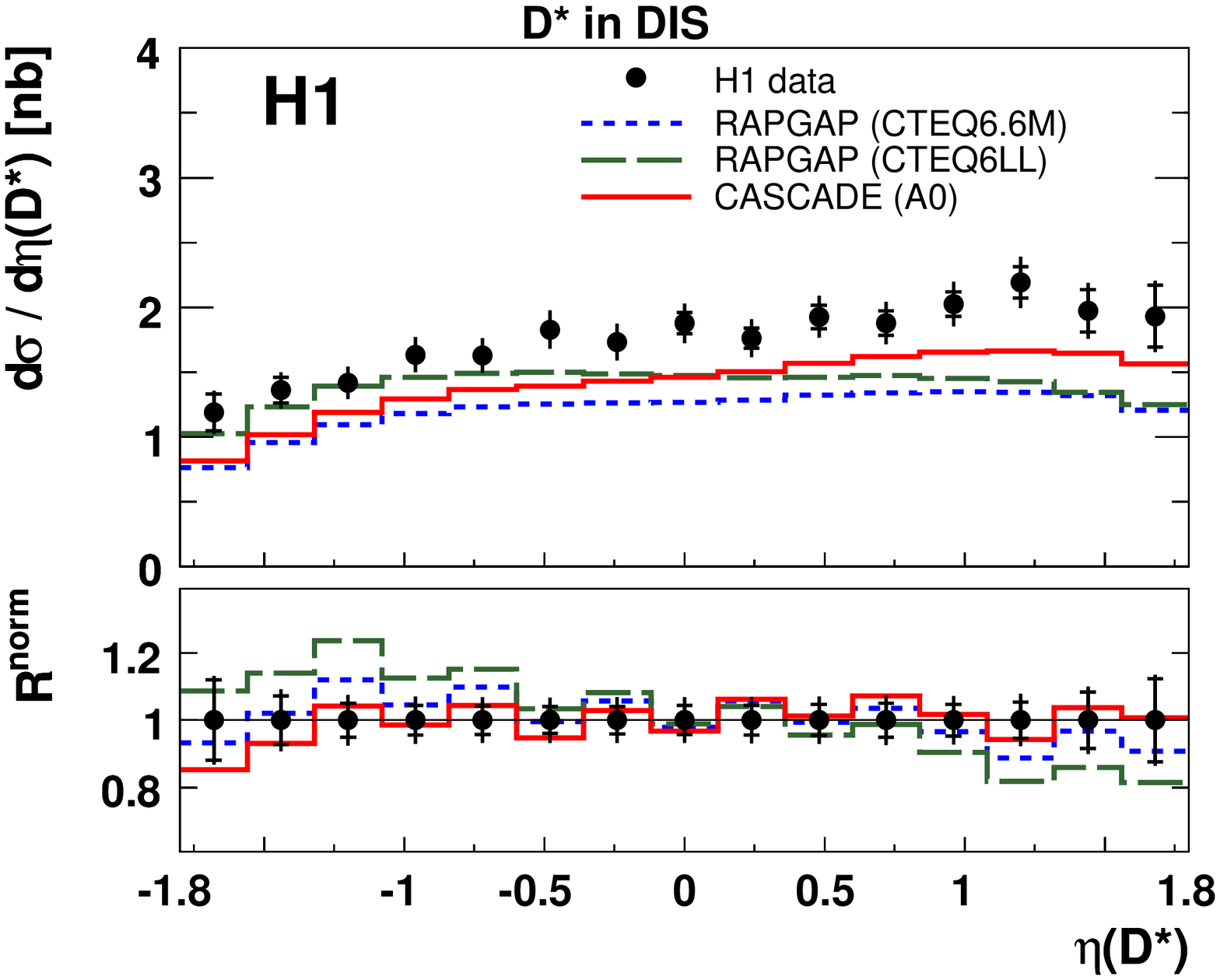}} 
\put(0,-1){\includegraphics[width=8cm, angle=0]{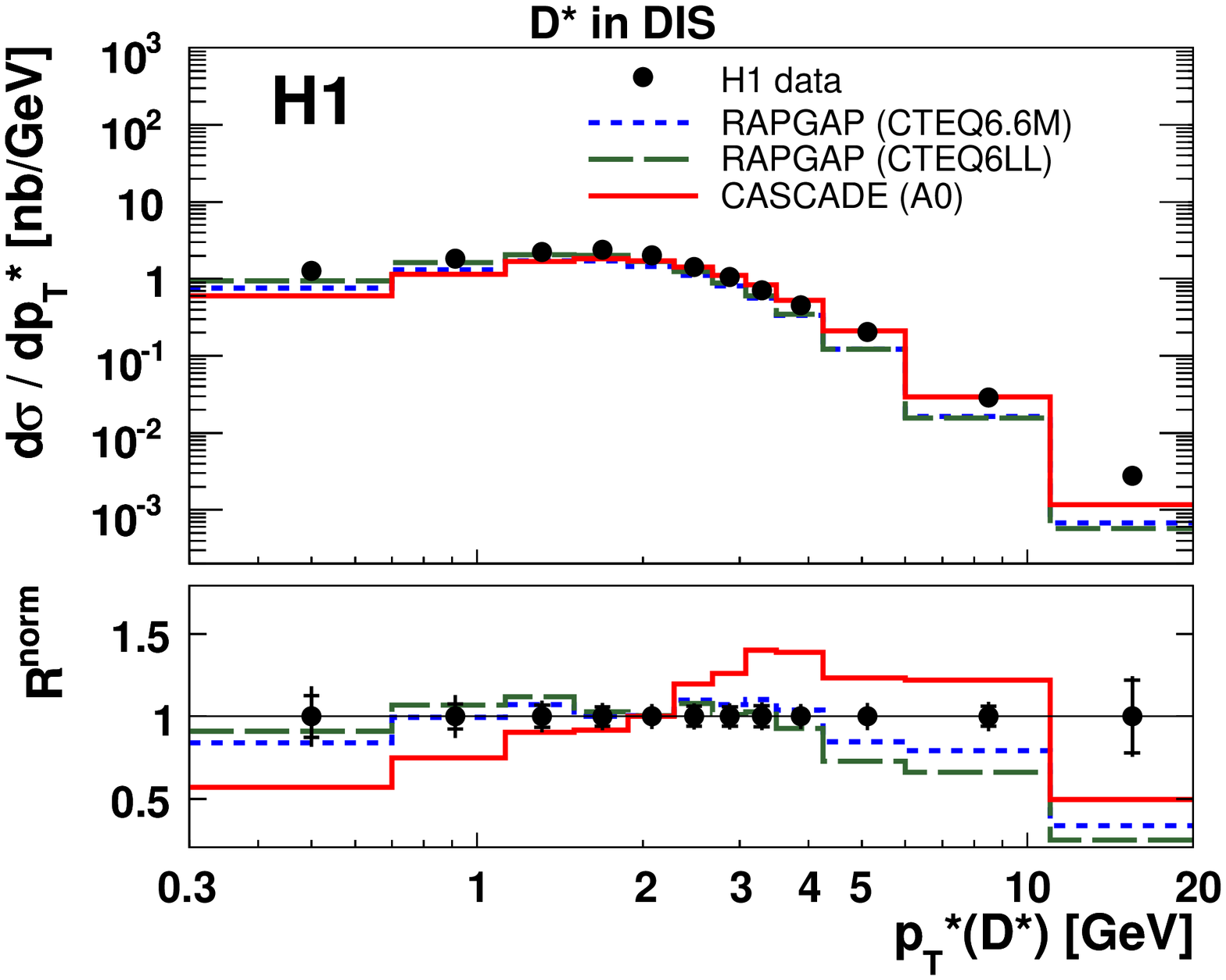}} 
\put(8,-1){\includegraphics[width=8cm, angle=0]{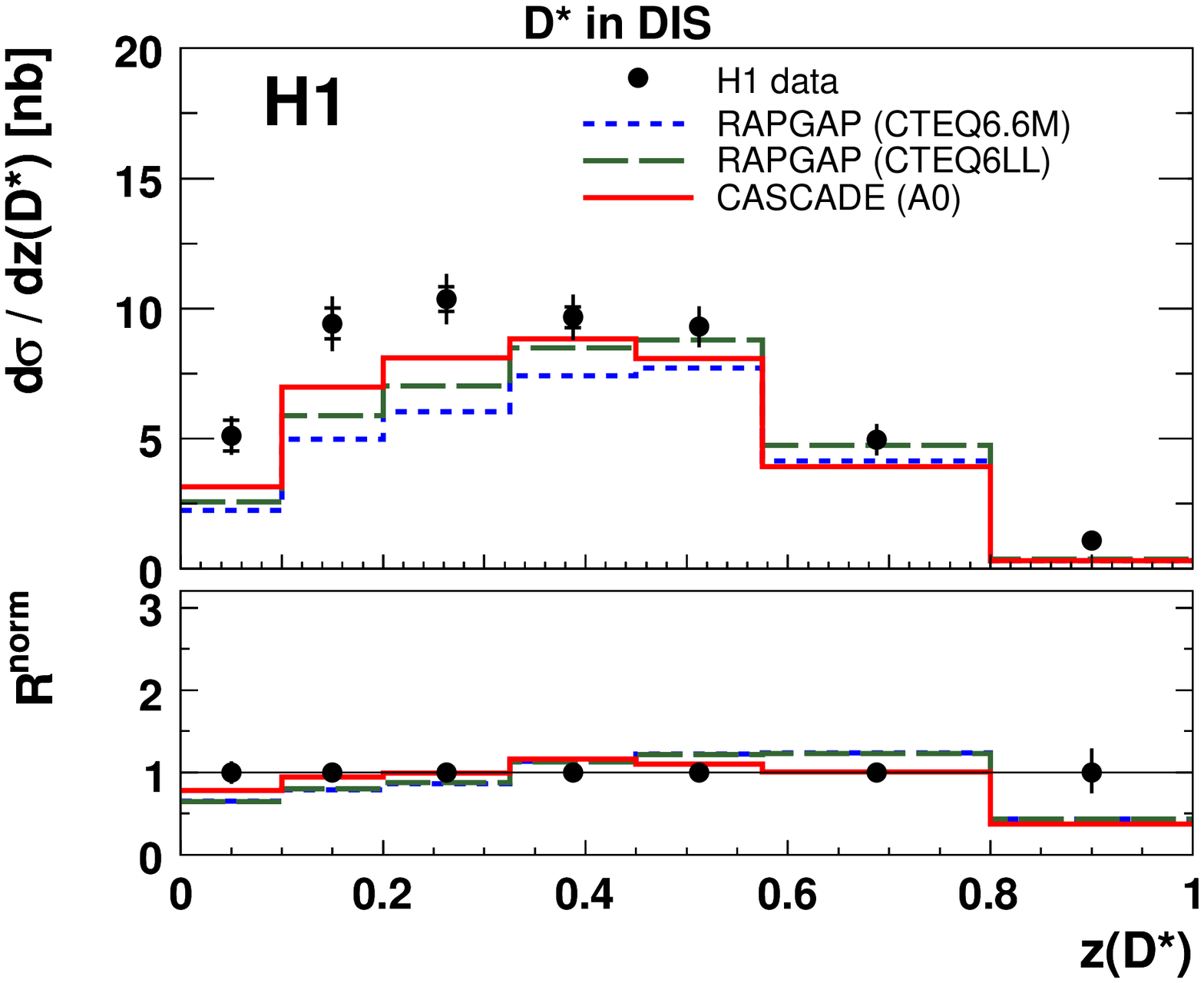}} 
\end{picture}
\caption{Differential \dstar\ cross section as a function of  
the transverse momentum $\pt(\dstar)$ and pseudo-rapidity $\eta(\dstar)$ 
in the laboratory frame, the transverse momentum $\pt^*(\dstar)$ in the 
$\gamma p$ centre-of-mass frame and the \dstar\ inelasticity $z(\dstar)$. 
The measurements correspond to the kinematic range of 
$5 < Q^2 < 100\ {\rm GeV}^2$, $0.02 < y < 0.7$ and $|\eta(\dstar)|<1.8$ and
$\pt(\dstar) > 1.25\ {\rm GeV}$. The data are shown as points, the 
inner error bars show the statistical 
error, the outer error bars represent the statistical and systematic errors 
added in quadrature. The data are compared to predictions by the
MC program RAPGAP with two different proton parton densities and by the
MC program CASCADE.
In the lower part of the figures the normalised ratio $R^{\rm norm}$ of theory 
to data (equation \ref{eqn:normXsecs}) is shown, which has reduced
normalisation uncertainties.}
\label{fig:XSectionDstarLO}
\end{figure}

\begin{figure}[htbp]
\unitlength1.0cm
\begin{picture}(16,14)
\put(0,6){\includegraphics[width=8cm, angle=0]{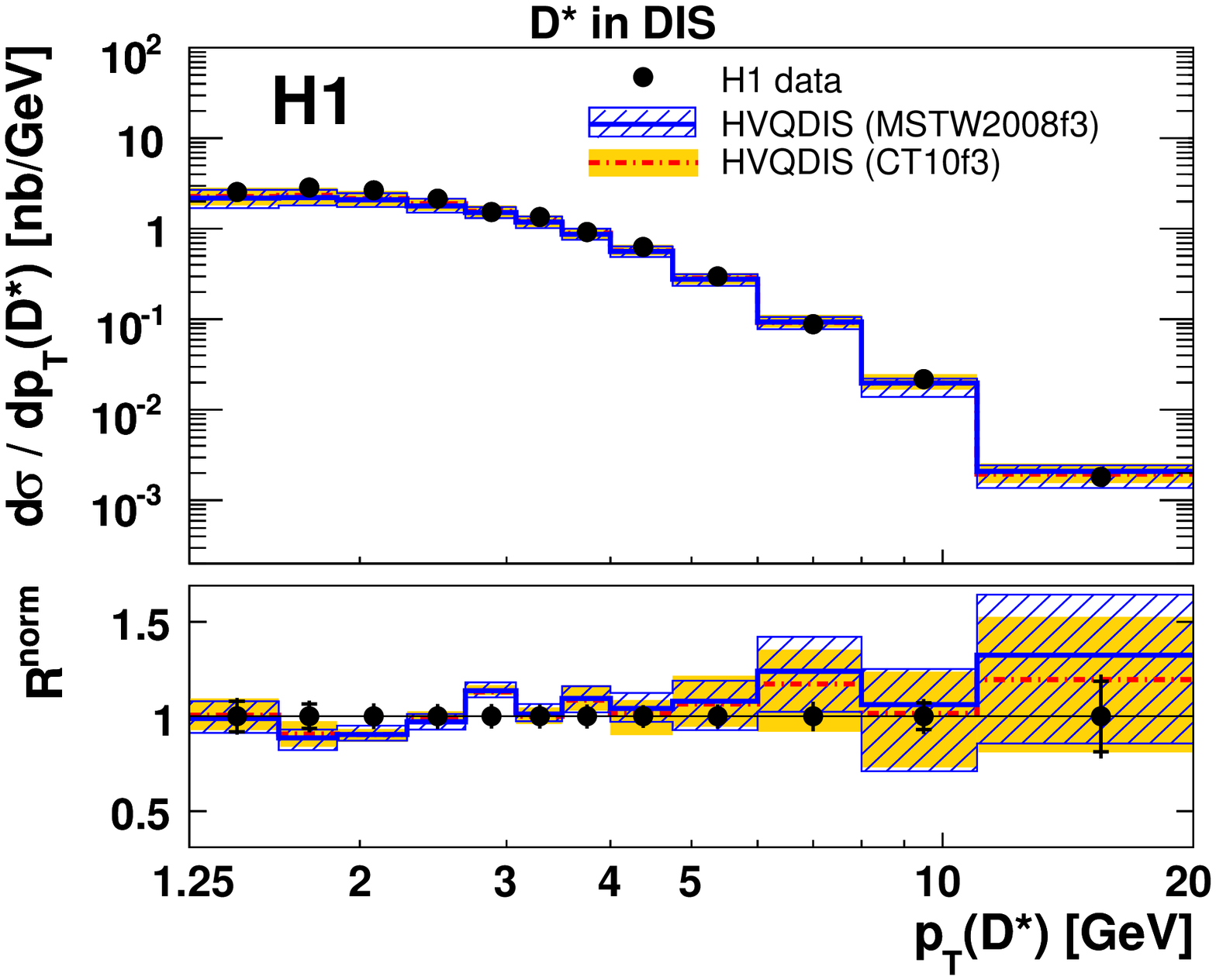}} 
\put(8,6){\includegraphics[width=8cm, angle=0]{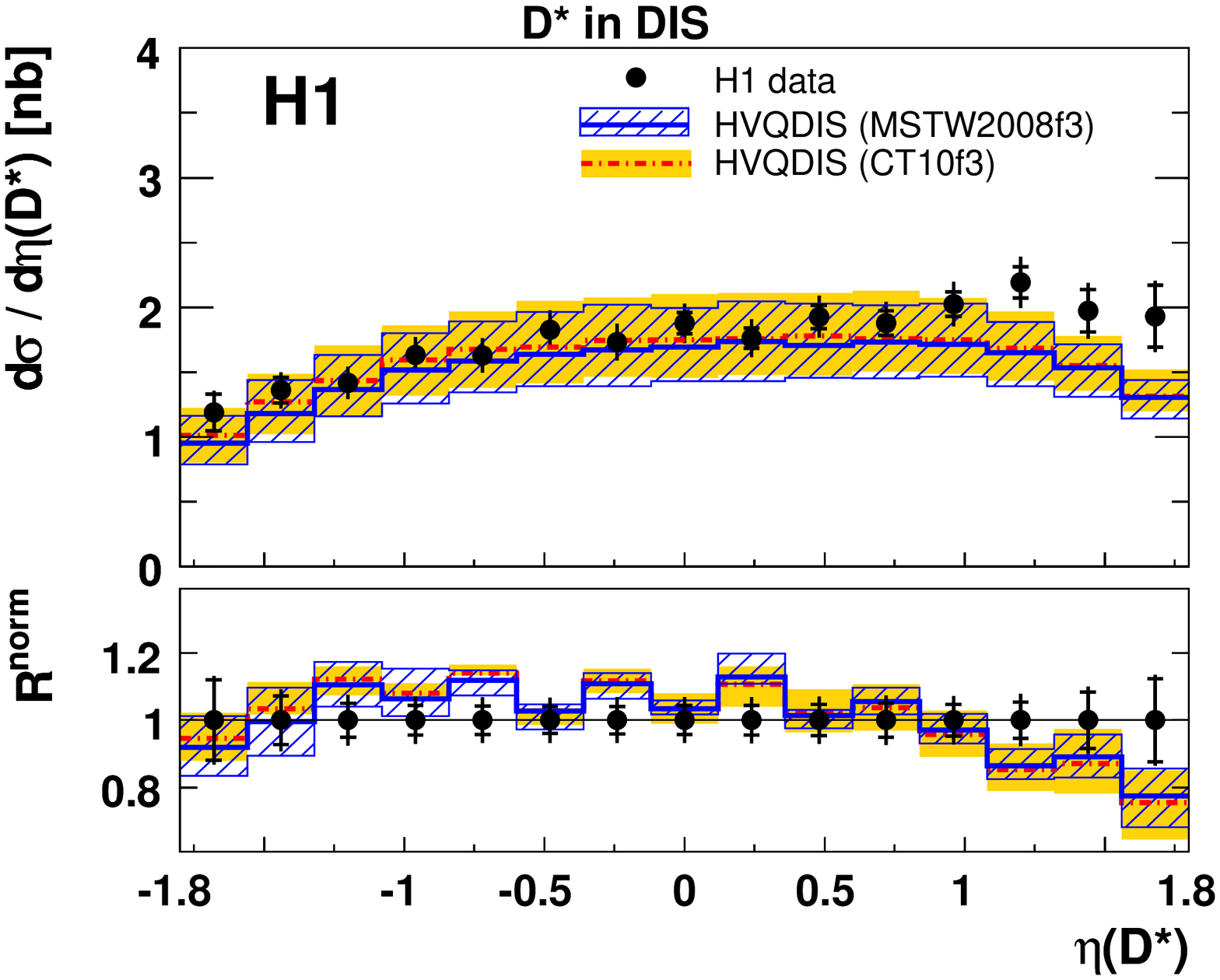}} 
\put(0,-1){\includegraphics[width=8cm, angle=0]{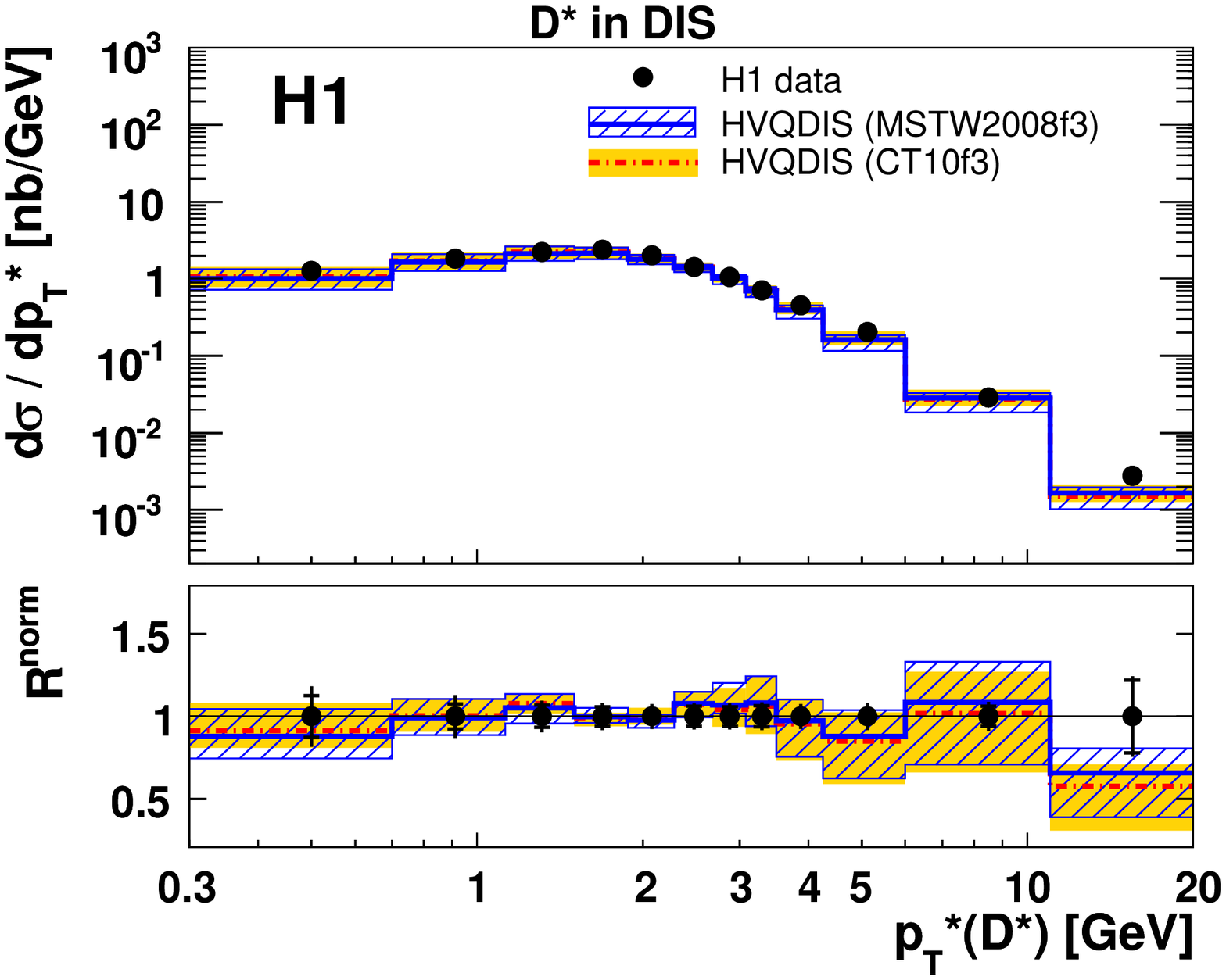}} 
\put(8,-1){\includegraphics[width=8cm, angle=0]{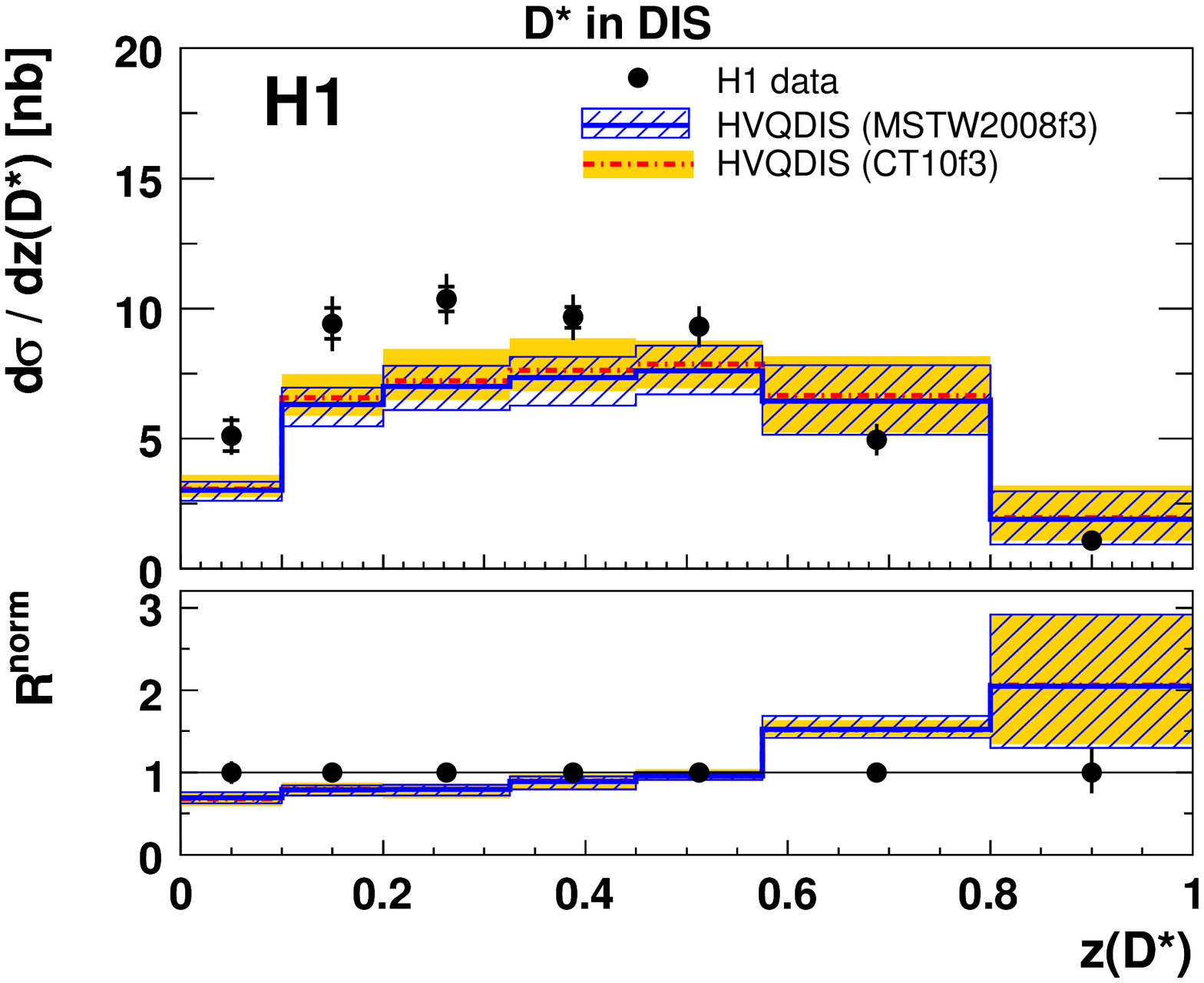}} 
\end{picture}
\caption{Differential \dstar\ cross section as a function of 
the transverse momentum $\pt(\dstar)$ and pseudo-rapidity $\eta(\dstar)$ 
in the laboratory frame, the transverse momentum $\pt^*(\dstar)$ in the 
$\gamma p$ centre-of-mass frame and the \dstar\ inelasticity $z(\dstar)$. 
The measurements correspond to the kinematic range of 
$5 < Q^2 < 100\ {\rm GeV}^2$, $0.02 < y < 0.7$ and $|\eta(\dstar)|<1.8$ and
$\pt(\dstar) > 1.25\ {\rm GeV}$. The data are shown as points, the 
inner error bars show the statistical 
error, the outer error bars represent the statistical and systematic errors 
added in quadrature. The data are compared to predictions by the
next-to-leading order calculation HVQDIS with two different proton parton
densities. The bands indicate the theoretical uncertainties 
(table~\ref{Tab_hvqdis_variation}).
In the lower part of the figures the normalised ratio $R^{\rm norm}$ of theory 
to data (equation \ref{eqn:normXsecs}) is shown, which has reduced
normalisation uncertainties.} 
\label{fig:XSectionDstarNLO}
\end{figure}

In table~\ref{Tab_xsec_dstar1} and figures~\ref{fig:XSectionDstarLO} and 
\ref{fig:XSectionDstarNLO}
the single differential cross sections are presented as a function of the 
kinematic variables of the \dstar~meson: the transverse momentum $\pt(\dstar)$
and pseudo-rapidity $\eta(\dstar)$ in the
laboratory frame, the transverse momentum $\pt^*(\dstar)$ 
in the $\gamma p$ centre-of-mass frame and the \dstar\ inelasticity 
$z(\dstar)$. The \dstar\ 
inelasticity $z(\dstar)$ corresponds to the fraction of the virtual photon 
momentum carried by the \dstar\ meson and 
is determined as $z(\dstar)=(P \cdot p_{\dstar})/(P \cdot q)=
(E-p_z)_{\dstar}/2yE_e$ 
where $E_e$ is the energy of the incoming electron and $P$, $q$ and $p_{\dstar}$ 
denote the four-momenta of the incoming proton, the exchanged photon and 
the \dstar\ meson, respectively. 
All predictions are able to describe the shape of the $\pt(\dstar)$ 
distribution of the data reasonably well, although RAPGAP has a tendency to 
underestimate the data at large $\pt(\dstar)$. The shape of the $\eta$ distribution 
shows sensitivity to the use of different parton densities in the RAPGAP MC. 
The prediction based on CTEQ6.6M agrees better with the data than 
the prediction based on CTEQ6LL. A very 
good description of the $\eta$ shape is obtained with the CASCADE MC. 
The HVQDIS calculations with CT10f3 and MSTW2008f3 both 
 describe the $\eta$ distribution reasonably well, but have a 
tendency to be too low in the positive $\eta$ (forward) 
region.  
For the transverse momentum $\pt^*(\dstar)$ in the $\gamma p$ 
centre-of-mass frame, the RAPGAP MC using either PDF is too steep at large
$\pt^*(\dstar)$, while the CASCADE prediction generally has a different
shape. The NLO predictions are in good agreement with the data.
The $z(\dstar)$ dependence is not described by any of the 
calculations, showing a deficit of all predictions at low $z(\dstar)$ values.

\begin{figure}[htbp]
\includegraphics[width=16cm, angle=0]{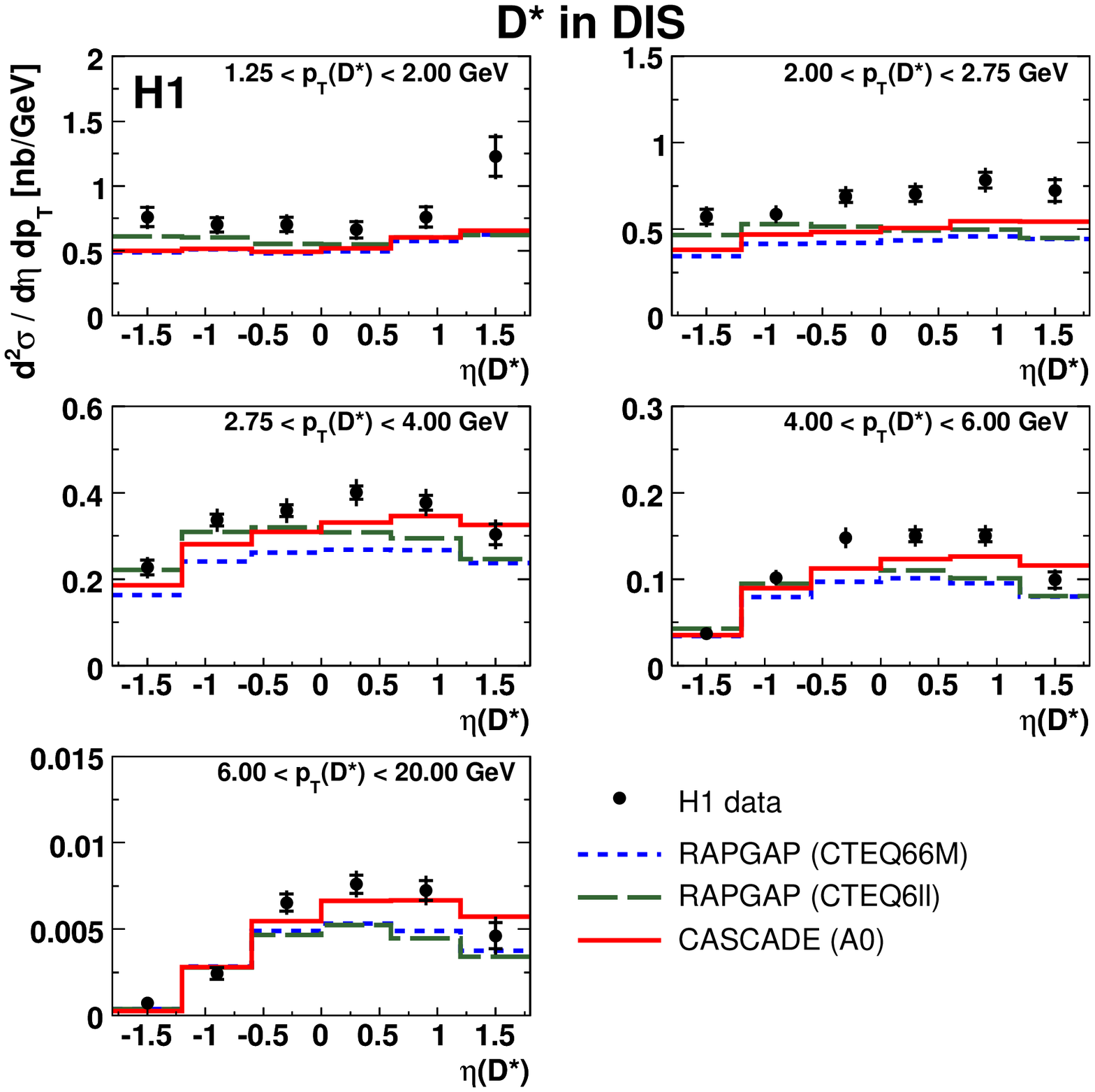} 
\caption{Double differential \dstar\ cross section as a function of 
the transverse momentum $\pt(\dstar)$
and pseudo-rapidity $\eta(\dstar)$ in the
laboratory frame. 
The measurements correspond to the kinematic range of 
$5 < Q^2 < 100\ {\rm GeV}^2$, $0.02 < y < 0.7$ and $|\eta(\dstar)|<1.8$ and
$\pt(\dstar) > 1.25\ {\rm GeV}$. The data are shown as points, the 
inner error bars show the statistical 
error, the outer error bars represent the statistical and systematic errors 
added in quadrature. The data are compared to predictions by the
MC program RAPGAP with two different proton parton densities and by the
MC program CASCADE.} 
\label{fig:XSectionPtEtaLO}
\end{figure}

\begin{figure}[htbp]
\includegraphics[width=16cm, angle=0]{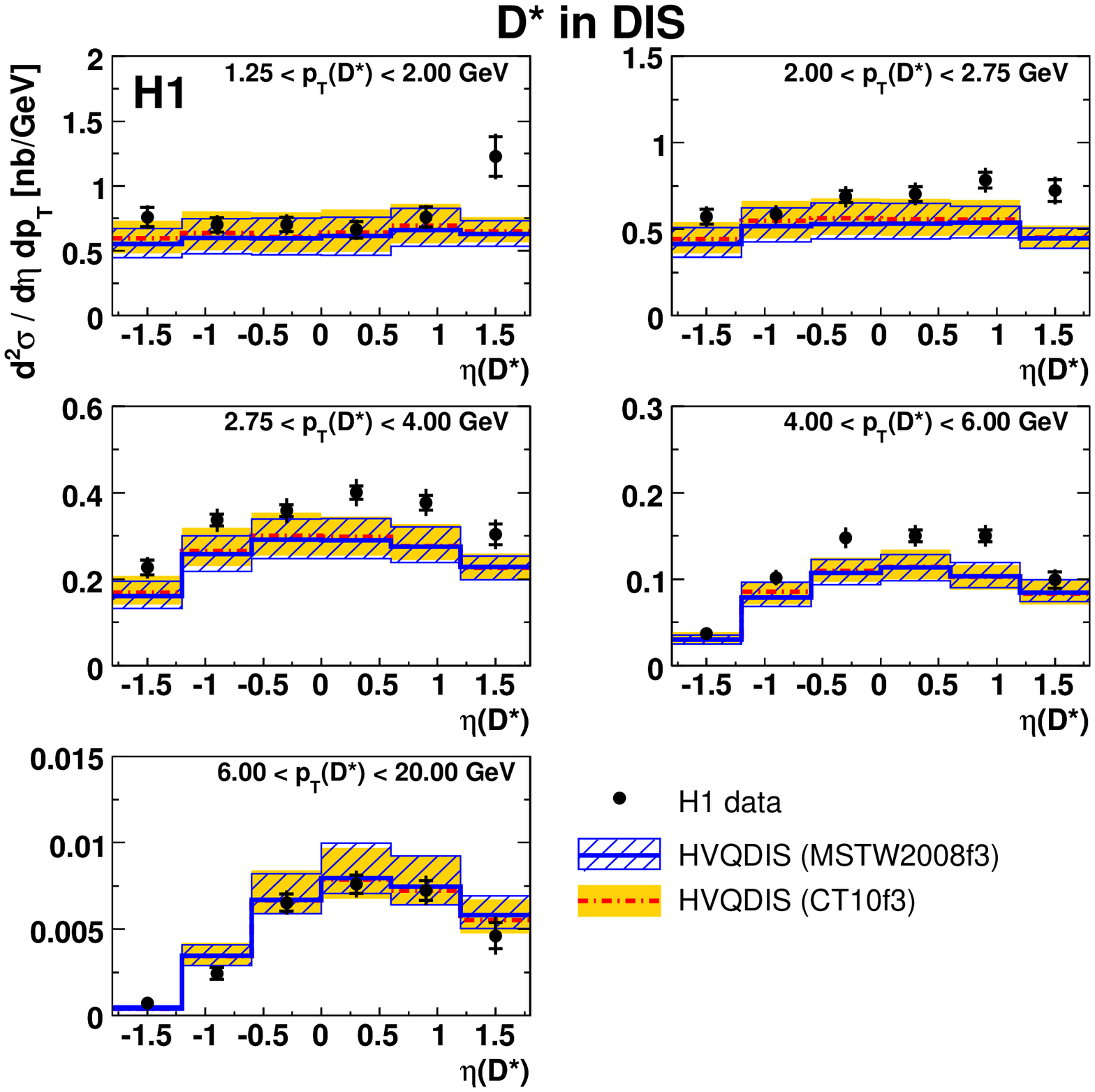} 
\caption{Double differential \dstar\ cross section as a function of 
the transverse momentum $\pt(\dstar)$
and pseudo-rapidity $\eta(\dstar)$ in the
laboratory frame. 
The measurements correspond to the kinematic range of 
$5 < Q^2 < 100\ {\rm GeV}^2$, $0.02 < y < 0.7$ and $|\eta(\dstar)|<1.8$ and
$\pt(\dstar) > 1.25\ {\rm GeV}$. The data are shown as points, the 
inner error bars show the statistical 
error, the outer error bars represent the statistical and systematic errors 
added in quadrature. The data are compared to predictions by the
next-to-leading order calculation HVQDIS with two different proton parton
densities. The bands indicate the theoretical uncertainties 
(table~\ref{Tab_hvqdis_variation}).} 
\label{fig:XSectionPtEtaNLO}
\end{figure}

In order to investigate the correlation between pseudo-rapidity and transverse
momentum, a double differential measurement in $\pt(\dstar)$ and 
$\eta(\dstar)$ is performed. In table~\ref{Tab_xsec_2d} and 
figures~\ref{fig:XSectionPtEtaLO} and \ref{fig:XSectionPtEtaNLO} 
the double differential cross section is presented as a function of the 
pseudo-rapidity $\eta(\dstar)$ in bins of the transverse momentum of 
the \dstar~meson $\pt(\dstar)$. In the backward direction almost 
no \dstar~mesons with large transverse momentum are produced. At low 
transverse momenta all predictions are below the data in the very forward 
direction. At $\pt(\dstar) > 6\ {\rm GeV}$ the CASCADE and
HVQDIS predictions give a good description of the data, while 
RAPGAP is too low. 

\begin{figure}[htbp]
\includegraphics[width=16cm, angle=0]{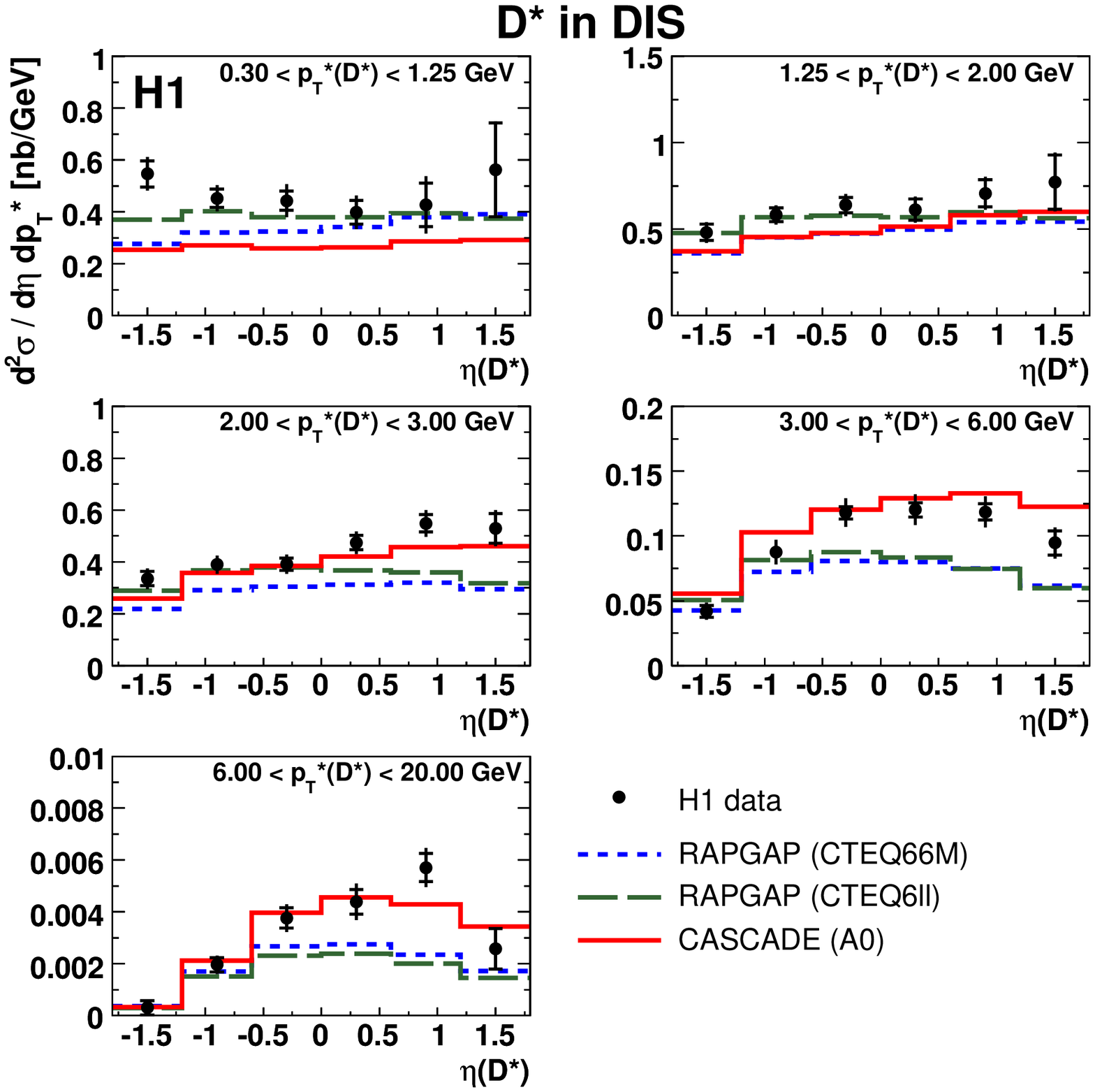} 
\caption{Double differential \dstar\ cross section as a function of 
the transverse 
momentum in the $\gamma p$ centre-of-mass frame $\pt^*(\dstar)$ and the
pseudo-rapidity $\eta(\dstar)$ in the
laboratory frame. 
The measurements correspond to the kinematic range of 
$5 < Q^2 < 100\ {\rm GeV}^2$, $0.02 < y < 0.7$ and $|\eta(\dstar)|<1.8$ and
$\pt(\dstar) > 1.25\ {\rm GeV}$. The data are shown as points, the 
inner error bars show the statistical 
error, the outer error bars represent the statistical and systematic errors 
added in quadrature. The data are compared to predictions by the
MC program RAPGAP with two different proton parton densities and by the
MC program CASCADE.} 
\label{fig:XSectionPtstarEtaLO}
\end{figure}

\begin{figure}[htbp]
\includegraphics[width=16cm, angle=0]{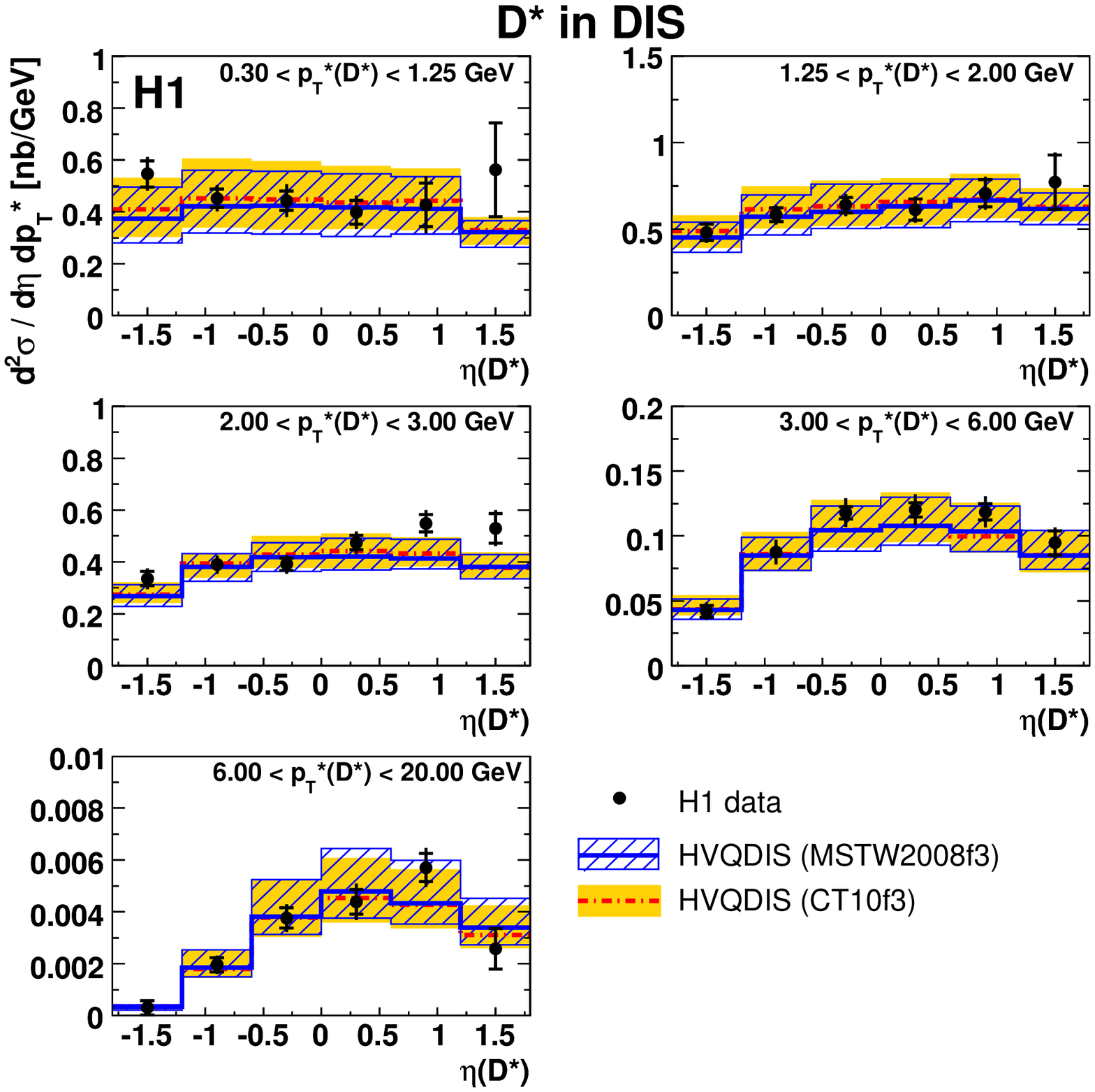} 
\caption{Double differential \dstar\ cross section as a function of 
the transverse 
momentum in the $\gamma p$ centre-of-mass frame $\pt^*(\dstar)$ and the
pseudo-rapidity $\eta(\dstar)$ in the
laboratory frame. 
The measurements correspond to the kinematic range of 
$5 < Q^2 < 100\ {\rm GeV}^2$, $0.02 < y < 0.7$ and $|\eta(\dstar)|<1.8$ and
$\pt(\dstar) > 1.25\ {\rm GeV}$. The data are shown as points, the 
inner error bars show the statistical 
error, the outer error bars represent the statistical and systematic errors 
added in quadrature. The data are compared to predictions by the
next-to-leading order calculation HVQDIS with two different proton parton
densities. The bands indicate the theoretical uncertainties 
(table~\ref{Tab_hvqdis_variation}).} 
\label{fig:XSectionPtstarEtaNLO}
\end{figure}

While the transverse momentum of the \dstar\ meson in the laboratory
frame is correlated with the photon virtuality $Q^2$, the transverse 
momentum in the $\gamma p$ centre-of-mass frame $\pt^*(\dstar)$ is 
directly related to the hard subprocess. The double differential 
cross section as a function of $\eta(\dstar)$ and  $\pt^*(\dstar)$ 
are presented in table~\ref{Tab_xsec_2d}
and figures~\ref{fig:XSectionPtstarEtaLO} and \ref{fig:XSectionPtstarEtaNLO}.
The distribution
shows similar behaviour to the double differential cross section 
as a function of $\eta(\dstar)$ and the transverse momentum $\pt(\dstar)$ 
in the laboratory frame. 
They are in general better described by the predictions of CASCADE and HVQDIS,
while RAPGAP underestimates the data for positive $\eta(\dstar)$ at large
$\pt^*(\dstar)$. 

\begin{figure}[htbp]
\includegraphics[width=16cm, angle=0]{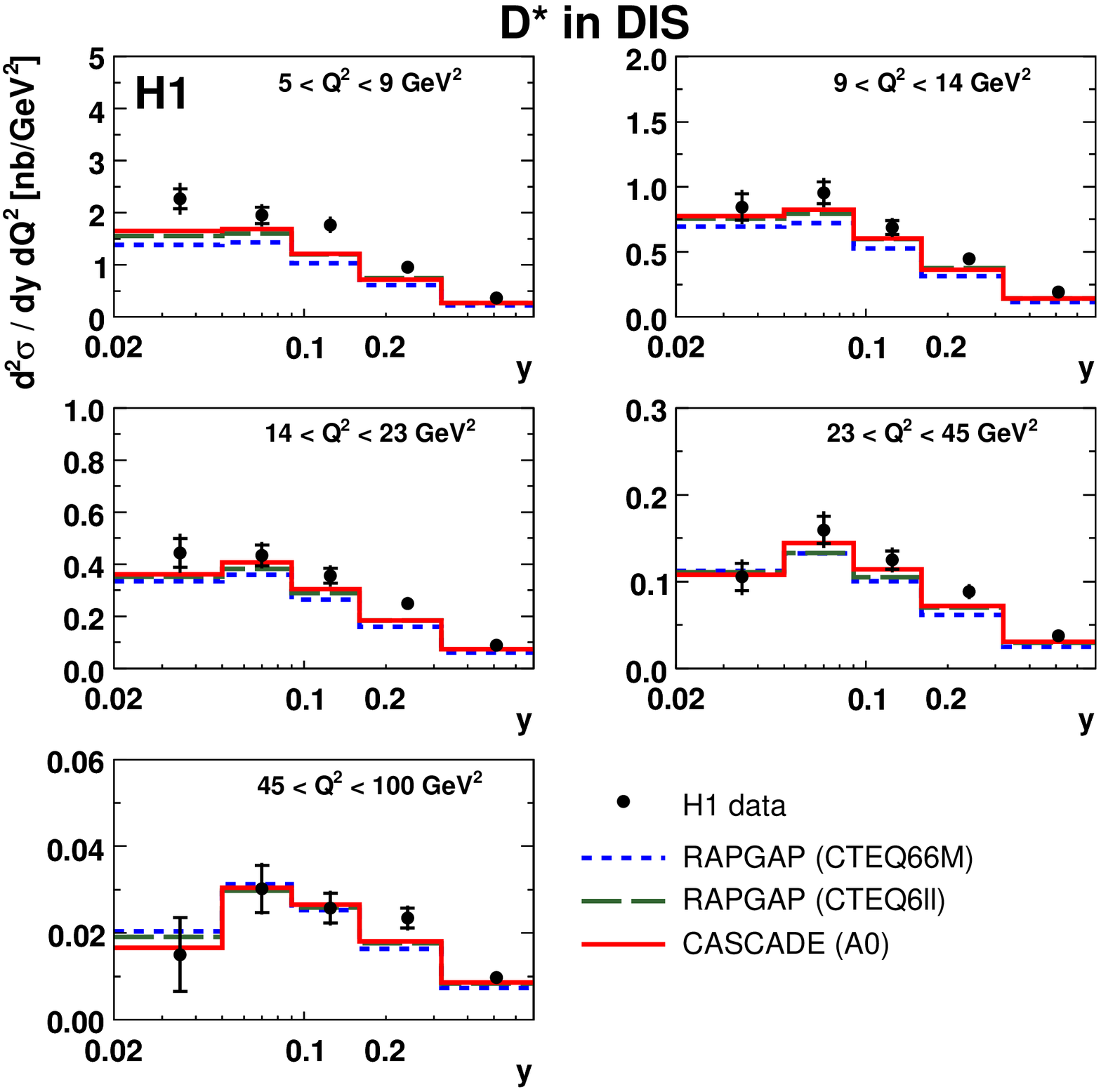} 
\caption{Double differential \dstar\ cross section as a function of the 
photon virtuality $Q^2$ 
and the inelasticity $y$. 
The measurements correspond to the kinematic range of 
$5 < Q^2 < 100\ {\rm GeV}^2$, $0.02 < y < 0.7$ and $|\eta(\dstar)|<1.8$ and
$\pt(\dstar) > 1.25\ {\rm GeV}$. The data are shown as points, the 
inner error bars show the statistical 
error, the outer error bars represent the statistical and systematic errors 
added in quadrature. The data are compared to predictions by the
MC program RAPGAP with two different proton parton densities and by the
MC program CASCADE.} 
\label{fig:XSectionYQ2LO}
\end{figure}

\begin{figure}[htbp]
\includegraphics[width=16cm, angle=0]{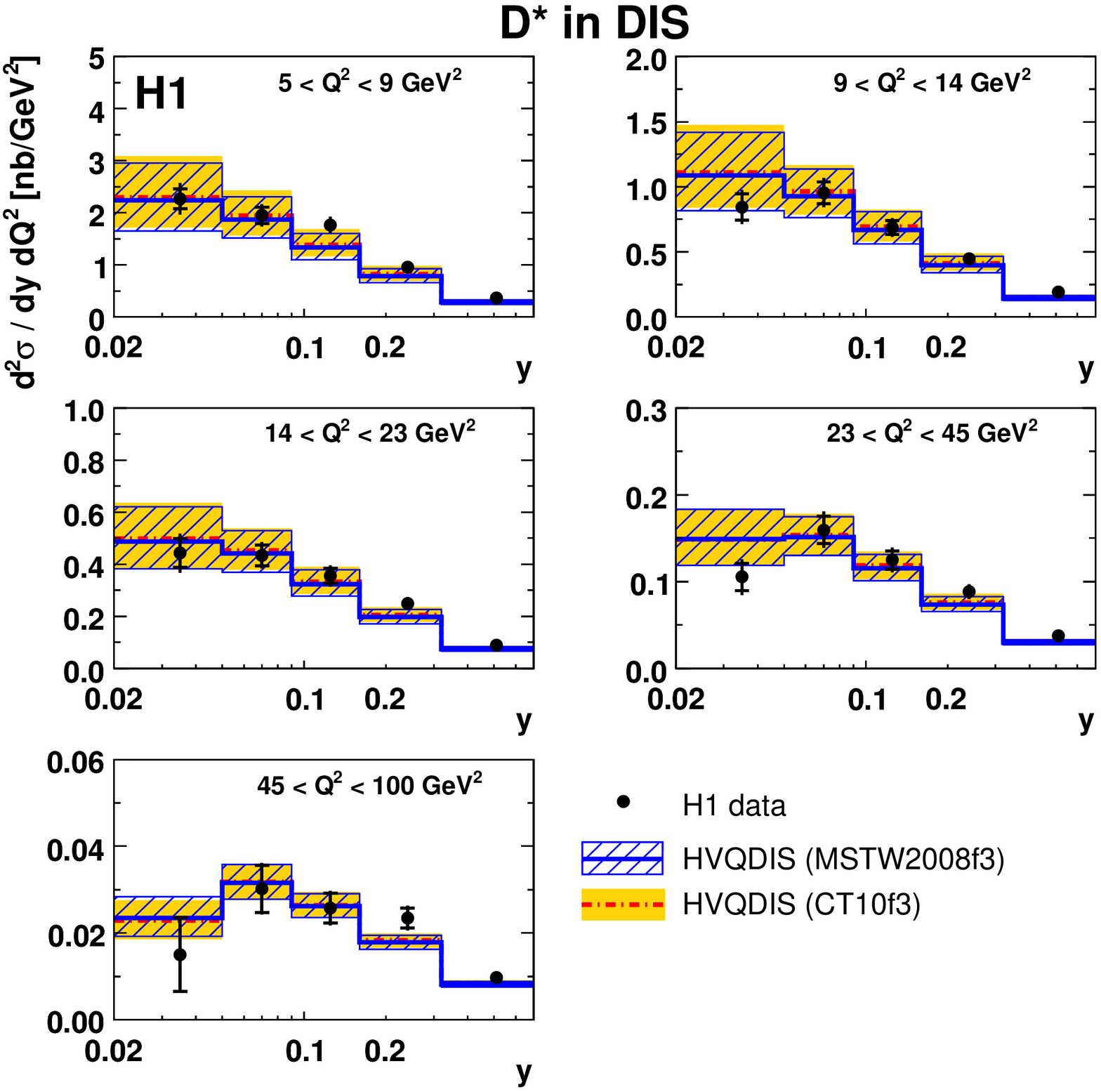} 
\caption{Double differential \dstar\ cross section as a function of 
photon virtuality $Q^2$ 
and the inelasticity $y$. 
The measurements correspond to the kinematic range of 
$5 < Q^2 < 100\ {\rm GeV}^2$, $0.02 < y < 0.7$ and $|\eta(\dstar)|<1.8$ and
$\pt(\dstar) > 1.25\ {\rm GeV}$. The data are shown as points, the 
inner error bars show the statistical 
error, the outer error bars represent the statistical and systematic errors 
added in quadrature. The data are compared to predictions by the
next-to-leading order calculation HVQDIS with two different proton parton
densities. The bands indicate the theoretical uncertainties 
(table~\ref{Tab_hvqdis_variation}).} 
\label{fig:XSectionYQ2NLO}
\end{figure}

The double differential cross section measurements in $y$ and $Q^2$ are 
presented in table~\ref{Tab_xsec_yq2} and figures~\ref{fig:XSectionYQ2LO} and 
\ref{fig:XSectionYQ2NLO}.  
All predictions are able to describe the distribution reasonably 
well, independent of the PDF choice. At low $Q^2$
RAPGAP as well as CASCADE has a tendency to be lower than the data.

\begin{figure}[htbp]
\unitlength1.0cm
\begin{picture}(16,14)
\put(0,6){\includegraphics[width=8cm, angle=0]{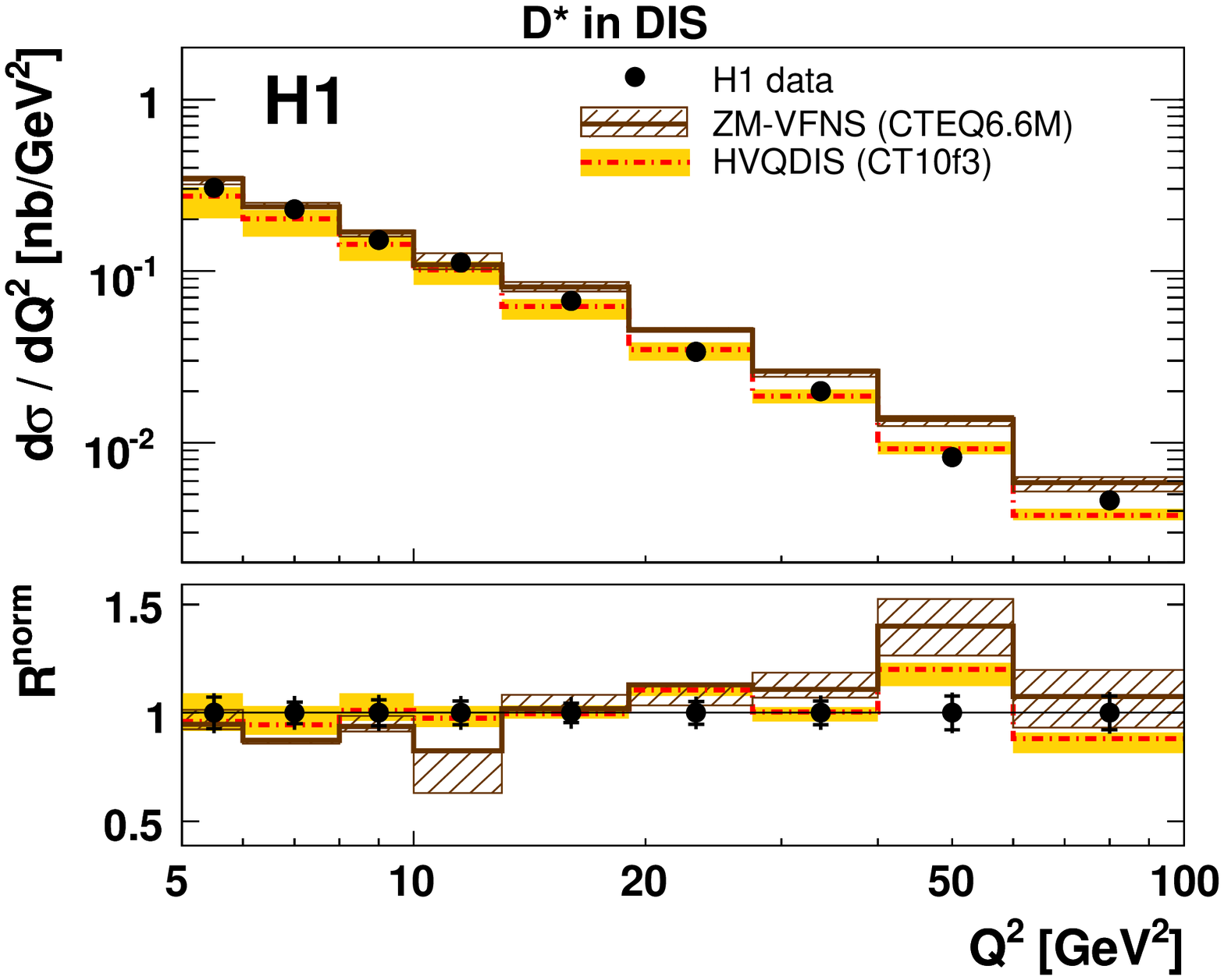}} 
\put(8,6){\includegraphics[width=8cm, angle=0]{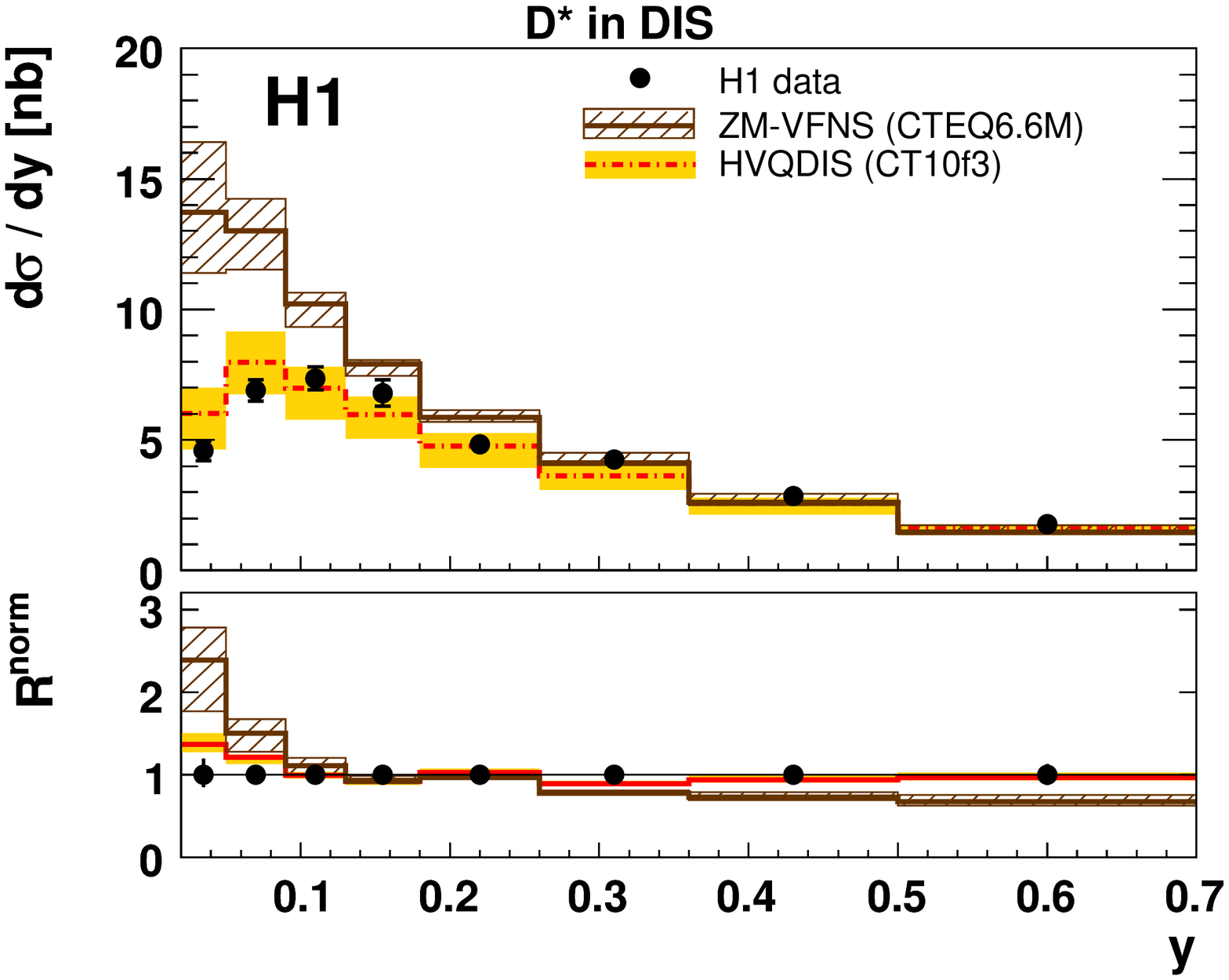}} 
\put(0,-1){\includegraphics[width=8cm, angle=0]{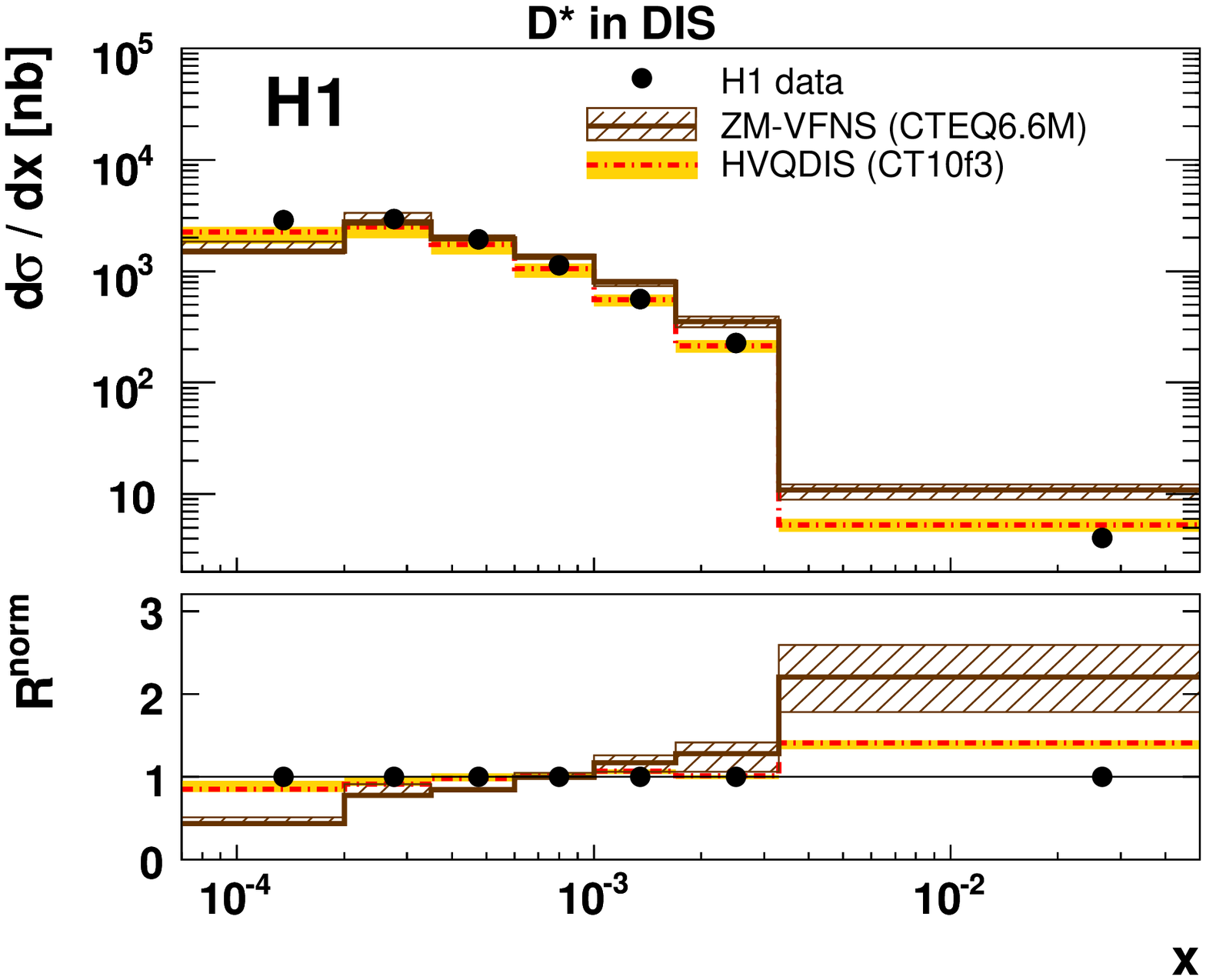}} 
\end{picture}
\caption{Differential \dstar\ cross section as a function of photon
virtuality $Q^2$, the inelasticity $y$ and Bjorken $x$. 
The measurements correspond to the kinematic range of 
$5 < Q^2 < 100\ {\rm GeV}^2$, $0.02 < y < 0.7$, $|\eta(\dstar)|<1.8$,
$\pt(\dstar) > 1.25\ {\rm GeV}$ with an additional cut on the \dstar\ 
transverse momentum in the $\gamma p$ centre-of-mass frame 
$\pt^*(\dstar) > 2.0\ {\rm GeV}$. The data are shown as points, the 
inner error bars show the statistical 
error, the outer error bars represent the statistical and systematic errors 
added in quadrature. The data are compared to a prediction to
next-to-leading order in the ZM-VFNS and to HVQDIS. The bands indicate the 
theoretical uncertainties (table~\ref{Tab_hvqdis_variation}). }
\label{fig:XSectionKinZM}
\end{figure}

\begin{figure}[htbp]
\unitlength1.0cm
\begin{picture}(16,14)
\put(0,6){\includegraphics[width=8cm, angle=0]{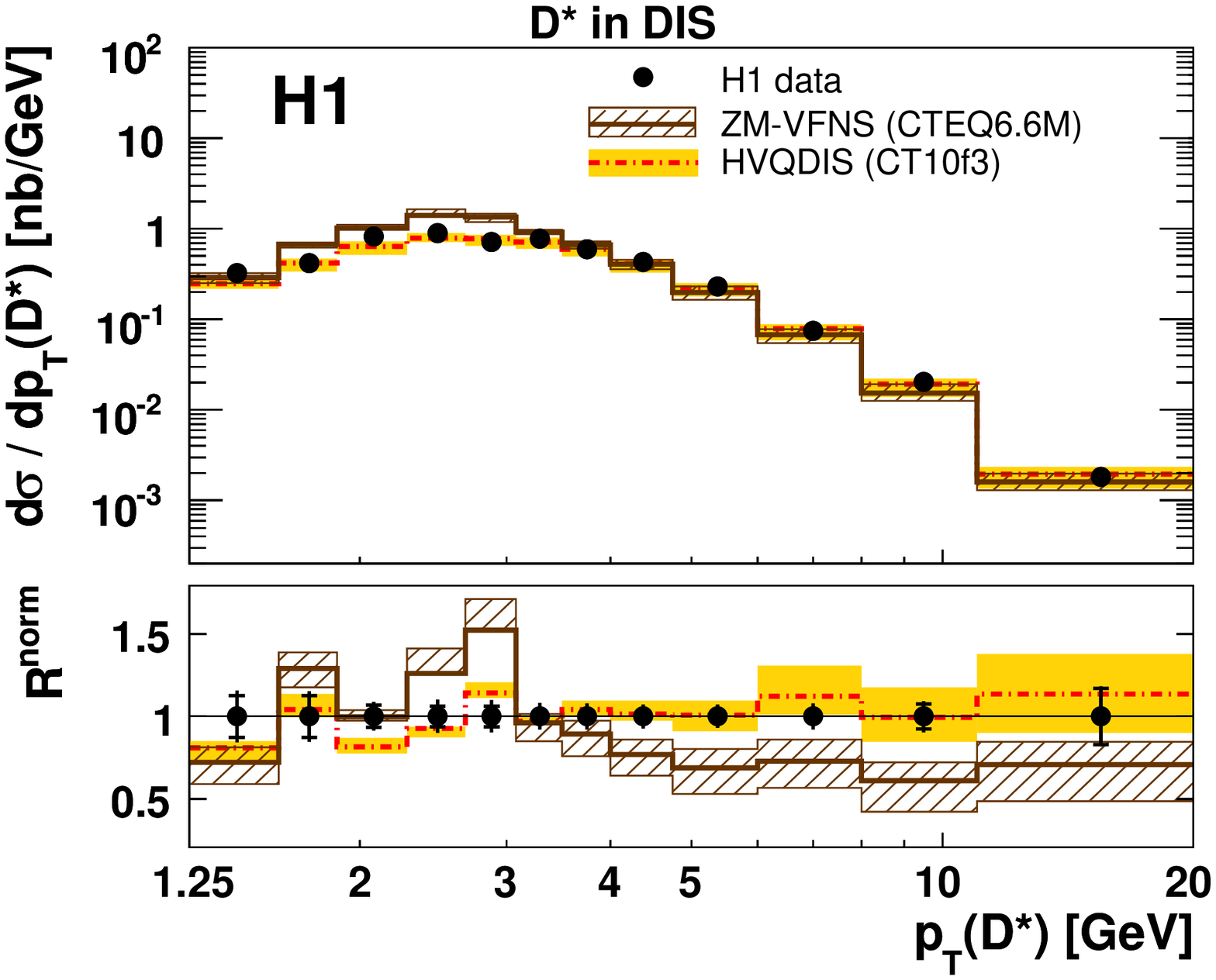}} 
\put(8,6){\includegraphics[width=8cm, angle=0]{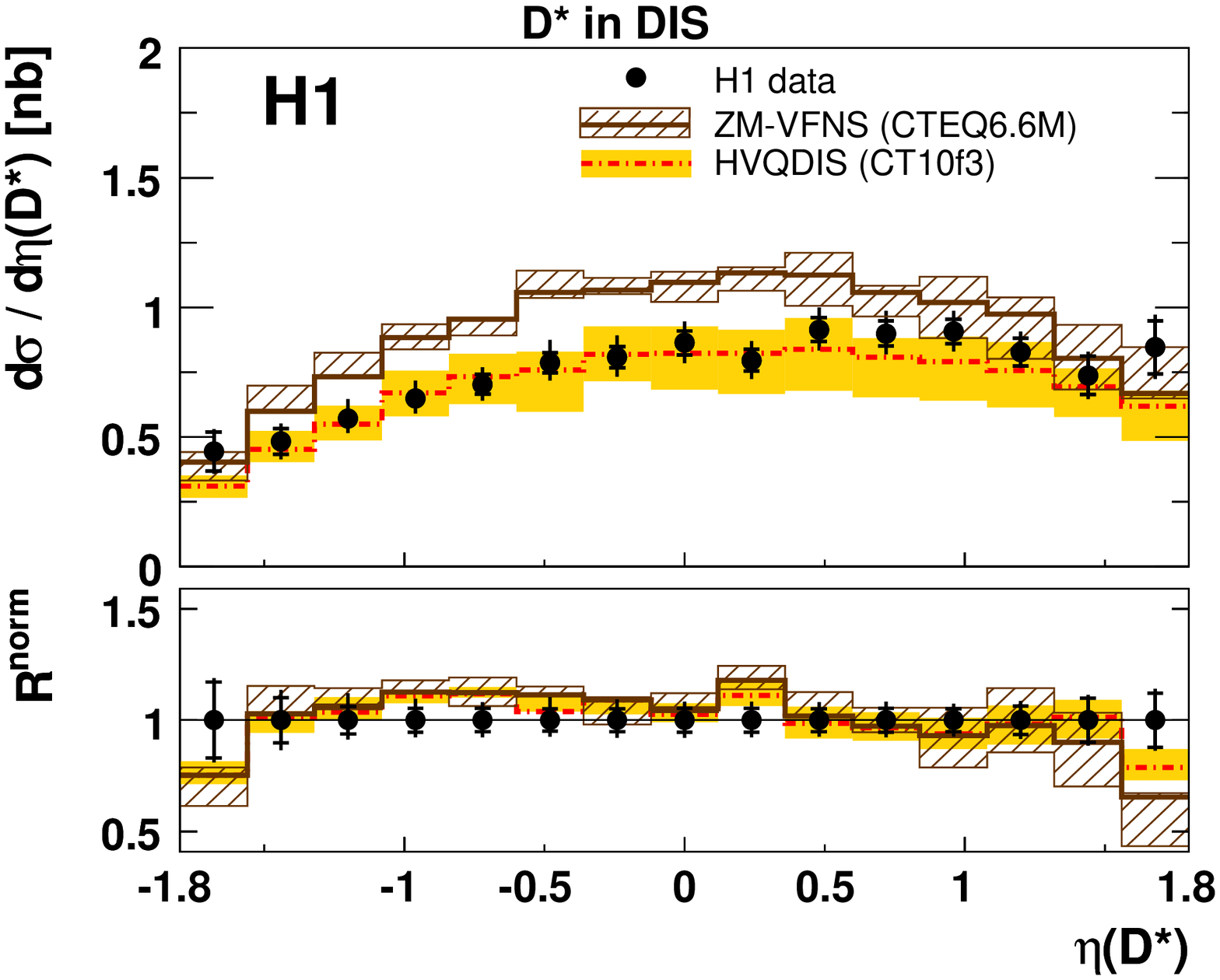}} 
\put(0,-1){\includegraphics[width=8cm, angle=0]{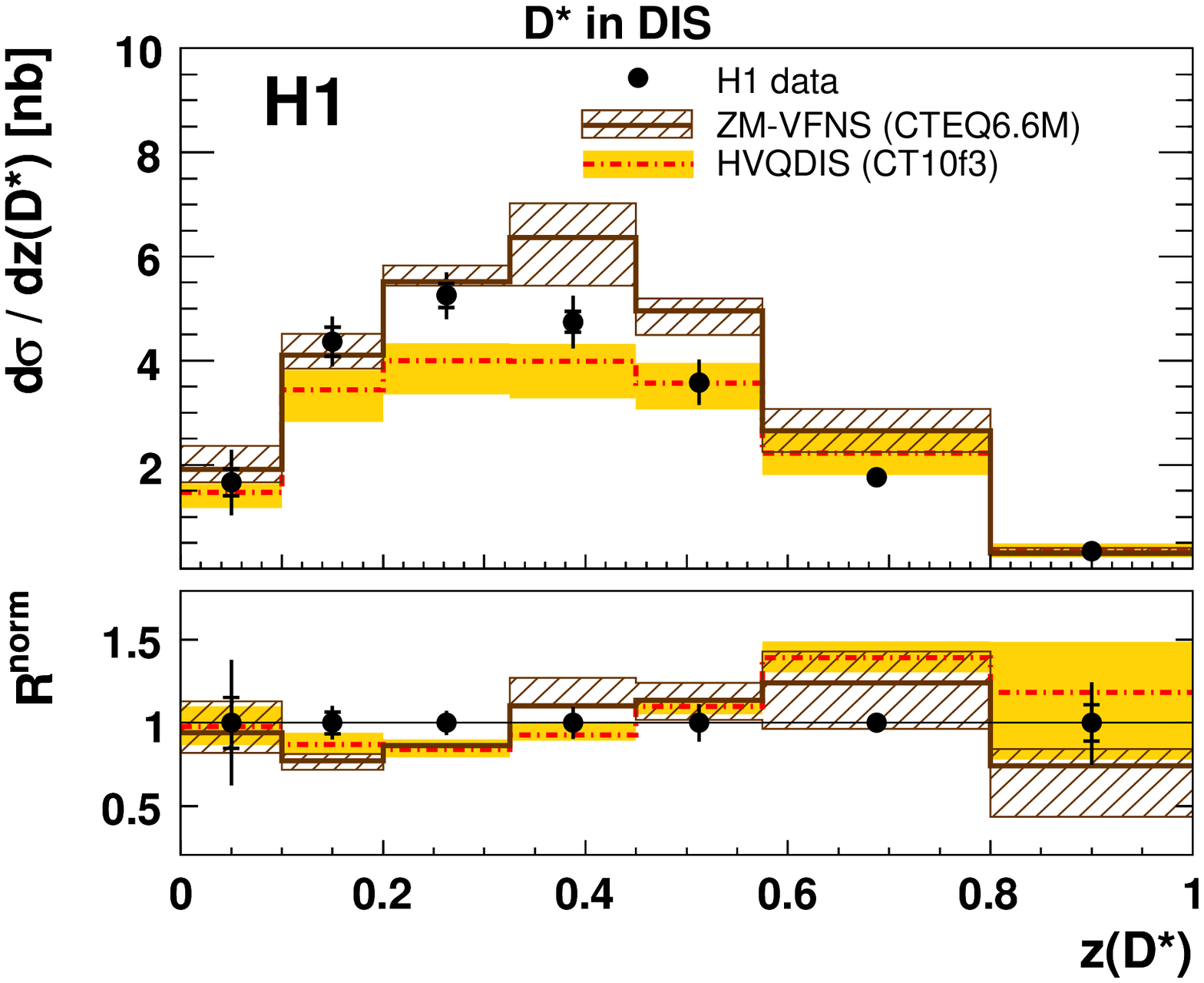}} 
\end{picture}
\caption{Differential \dstar\ cross section as a function of the 
transverse momentum $\pt(\dstar)$
and pseudo-rapidity $\eta(\dstar)$ in the
laboratory frame and the \dstar\ inelasticity $z(\dstar)$. 
The measurements correspond to the kinematic range of $5 < Q^2 < 100\ {\rm GeV}^2$,
$0.02 < y < 0.7$, $|\eta(\dstar)|<1.8$, 
$\pt(\dstar) > 1.25\ {\rm GeV}$ with an additional cut on the \dstar\ 
transverse momentum in the $\gamma p$ centre-of-mass frame 
$\pt^*(\dstar) > 2.0\ {\rm GeV}$. The data are shown as points, the 
inner error bars show the statistical 
error, the outer error bars represent the statistical and systematic errors 
added in quadrature. The data are compared to a prediction to
next-to-leading order in the ZM-VFNS and to HVQDIS. The bands indicate the 
theoretical uncertainties (table~\ref{Tab_hvqdis_variation}). }
\label{fig:XSectionDstarZM}
\end{figure}

To allow a comparison to the ZM-VFNS predictions, the cross 
sections are also measured with an additional cut $\pt^*(\dstar) >2\ {\rm GeV}$.
The measurements are shown in tables~\ref{Tab_xsec_kin} and 
\ref{Tab_xsec_dstar1} and in figures~\ref{fig:XSectionKinZM} and
\ref{fig:XSectionDstarZM}. The ZM-VFNS calculation overshoots the data at low
$y$. The $x$ dependence is less steep than for the data, and it has a different 
shape in $\pt(\dstar)$. The dependence of the cross section on the other
variables is described reasonably well. In general the 
ZM-VFNS prediction describes the data worse than the
NLO FFNS calculation HVQDIS. Also at higher $Q^2 > 100\ {\rm GeV}^2$ the 
ZM-VFNS prediction fails to describe the 
\dstar\ production~\cite{h1dstarhighQ2}.

\begin{figure}[htbp]
\unitlength1.0cm
\begin{center}
\begin{picture}(12,19)
\put(0,8){\includegraphics[width=12cm, angle=0]{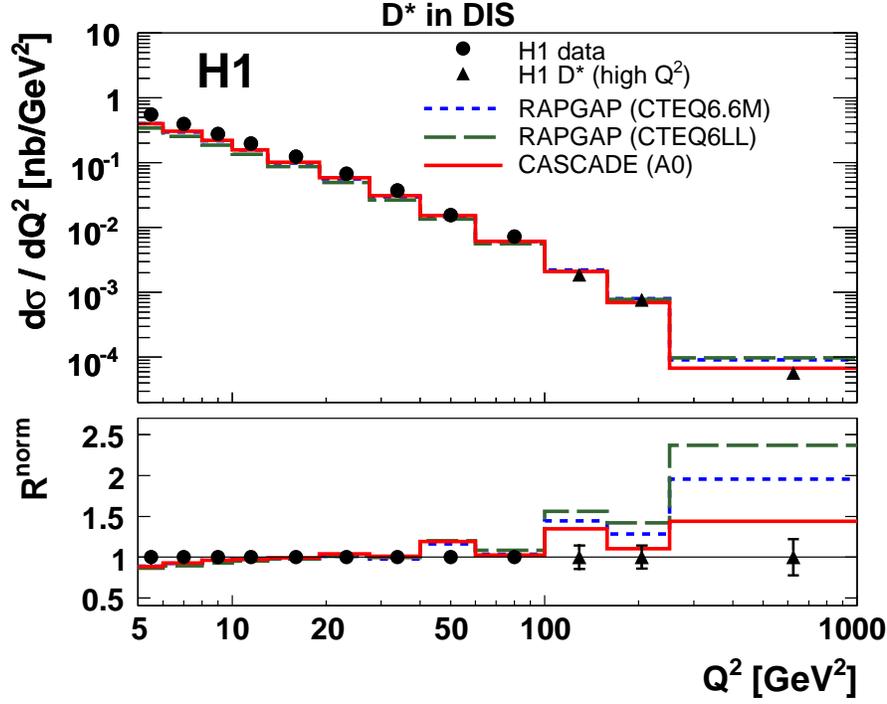}} 
\put(0,-2){\includegraphics[width=12cm, angle=0]{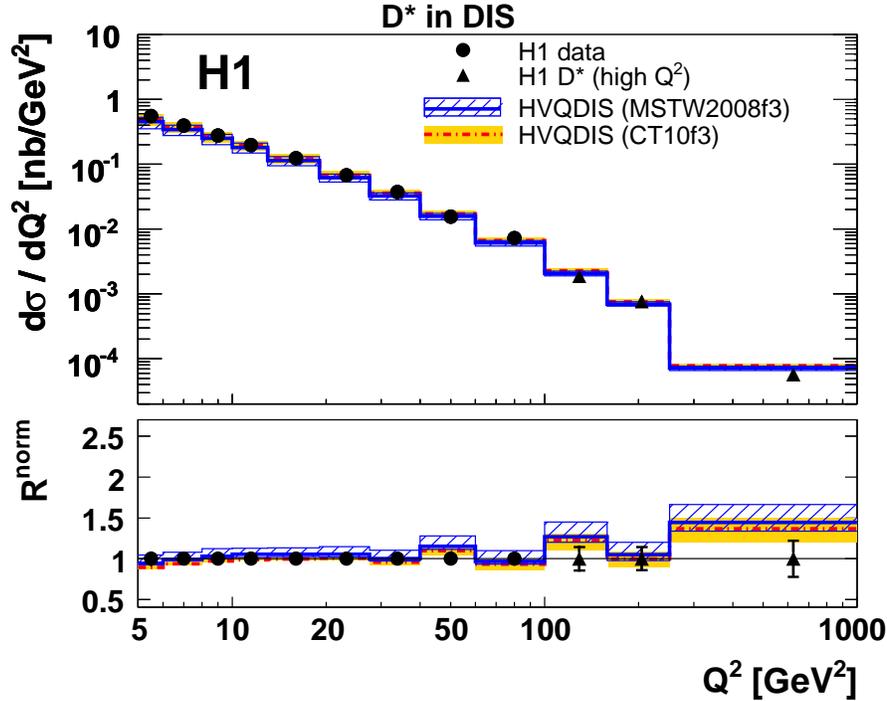}} 
\end{picture}
\end{center}
\caption{Differential \dstar\ cross section as a function of the photon
virtuality $Q^2$. 
The measurements correspond to  
the kinematic range of $0.02 < y < 0.7$, $|\eta(\dstar)|<1.5$ and
$\pt(\dstar) > 1.5\ {\rm GeV}$. The data of this measurement (points)
are shown in a phase space with stronger restrictions on $\eta(\dstar)$ and 
$\pt(\dstar)$ to be comparable to a previous measurement at 
higher $Q^2$~\cite{h1dstarhighQ2} (triangles). The inner error 
bars show the statistical error, the outer error bars represent the 
statistical and systematic errors added in quadrature. The data are 
compared to predictions by the MC program RAPGAP with two different proton 
PDFs and by the MC program CASCADE (left) and to predictions 
by the next-to-leading order calculation HVQDIS with two different proton 
PDFs (right).}
\label{fig:XSectionWholeQ2}
\end{figure}

In order to facilitate the comparison with previous measurements the cross
sections are also measured in a reduced phase space of the
\dstar~meson: $\pt(\dstar) > 1.5\ {\rm GeV}$ and $|\eta(\dstar)|<1.5$. They are
listed in tables~\ref{Tab_xsec_kin_redphasespace} and
\ref{Tab_xsec_dstar1_redphasespace}. 
In figure~\ref{fig:XSectionWholeQ2} these measurements
are shown as a function of $Q^2$ together with the results of the
measurement at high $Q^2$~\cite{h1dstarhighQ2}. These measurements
span over almost three orders of magnitude in $Q^2$. The data are well 
described by CASCADE and the HVQDIS predictions
with both PDF sets in the whole $Q^2$ range, while RAPGAP overshoots the data 
at high $Q^2$.

\section{Charm Contribution to the Proton Structure Function}

The charm contribution $\ftwocc(x,Q^2)$ to the proton structure function
$F_2(x,Q^2)$ is related to the charm cross section in the
one photon exchange approximation by:
\begin{equation}
\displaystyle
\frac{{\rm d}^2\sigma^{c\bar{c}}}{{\rm d}x\,{\rm d}Q^2}=\frac{2\pi\alpha_{em}^2}{Q^4x}
\left( [1+\left(1-y\right)^2]\;\ftwocc(x,Q^2)-y^2 F^{c\bar{c}}_L(x,Q^2)
\right) ~~.
\end{equation}
Weak interaction effects are neglected and $\alpha_{em}$ denotes the 
electromagnetic coupling constant.
The contribution from the structure function $F^{c\bar{c}}_L$ is less 
than $4\%$ in the present $x-Q^2$ range. Assuming the ratio 
$F^{c\bar{c}}_L/\ftwocc$ is predicted correctly within a model, 
the measured inclusive \dstarpm\ cross sections
$\sigma_{\rm{vis}}^{\rm{exp}}(y,Q^2)$ in bins of $y$ and $Q^2$ are
converted to a bin centre corrected
$\ftwocc\,^{\rm{exp}}(\langle x\rangle,\langle Q^2\rangle)$ 
using the relation $Q^2 = xys$ and 
extrapolating $\sigma_{\rm{vis}}$ to the full phase space:
\begin{equation}
\ftwocc\,^{\rm{exp}}(\langle x\rangle,\langle Q^2\rangle)=
\frac{\displaystyle \sigma_{\rm{vis}}^{\rm{exp}}(y,Q^2)}
{\displaystyle \sigma_{\rm{vis}}^{\rm{theo}}(y,Q^2)}\cdot
\ftwocc\,^{\rm{theo}}(\langle x\rangle,\langle Q^2\rangle)~~.
\label{f2cexp}
\end{equation}
Here $\sigma_{\rm{vis}}^{\rm{theo}}$ and $\ftwocc\,^{\rm{theo}}$ are the
theoretical  predictions. 
The HVQDIS program is used to calculate 
$\sigma_{\rm{vis}}^{\rm{theo}}$ and
$\ftwocc\,^{\rm{theo}}$ in the NLO DGLAP scheme.
In the kinematic range of the current analysis
the beauty contribution to the \dstar\ cross section is small. It is estimated
with HVQDIS and subtracted for the determination of \ftwocc.

\begin{figure}[htbp]
\begin{center}
\includegraphics[width=12cm, angle=0]{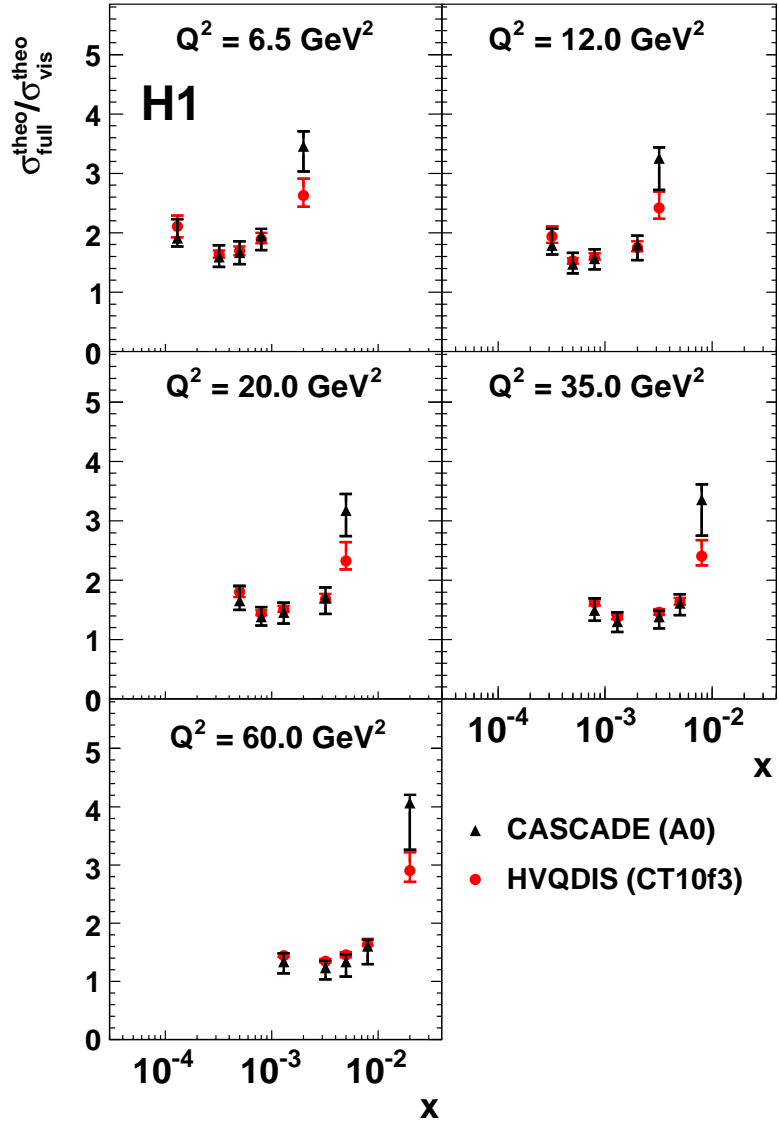} 
\end{center}
\caption{Extrapolation factors from the visible phase space 
(table~\ref{tab:range}) 
to the total phase space for the \dstar\ meson as determined from 
HVQDIS and CASCADE. The error bars show the extrapolation uncertainty which
is determined by varying the theory parameters listed in 
table~\ref{Tab_hvqdis_variation} for HVQDIS and in
table~\ref{Tab_MC_parameters} for CASCADE.}
\label{factors}
\end{figure}

The measurement covers about $50\%$ of the total phase space for
charm production. The extrapolation factor to the full phase space is
model dependent. Since CASCADE also
provides a reasonable description of the cross sections in the phase space 
covered in this analysis, 
it is used as an alternative model to determine $\ftwocc\,^{\rm{exp}}$ in order 
to investigate this model dependence. The extrapolation factors in the present
analysis, defined as the ratio of the full cross section 
$\sigma_{\rm{full}}^{\rm{theo}}$ to the cross section
$\sigma_{\rm{vis}}^{\rm{theo}}$ in the visible phase space of the \dstar\ meson,
determined with HVQDIS and CASCADE, are compared in figure~\ref{factors}. They
differ by about $10\%$ at medium $x$. In the largest $x$ bin the 
extrapolation factor as well as the difference between the two models
increases significantly. In general the extrapolation factor determined
with HVQDIS has smaller uncertainties than the one from CASCADE.
Due to the 
larger phase space of the \dstar\ meson in 
the present analysis 
compared to previous measurements, where the phase space coverage amounted 
to about $30\%$, the extrapolation factor to the full phase space is 
considerably smaller, although the model dependence of the extrapolation 
remains sizable. 

\begin{figure}[htbp]
\begin{center}
\includegraphics[width=12cm, angle=0]{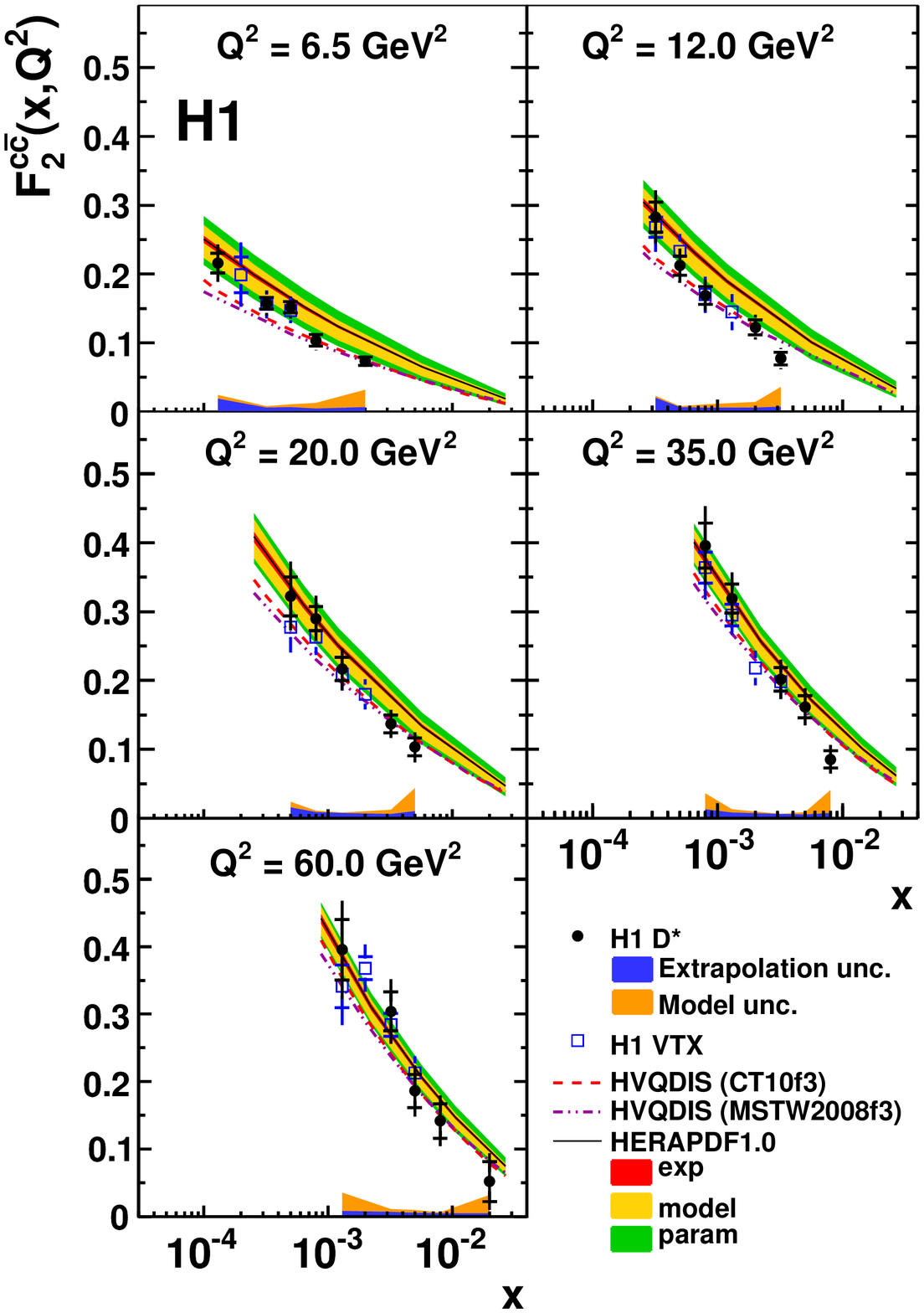} 
\end{center}
\caption{\ftwocc\ as derived from \dstar\ data with HVQDIS (points). 
The inner error bars show the statistical uncertainty, the outer error bar the 
statistical and experimental systematic uncertainty added in quadrature. 
The extrapolation uncertainty within the HVQDIS model is shown as
blue band in the bottom of the plots. The outer (orange) band shows the
model uncertainty obtained from the difference in \ftwocc\ 
determined with HVQDIS and CASCADE. The data are compared to the measurement
of \ftwocc\ with the H1 vertex detector~\cite{h1vertex09} (open squares), 
to NLO DGLAP predictions from HVQDIS with two different proton PDFs, and to 
the \ftwocc\ prediction of HERAPDF1.0.}
\label{f2cc_hvqdis}
\end{figure}

In table~\ref{Tab_f2cc} and figure~\ref{f2cc_hvqdis} the resulting \ftwocc\ 
extracted from the inclusive \dstarpm\ cross sections with HVQDIS 
is shown as a function of $x$ for different values of $Q^2$. 
In addition to the experimental systematic uncertainties 
described in section~\ref{difwq} the extrapolation (equation~\ref{f2cexp})
leads to an uncertainty. This extrapolation
uncertainty is determined by varying the theory parameters listed in 
table~\ref{Tab_hvqdis_variation} simultaneously in the calculation of
$\sigma_{\rm{vis}}^{\rm{theo}}$ and $\ftwocc\,^{\rm{theo}}$. The resulting 
uncertainties on \ftwocc\ are shown separately in figure~\ref{f2cc_hvqdis}. 
HVQDIS and CASCADE both give a reasonable description of the measured cross
sections and can be used to extract \ftwocc. The differences obtained 
in $\ftwocc\,^{\rm{exp}}$ for the two models are used to define the 
model uncertainty on \ftwocc, which is also given in figure~\ref{f2cc_hvqdis}.

The results of a \ftwocc\ measurement based on lifetime information 
determined with the H1 silicon vertex detector CST~\cite{h1vertex09} is 
compared to the present measurement in figure~\ref{f2cc_hvqdis}.
The two measurements are based on independent
methods with similar precision and agree very well.
The \ftwocc\ determined with \dstar\ mesons 
covers a larger range in $x$ due to the larger $\eta$ coverage of the 
CJCs compared to the CST. It also has smaller uncertainties 
at low $Q^2$, where the uncertainty of 
the lifetime based measurement is dominated by the light quark background.

Figure~\ref{f2cc_hvqdis} also compares the FFNS NLO calculation of \ftwocc\ 
to the measurement using the MSTW2008f3 and CT10f3 proton PDFs. Both
calculations give a reasonable description of the data. 
The \ftwocc\ measurement is also compared to the \ftwocc\ prediction for 
HERAPDF1.0~\cite{herapdf}, which has been extracted from the H1 and ZEUS 
combined inclusive proton structure function data. A general-mass 
variable-flavour-number scheme~\cite{RTGMVFNS} has been used 
which interpolates between the FFNS and the ZM-VFNS. 
The uncertainty on \ftwocc\ for the HERAPDF1.0 prediction is dominated by 
the variation of the charm mass in the PDF fit, which is included in the 
model uncertainty of the prediction. In general the prediction agrees with
the \ftwocc\ measurement, showing that the gluon density determined from 
the scaling violations of the inclusive DIS cross section is 
consistent with that observed in charm production. At large $x$ the central 
value of the HERAPDF1.0 prediction has a tendency to lie above the 
\ftwocc\ data, which may indicate a preference for a larger charm mass than 
the central value used for HERAPDF1.0, $m_c=1.4\ {\rm GeV}$. 

\begin{figure}[htbp]
\begin{center}
\includegraphics[width=12cm, angle=0]{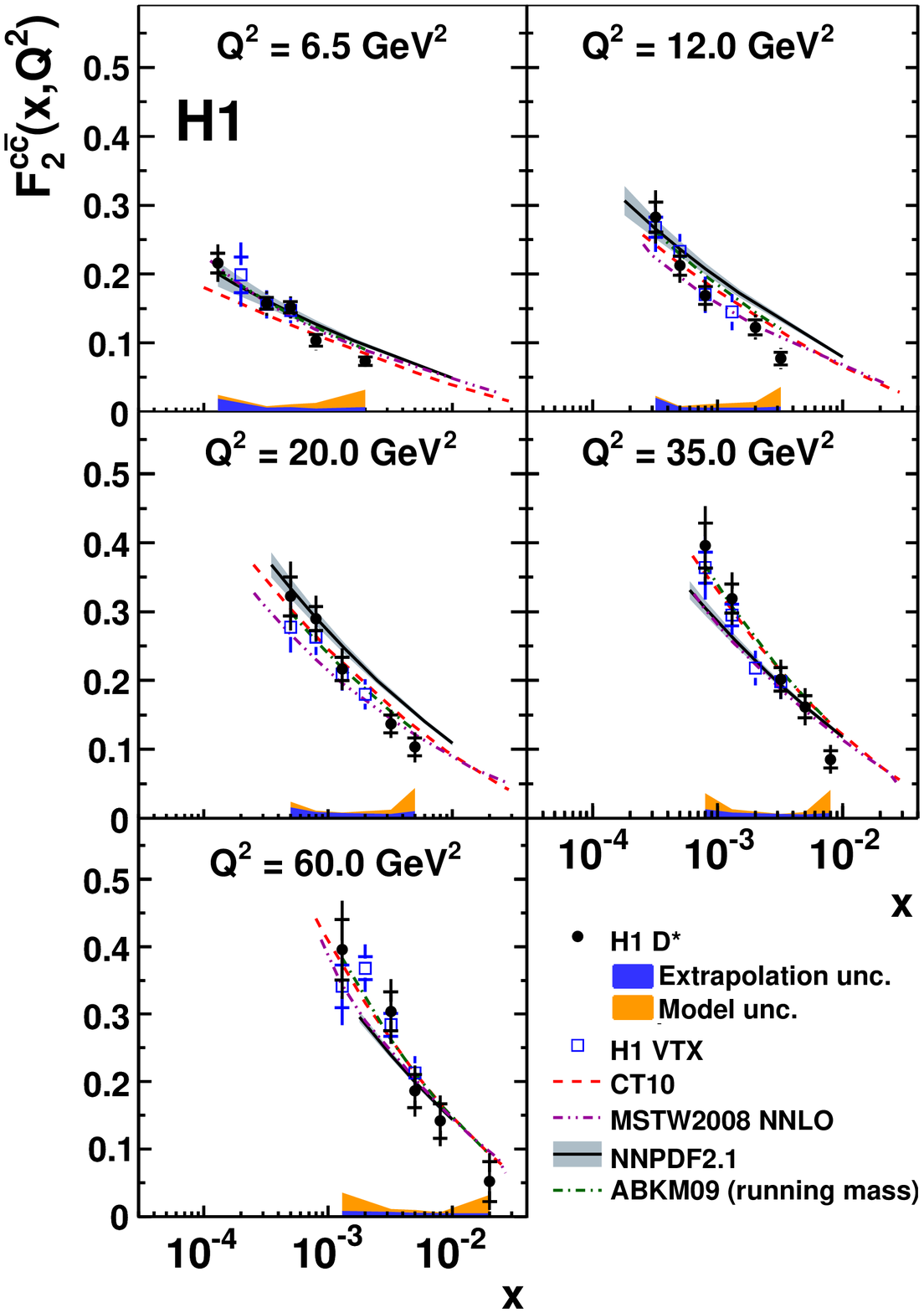} 
\end{center}
\caption{\ftwocc\ as derived from \dstar\ data with HVQDIS (points). 
The inner error bars show the statistical uncertainty, the outer error bar the 
statistical and experimental systematic uncertainty added in quadrature. 
The extrapolation uncertainty within the HVQDIS model is shown as
blue band in the bottom of the plots. The outer (orange) band shows the
model uncertainty obtained from the difference in \ftwocc\ 
determined with HVQDIS and CASCADE. The data are compared to the measurement
of \ftwocc\ with the H1 vertex detector~\cite{h1vertex09} (open squares)
and to predictions from the global PDF fits CT10 (dashed line),
MSTW08 at NNLO (dark dashed-dotted line), NNPDF2.1 (shaded band) and 
ABKM (light dashed-dotted line).}
\label{f2cc_pdfs}
\end{figure}

The \ftwocc\ measurement is compared to predictions from global PDF fits
in figure~\ref{f2cc_pdfs}:
CT10~\cite{Lai:2010vv}, MSTW2008 NNLO~\cite{mstw2008} and NNPDF2.1~\cite{nnpdf} 
have been derived in general-mass 
variable-flavour-number schemes, while for the ABKM fit~\cite{abkm_rm}   
the FFNS including higher order radiative corrections in QCD adopting 
the running $\overline{\mbox{MS}}$ mass has been used. 
 All predictions give a reasonable description
of the measurement. At low $Q^2$ all predictions have
a tendency to decrease less steeply with $x$ than the data.

\begin{figure}[htbp]
\begin{center}
\includegraphics[width=12cm,angle=0]{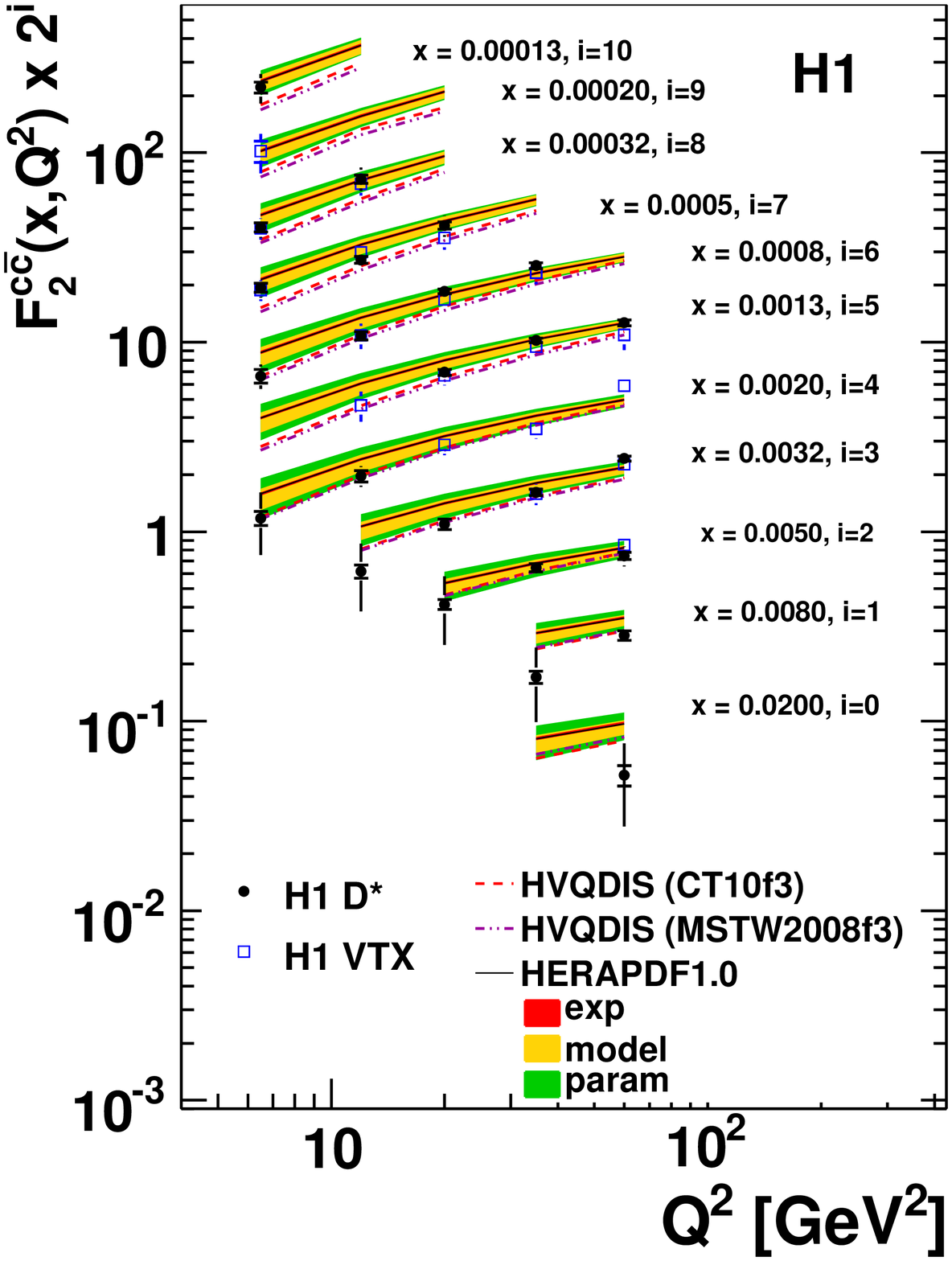} 
\end{center}
\caption{\ftwocc\ as a function of $Q^2$ for different $x$, as derived 
from \dstar\ data with HVQDIS (points). The inner
error bars show the statistical uncertainty, the outer error bar the 
total uncertainty, including statistical, experimental systematic, extrapolation
and model uncertainty added in quadrature. 
The data are compared to the measurement
of \ftwocc\ with the H1 vertex detector~\cite{h1vertex09} (open squares), to 
NLO DGLAP predictions from HVQDIS with two different proton PDFs, and to the 
\ftwocc\ prediction of HERAPDF1.0.}
\label{f2cc_scaling}
\end{figure}

The measured \ftwocc\ as a function of $Q^2$ for different
values of $x$ is shown in figure~\ref{f2cc_scaling}. 
Scaling violations are observed.
The $Q^2$ dependence of the data is well reproduced by the FFNS NLO 
calculation, but at low $x$ the
predictions are below the data, an effect which is larger if the 
MSTW2008f3 set is used as proton PDF.
The HERAPDF1.0 prediction is 
in agreement with the data.

\section{Conclusion}
A measurement of \dstar~meson production in deep-inelastic scattering is 
performed with a tenfold increase in statistics and a significantly 
enlarged phase space compared to the previous H1 measurement. 
Single and double differential cross sections are determined as a function 
of variables describing the kinematics of the event as well as
of the \dstar\ meson. The measurements are found to be reasonably well 
described by predictions based on the fixed-flavour-number scheme, namely
the leading order Monte Carlo simulations RAPGAP and CASCADE 
as well as the next-to-leading order calculation HVQDIS. The data are also
compared to a next-to-leading order calculation in the zero-mass 
variable-flavour-number scheme, which in general describes the data less 
well and is particularly high at low $y$.

The double differential cross section as a function of $Q^2$ and $y$ is 
used to determine the charm contribution \ftwocc\ to the proton structure 
function $F_2$. The extrapolation to the full phase space is done with
two different models, using
the next-to-leading order calculation HVQDIS and the Monte Carlo program 
CASCADE based on leading order matrix elements and parton showers.
The results for \ftwocc\ in these two models are very similar except for 
the highest $x$ values. The results agree well with a measurement based 
on lifetime information determined with the H1 vertex 
detector. 
The data are well described by next-to-leading
order calculations using different PDFs, showing that the gluon density 
determined from the scaling violations of the inclusive DIS cross sections is 
consistent with the one observed in charm production.

\section*{Acknowledgements}

We are grateful to the HERA machine group whose outstanding efforts have 
made this experiment possible. We thank the engineers and technicians 
for their work in constructing and maintaining the H1 detector, our 
funding agencies for financial support, the DESY technical staff for 
continual assistance and the DESY directorate for support and for the 
hospitality which they extend to the non-DESY members of the collaboration.

\clearpage


\renewcommand{\arraystretch}{1.22}
\begin{table}[htbp]
\begin{center}
{\scriptsize
\begin{tabular}{|r r|d l l|d l l|}
\hline
\multicolumn{2}{| c |}{} & 
\multicolumn{3}{ c |}{} &
\multicolumn{3}{ c |}{${{\rm d}\sigma}/{{\rm d}  Q^2} \
[{{\rm nb}}/{\rm GeV^2}]$} \\ 
\multicolumn{2}{| c |}{\rb{$Q^2\ {\rm range} \  [{\rm GeV}^2]$}} & 
\multicolumn{3}{ c |}{\rb{${{\rm d}\sigma}/{{\rm d}  Q^2} \
[{{\rm nb}}/{\rm GeV^2}]$}} &
\multicolumn{3}{ c |}{$ {\rm\ for\ } \pt^*(\dstar) > 2.0\ {\rm GeV} $} 
\\ \hline
$  5.0$ & $  6.0$ & 0.782  & $\pm 0.048  $ & $ \pm ^{0.058}  _{0.057}  $ & 0.317   & $\pm 0.023   $ & $ \pm ^{0.023} _{0.022} $\\  
$  6.0$ & $  8.0$ & 0.538  & $\pm 0.022  $ & $ \pm ^{0.039}  _{0.039}  $ & 0.237   & $\pm 0.012   $ & $ \pm ^{0.018} _{0.018} $\\  
$  8.0$ & $ 10.0$ & 0.384  & $\pm 0.018  $ & $ \pm ^{0.028}  _{0.028}  $ & 0.1568  & $\pm 0.0094  $ & $ \pm ^{0.0115} _{0.0110} $\\  
$ 10.0$ & $ 13.0$ & 0.249  & $\pm 0.011  $ & $ \pm ^{0.018}  _{0.018}  $ & 0.1156  & $\pm 0.0063  $ & $ \pm ^{0.0085} _{0.0084} $\\  
$ 13.0$ & $ 19.0$ & 0.1549 & $\pm 0.0057 $ & $ \pm ^{0.0109} _{0.0110} $ & 0.0695  & $\pm 0.0031  $ & $ \pm ^{0.0051} _{0.0050} $\\  
$ 19.0$ & $ 27.5$ & 0.0874 & $\pm 0.0038 $ & $ \pm ^{0.0062} _{0.0062} $ & 0.0350  & $\pm 0.0019  $ & $ \pm ^{0.0025} _{0.0025} $\\  
$ 27.5$ & $ 40.0$ & 0.0463 & $\pm 0.0022 $ & $ \pm ^{0.0032} _{0.0032} $ & 0.0206  & $\pm 0.0011  $ & $ \pm ^{0.0015} _{0.0015} $\\  
$ 40.0$ & $ 60.0$ & 0.0188 & $\pm 0.0013 $ & $ \pm ^{0.0014} _{0.0014} $ & 0.00856 & $\pm 0.00067 $ & $ \pm ^{0.00059} _{0.00059} $\\ 
$ 60.0$ & $100.0$ & 0.00824& $\pm 0.00057$ & $ \pm ^{0.00058}_{0.00057}$ & 0.00478 & $\pm 0.00037 $ & $ \pm ^{0.00034} _{0.00034} $\\ 

\hline\hline
\multicolumn{2}{| c |}{ } & 
\multicolumn{3}{ c |}{ } &
\multicolumn{3}{ c |}{${{\rm d}\sigma}/{{\rm d}  y}  \
[{{\rm nb}}] $} \\
\multicolumn{2}{| c |}{\rb{$y\ {\rm range}$}} & 
\multicolumn{3}{ c |}{\rb{${{\rm d}\sigma}/{{\rm d}  y} \
[{{\rm nb}}]$}} &
\multicolumn{3}{ c |}{$ {\rm\ for\ } \pt^*(\dstar) > 2.0\ {\rm GeV} $} 
\\ \hline
$ 0.02$ & $0.05$ & 21.67 & $\pm 1.04 $ & $ \pm ^{2.53} _{2.83} $ & 4.75 & $\pm 0.40 $ & $ \pm ^{0.81} _{0.65} $\\ 
$ 0.05$ & $0.09$ & 20.97 & $\pm 0.94 $ & $ \pm ^{1.51} _{1.79} $ & 7.14 & $\pm 0.41 $ & $ \pm ^{0.68} _{0.71} $\\ 
$ 0.09$ & $0.13$ & 20.05 & $\pm 0.97 $ & $ \pm ^{1.48} _{1.58} $ & 7.61 & $\pm 0.46 $ & $ \pm ^{0.68} _{0.65} $\\ 
$ 0.13$ & $0.18$ & 14.63 & $\pm 0.80 $ & $ \pm ^{1.04} _{1.03} $ & 6.79 & $\pm 0.50 $ & $ \pm ^{0.52} _{0.52} $\\ 
$ 0.18$ & $0.26$ & 12.61 & $\pm 0.54 $ & $ \pm ^{0.91} _{0.90} $ & 5.01 & $\pm 0.25 $ & $ \pm ^{0.39} _{0.38} $\\ 
$ 0.26$ & $0.36$ &  8.39 & $\pm 0.43 $ & $ \pm ^{0.72} _{0.63} $ & 4.25 & $\pm 0.20 $ & $ \pm ^{0.32} _{0.32} $\\ 
$ 0.36$ & $0.50$ &  5.87 & $\pm 0.31 $ & $ \pm ^{0.63} _{0.47} $ & 2.96 & $\pm 0.17 $ & $ \pm ^{0.25} _{0.24} $\\ 
$ 0.50$ & $0.70$ &  3.00 & $\pm 0.27 $ & $ \pm ^{0.32} _{0.30} $ & 1.83 & $\pm 0.16 $ & $ \pm ^{0.19} _{0.18} $\\ 

\hline\hline
\multicolumn{2}{| c |}{ } & 
\multicolumn{3}{ c |}{ } &
\multicolumn{3}{ c |}{${{\rm d}\sigma}/{{\rm d}  x}  \
[{{\rm nb}}] $} \\
\multicolumn{2}{| c |}{\rb{$x\ {\rm range}$}} & 
\multicolumn{3}{ c |}{\rb{${{\rm d}\sigma}/{{\rm d}  x} \
[{{\rm nb}}]$}} &
\multicolumn{3}{ c |}{$ {\rm\ for\ } \pt^*(\dstar) > 2.0\ {\rm GeV} $} 
\\ \hline
 $0.00007$ & $0.00020$ & 4990 & $\pm 300 $ & $ \pm ^{440} _{390} $ & 2970 & $\pm 180 $ & $ \pm ^{230} _{210} $\\ 
 $0.00020$ & $0.00035$ & 6020 & $\pm 280 $ & $ \pm ^{460} _{440} $ & 3060 & $\pm 160 $ & $ \pm ^{220} _{210} $\\ 
 $0.00035$ & $0.00060$ & 4180 & $\pm 170 $ & $ \pm ^{320} _{310} $ & 1994 & $\pm 93 $ & $ \pm ^{143} _{141} $\\ 
 $0.00060$ & $0.00100$ & 2631 & $\pm 109 $ & $ \pm ^{190} _{188} $ & 1172 & $\pm 57 $ & $ \pm ^{87} _{86} $\\ 
 $0.00100$ & $0.00170$ & 1540 & $\pm 61 $  & $ \pm ^{107} _{108} $ & 586 & $\pm 31 $ & $ \pm ^{42} _{42} $\\ 
 $0.00170$ & $0.00330$ &  579 & $\pm 24 $  & $ \pm ^{44} _{45}   $ & 235 & $\pm 12 $ & $ \pm ^{22} _{22} $\\ 
 $0.00330$ & $0.05000$ & 13.24 &$\pm 0.61 $& $ \pm ^{1.17}_{1.33}$ & 4.20 & $\pm 0.24 $ & $ \pm ^{0.45} _{0.45} $\\

\hline
\end{tabular}
}
\end{center}
\caption{Differential \dstar\ cross section as a function of  $Q^2$, 
$y$ and $x$ in the kinematic range of $5 < Q^2 < 100\ {\rm GeV}^2$,
$0.02 < y < 0.7$, $|\eta(\dstar)|<1.8$ and 
$\pt(\dstar) > 1.25\ {\rm GeV}$.
The first quoted uncertainty is statistical and the second is systematic.
}
\label{Tab_xsec_kin}
\end{table}
\begin{table}[htbp]
\begin{center}
{\scriptsize
\begin{tabular}{|r r|d l l|d l l|}
\hline
\multicolumn{2}{| c |}{ } & 
\multicolumn{3}{ c |}{ } &
\multicolumn{3}{ c |}{${{\rm d}\sigma}/{{\rm d}  p_T} \
[{{\rm nb}}/{\rm GeV}] $} \\
\multicolumn{2}{| c |}{\rb{$p_T\ {\rm range} \  [{\rm GeV}]$}} & 
\multicolumn{3}{ c |}{\rb{${{\rm d}\sigma}/{{\rm d}  p_T} \
[{{\rm nb}}/{\rm GeV}]$}} &
\multicolumn{3}{ c |}{$ {\rm\ for\ } \pt^*(\dstar) > 2.0\ {\rm GeV} $} 
\\ \hline
$ 1.25$ & $\ \ 1.60$ & 2.55   & $\pm 0.21   $ & $ \pm ^{0.18} _{0.18}     $ & 0.334 & $\pm 0.042 $ & $ \pm ^{0.035} _{0.035} $\\
$ 1.60$ & $\ \ 1.88$ & 2.88   & $\pm 0.19   $ & $ \pm ^{0.20} _{0.20}     $ & 0.436 & $\pm 0.055 $ & $ \pm ^{0.041} _{0.042} $\\
$ 1.88$ & $\ \ 2.28$ & 2.68   & $\pm 0.11   $ & $ \pm ^{0.19} _{0.19}     $ & 0.853 & $\pm 0.057 $ & $ \pm ^{0.061} _{0.062} $\\
$ 2.28$ & $\ \ 2.68$ & 2.147  & $\pm 0.086  $ & $ \pm ^{0.149} _{0.149}   $ & 0.935 & $\pm 0.057 $ & $ \pm ^{0.086} _{0.086} $\\
$ 2.68$ & $\ \ 3.08$ & 1.538  & $\pm 0.058  $ & $ \pm ^{0.107} _{0.107}   $ & 0.744 & $\pm 0.047 $ & $ \pm ^{0.065} _{0.065} $\\
$ 3.08$ & $\ \ 3.50$ & 1.362  & $\pm 0.047  $ & $ \pm ^{0.094} _{0.094}   $ & 0.806 & $\pm 0.042 $ & $ \pm ^{0.061} _{0.061} $\\
$ 3.50$ & $\ \ 4.00$ & 0.924  & $\pm 0.032  $ & $ \pm ^{0.064} _{0.064}   $ & 0.620 & $\pm 0.033 $ & $ \pm ^{0.046} _{0.046} $\\
$ 4.00$ & $\ \ 4.75$ & 0.630  & $\pm 0.020  $ & $ \pm ^{0.043} _{0.043}   $ & 0.443 & $\pm 0.022 $ & $ \pm ^{0.031} _{0.031} $\\
$ 4.75$ & $\ \ 6.00$ & 0.2987 & $\pm 0.0098 $ & $ \pm ^{0.0209} _{0.0208} $ & 0.239 & $\pm 0.012 $ & $ \pm ^{0.016} _{0.016} $\\
$ 6.00$ & $\ \ 8.00$ & 0.0883 & $\pm 0.0039 $ & $ \pm ^{0.0070} _{0.0067} $ & 0.0769 & $\pm 0.0042 $ & $ \pm ^{0.0056} _{0.0054} $\\
$ 8.00$ & $   11.00$ & 0.0217 & $\pm 0.0015 $ & $ \pm ^{0.0016} _{0.0016} $ & 0.0210 & $\pm 0.0016 $ & $ \pm ^{0.0015} _{0.0014} $\\
$11.00$ & $   20.00$ & 0.00183& $\pm 0.00034$ & $ \pm ^{0.00023}_{0.00022}$ & 0.00188 & $\pm 0.00032 $ & $ \pm ^{0.00015} _{0.00013} $\\

\hline\hline
\multicolumn{2}{| c |}{ } & 
\multicolumn{3}{ c |}{ } &
\multicolumn{3}{ c |}{${{\rm d}\sigma}/{{\rm d}  \eta} \
[{{\rm nb}}]$} \\ 
\multicolumn{2}{| c |}{\rb{$\eta\ {\rm range}$}} & 
\multicolumn{3}{ c |}{\rb{${{\rm d}\sigma}/{{\rm d}  \eta} \
[{{\rm nb}}]$}} &
\multicolumn{3}{ c |}{$ {\rm\ for\ } \pt^*(\dstar) > 2.0\ {\rm GeV} $} 
\\ \hline
$ -1.80$ & $-1.56$ & 1.19  & $\pm 0.14  $ & $ \pm ^{0.09}  _{0.09}  $ & 0.460 & $\pm 0.078 $ & $ \pm ^{0.061} _{0.047} $\\
$ -1.56$ & $-1.32$ & 1.362 & $\pm 0.097 $ & $ \pm ^{0.101} _{0.102} $ & 0.500 & $\pm 0.051 $ & $ \pm ^{0.050} _{0.042} $\\
$ -1.32$ & $-1.08$ & 1.418 & $\pm 0.071 $ & $ \pm ^{0.102} _{0.100} $ & 0.592 & $\pm 0.037 $ & $ \pm ^{0.068} _{0.043} $\\
$ -1.08$ & $-0.84$ & 1.635 & $\pm 0.071 $ & $ \pm ^{0.118} _{0.116} $ & 0.672 & $\pm 0.036 $ & $ \pm ^{0.062} _{0.047} $\\
$ -0.84$ & $-0.60$ & 1.629 & $\pm 0.069 $ & $ \pm ^{0.115} _{0.115} $ & 0.728 & $\pm 0.038 $ & $ \pm ^{0.056} _{0.053} $\\
$ -0.60$ & $-0.36$ & 1.829 & $\pm 0.073 $ & $ \pm ^{0.130} _{0.130} $ & 0.814 & $\pm 0.041 $ & $ \pm ^{0.083} _{0.058} $\\
$ -0.36$ & $-0.12$ & 1.731 & $\pm 0.071 $ & $ \pm ^{0.123} _{0.121} $ & 0.836 & $\pm 0.042 $ & $ \pm ^{0.077} _{0.064} $\\
$ -0.12$ & $ 0.12$ & 1.878 & $\pm 0.081 $ & $ \pm ^{0.131} _{0.131} $ & 0.894 & $\pm 0.048 $ & $ \pm ^{0.070} _{0.067} $\\
$  0.12$ & $ 0.36$ & 1.763 & $\pm 0.078 $ & $ \pm ^{0.123} _{0.123} $ & 0.824 & $\pm 0.044 $ & $ \pm ^{0.065} _{0.060} $\\
$  0.36$ & $ 0.60$ & 1.927 & $\pm 0.090 $ & $ \pm ^{0.136} _{0.136} $ & 0.947 & $\pm 0.048 $ & $ \pm ^{0.074} _{0.068} $\\
$  0.60$ & $ 0.84$ & 1.880 & $\pm 0.095 $ & $ \pm ^{0.134} _{0.133} $ & 0.931 & $\pm 0.050 $ & $ \pm ^{0.075} _{0.066} $\\
$  0.84$ & $ 1.08$ & 2.025 & $\pm 0.097 $ & $ \pm ^{0.144} _{0.142} $ & 0.939 & $\pm 0.049 $ & $ \pm ^{0.067} _{0.065} $\\
$  1.08$ & $ 1.32$ & 2.19  & $\pm 0.12  $ & $ \pm ^{0.16}  _{0.16}  $ & 0.856 & $\pm 0.056 $ & $ \pm ^{0.062} _{0.059} $\\
$  1.32$ & $ 1.56$ & 1.97  & $\pm 0.17  $ & $ \pm ^{0.14}  _{0.14}  $ & 0.764 & $\pm 0.077 $ & $ \pm ^{0.055} _{0.055} $\\
$  1.56$ & $ 1.80$ & 1.93  & $\pm 0.24  $ & $ \pm ^{0.14}  _{0.14}  $ & 0.876 & $\pm 0.107 $ & $ \pm ^{0.075} _{0.069} $\\

\hline
\multicolumn{8}{c}{ }\\[-2.5ex]
\cline{1-5}
\multicolumn{2}{| c |}{$p_T^*\ {\rm range} \  [{\rm GeV}]$} & 
\multicolumn{3}{ c |}{${{\rm d}\sigma}/{{\rm d}  p_T^*} \
[{{\rm nb}}/{\rm GeV}]$} \\ \cline{1-5}
$ 0.300$ & $ 0.700$ & 1.26 &   $\pm 0.16   $ & $ \pm ^{0.18} _{0.18} $\\ 
$ 0.700$ & $ 1.125$ & 1.83 &   $\pm 0.14   $ & $ \pm ^{0.21} _{0.21} $\\ 
$ 1.125$ & $ 1.500$ & 2.22 &   $\pm 0.15   $ & $ \pm ^{0.18} _{0.19} $\\ 
$ 1.500$ & $ 1.880$ & 2.39 &   $\pm 0.14   $ & $ \pm ^{0.17} _{0.17} $\\ 
$ 1.880$ & $ 2.280$ & 2.02 &   $\pm 0.11   $ & $ \pm ^{0.14} _{0.14} $\\ 
$ 2.280$ & $ 2.680$ & 1.417 &  $\pm 0.086  $ & $ \pm ^{0.099} _{0.099} $\\ 
$ 2.680$ & $ 3.080$ & 1.055 &  $\pm 0.063  $ & $ \pm ^{0.074} _{0.074} $\\ 
$ 3.080$ & $ 3.500$ & 0.711 &  $\pm 0.045  $ & $ \pm ^{0.051} _{0.050} $\\ 
$ 3.500$ & $ 4.250$ & 0.453 &  $\pm 0.022  $ & $ \pm ^{0.033} _{0.033} $\\ 
$ 4.250$ & $ 6.000$ & 0.2028 & $\pm 0.0080 $ & $ \pm ^{0.0173} _{0.0162} $\\ 
$ 6.000$ & $11.000$ & 0.0287 & $\pm 0.0017 $ & $ \pm ^{0.0023} _{0.0022} $\\ 
$11.000$ & $20.000$ & 0.00278 &$\pm 0.00062$ & $ \pm ^{0.00030} _{0.00028} $\\ 

\cline{1-5}
\multicolumn{8}{c}{ }\\[-2.5ex]
\hline
\multicolumn{2}{| c |}{ } & 
\multicolumn{3}{ c |}{ } &
\multicolumn{3}{ c |}{${{\rm d}\sigma}/{{\rm d}  z} \
[{{\rm nb}}]$} \\
\multicolumn{2}{| c |}{\rb{$z\ {\rm range}$}} & 
\multicolumn{3}{ c |}{\rb{${{\rm d}\sigma}/{{\rm d}  z} \
[{{\rm nb}}]$}} &
\multicolumn{3}{ c |}{$ {\rm\ for\ } \pt^*(\dstar) > 2.0\ {\rm GeV} $} 
\\ \hline
 $0.000$ & $0.100$ & 5.12 & $\pm 0.59 $ & $ \pm ^{0.45} _{0.40}  $ & 1.72 & $\pm 0.26 $ & $ \pm ^{0.59} _{0.60} $\\
 $0.100$ & $0.200$ & 9.42 & $\pm 0.59 $ & $ \pm ^{0.86} _{0.80}  $ & 4.52 & $\pm 0.29 $ & $ \pm ^{0.41} _{0.40} $\\
 $0.200$ & $0.325$ &10.36 & $\pm 0.48 $ & $ \pm ^{0.86} _{0.78}  $ & 5.44 & $\pm 0.24 $ & $ \pm ^{0.40} _{0.40} $\\
 $0.325$ & $0.450$ & 9.66 & $\pm 0.41 $ & $ \pm ^{0.77} _{0.72}  $ & 4.91 & $\pm 0.21 $ & $ \pm ^{0.48} _{0.47} $\\
 $0.450$ & $0.575$ & 9.30 & $\pm 0.36 $ & $ \pm ^{0.68} _{0.71}  $ & 3.71 & $\pm 0.16 $ & $ \pm ^{0.42} _{0.42} $\\
 $0.575$ & $0.800$ & 4.97 & $\pm 0.16 $ & $ \pm ^{0.46} _{0.61}  $ & 1.156& $\pm 0.066$ & $ \pm ^{0.167}_{0.163} $\\
 $0.800$ & $1.000$ & 1.086& $\pm 0.082$ & $ \pm ^{0.305}_{0.266} $ & 0.347& $\pm 0.038$ & $ \pm ^{0.077}_{0.080} $\\

\hline
\end{tabular}
}
\end{center}
\caption{Differential \dstar\ cross section as a function of $\pt(\dstar)$,
$\eta(\dstar)$, $\pt^*(\dstar)$ and  $z(\dstar)$ 
in the kinematic range of $5 < Q^2 < 100\ {\rm GeV}^2$,
$0.02 < y < 0.7$, $|\eta(\dstar)|<1.8$, 
$\pt(\dstar) > 1.25\ {\rm GeV}$.
The first quoted uncertainty is statistical and the second is systematic.
}
\label{Tab_xsec_dstar1}
\end{table}

\begin{table}[htbp]
\begin{center}
{\scriptsize
\begin{tabular}{|r r|d l l|d l l|}
\hline
\multicolumn{2}{| c |}{$\eta\ {\rm range}$} & 
\multicolumn{3}{ c |}{${{\rm d}^2\sigma}/{{\rm d}  \eta}{{\rm d}  p_T} \
[{{\rm nb}}/{\rm GeV}]$} &
\multicolumn{3}{ c |}{${{\rm d}^2\sigma}/{{\rm d}  \eta}{{\rm d}  p^*_T} \
[{{\rm nb}}/{\rm GeV}]$} 
\\ \hline
& & \multicolumn{3}{ c |}{$ 1.25 < p_T < 2.00\ {\rm GeV}$} 
& \multicolumn{3}{ c |}{$ 0.30 < p^*_T < 1.25\ {\rm GeV}$} 
\\ \hline
$ -1.8$ &$ -1.2   $ & 0.760 & $\pm 0.074 $ & $ \pm ^{0.058} _{0.055} $ & 0.547 & $\pm 0.051 $ & $ \pm ^{0.041} _{0.041} $ \\ 
$ -1.2$ &$ -0.6   $ & 0.701 & $\pm 0.055 $ & $ \pm ^{0.052} _{0.052} $ & 0.453 & $\pm 0.035 $ & $ \pm ^{0.037} _{0.037} $ \\ 
$-0.6 $ &$\ \ \ 0.0$ & 0.704 & $\pm 0.057 $ & $ \pm ^{0.052} _{0.052} $ & 0.443 & $\pm 0.038 $ & $ \pm ^{0.051} _{0.051} $ \\ 
$ 0.0 $ &$\ \ \ 0.6$ & 0.663 & $\pm 0.062 $ & $ \pm ^{0.049} _{0.049} $ & 0.398 & $\pm 0.046 $ & $ \pm ^{0.046} _{0.047} $ \\ 
$ 0.6 $ &$\ \ \ 1.2$ & 0.760 & $\pm 0.078 $ & $ \pm ^{0.053} _{0.054} $ & 0.427 & $\pm 0.084 $ & $ \pm ^{0.072} _{0.070} $ \\ 
$ 1.2 $ &$\ \ \ 1.8$ & 1.23  & $\pm 0.15  $ & $ \pm ^{0.09} _{0.09}   $ & 0.56  & $\pm 0.18  $ & $ \pm ^{0.04}  _{0.05}  $ \\ 
\hline\hline
& & \multicolumn{3}{ c |}{$ 2.00 < p_T < 2.75\ {\rm GeV}$} 
& \multicolumn{3}{ c |}{$ 1.25 < p^*_T < 2.00\ {\rm GeV}$}
\\ \hline
$ -1.8$ &$ -1.2   $ & 0.573 & $\pm 0.042 $ & $ \pm ^{0.044} _{0.044} $ & 0.481 & $\pm 0.047 $ & $ \pm ^{0.038} _{0.035} $ \\ 
$ -1.2$ &$ -0.6   $ & 0.586 & $\pm 0.030 $ & $ \pm ^{0.042} _{0.041} $ & 0.583 & $\pm 0.040 $ & $ \pm ^{0.041} _{0.041} $ \\ 
$-0.6 $ &$\ \ \ 0.0$ & 0.688 & $\pm 0.034 $ & $ \pm ^{0.051} _{0.051} $ & 0.640 & $\pm 0.044 $ & $ \pm ^{0.047} _{0.045} $ \\ 
$ 0.0 $ &$\ \ \ 0.6$ & 0.703 & $\pm 0.043 $ & $ \pm ^{0.050} _{0.050} $ & 0.614 & $\pm 0.060 $ & $ \pm ^{0.043} _{0.048} $ \\ 
$ 0.6 $ &$\ \ \ 1.2$ & 0.783 & $\pm 0.045 $ & $ \pm ^{0.055} _{0.055} $ & 0.708 & $\pm 0.079 $ & $ \pm ^{0.051} _{0.057} $ \\ 
$ 1.2 $ &$\ \ \ 1.8$ & 0.723 & $\pm 0.062 $ & $ \pm ^{0.051} _{0.052} $ & 0.77  & $\pm 0.16  $ & $ \pm ^{0.06}  _{0.06}  $ \\ 
\hline\hline
& & \multicolumn{3}{ c |}{$ 2.75 < p_T < 4.00\ {\rm GeV}$} 
& \multicolumn{3}{ c |}{$ 2.00 < p^*_T < 3.00\ {\rm GeV}$} 
\\ \hline
$ -1.8$ &$ -1.2   $ & 0.227 & $\pm 0.017 $ & $ \pm ^{0.018} _{0.017} $ & 0.336 & $\pm 0.028 $ & $ \pm ^{0.029} _{0.029} $ \\
$ -1.2$ &$ -0.6   $ & 0.336 & $\pm 0.014 $ & $ \pm ^{0.023} _{0.023} $ & 0.390 & $\pm 0.021 $ & $ \pm ^{0.029} _{0.027} $ \\
$-0.6 $ &$\ \ \ 0.0$ & 0.359 & $\pm 0.014 $ & $ \pm ^{0.025} _{0.025} $ & 0.392 & $\pm 0.024 $ & $ \pm ^{0.028} _{0.028} $ \\
$ 0.0 $ &$\ \ \ 0.6$ & 0.401 & $\pm 0.016 $ & $ \pm ^{0.028} _{0.028} $ & 0.474 & $\pm 0.028 $ & $ \pm ^{0.034} _{0.035} $ \\
$ 0.6 $ &$\ \ \ 1.2$ & 0.377 & $\pm 0.017 $ & $ \pm ^{0.027} _{0.026} $ & 0.549 & $\pm 0.033 $ & $ \pm ^{0.040} _{0.040} $ \\
$ 1.2 $ &$\ \ \ 1.8$ & 0.304 & $\pm 0.024 $ & $ \pm ^{0.024} _{0.024} $ & 0.530 & $\pm 0.057 $ & $ \pm ^{0.038} _{0.039} $ \\
\hline\hline
& & \multicolumn{3}{ c |}{$ 4.00 < p_T < 6.00\ {\rm GeV}$}  
& \multicolumn{3}{ c |}{$ 3.00 < p^*_T < 6.00\ {\rm GeV}$}
\\ \hline
$ -1.8$ &$ -1.2   $ & 0.0368 & $\pm 0.0051 $ & $ \pm ^{0.0034} _{0.0034} $ & 0.0419 & $\pm 0.0047 $ & $ \pm ^{0.0046} _{0.0040} $ \\ 
$ -1.2$ &$ -0.6   $ & 0.1017 & $\pm 0.0051 $ & $ \pm ^{0.0072} _{0.0074} $ & 0.0875 & $\pm 0.0042 $ & $ \pm ^{0.0085} _{0.0068} $ \\ 
$-0.6 $ &$\ \ \ 0.0$ & 0.1480 & $\pm 0.0059 $ & $ \pm ^{0.0104} _{0.0104} $ & 0.1179 & $\pm 0.0048 $ & $ \pm ^{0.0099} _{0.0095} $ \\ 
$ 0.0 $ &$\ \ \ 0.6$ & 0.1502 & $\pm 0.0068 $ & $ \pm ^{0.0107} _{0.0108} $ & 0.1203 & $\pm 0.0055 $ & $ \pm ^{0.0106} _{0.0103} $ \\ 
$ 0.6 $ &$\ \ \ 1.2$ & 0.1503 & $\pm 0.0068 $ & $ \pm ^{0.0106} _{0.0109} $ & 0.1185 & $\pm 0.0063 $ & $ \pm ^{0.0103} _{0.0095} $ \\ 
$ 1.2 $ &$\ \ \ 1.8$ & 0.0991 & $\pm 0.0095 $ & $ \pm ^{0.0075} _{0.0079} $ & 0.0947 & $\pm 0.0095 $ & $ \pm ^{0.0077} _{0.0072} $ \\ 
\hline\hline
& & \multicolumn{3}{ c |}{$ 6.00 < p_T < 20.00\ {\rm GeV}$}  
& \multicolumn{3}{ c |}{$ 6.00 < p^*_T < 20.00\ {\rm GeV}$}
\\ \hline
$ -1.8$ &$ -1.2   $ & 0.00073 & $\pm 0.00030 $ & $ \pm ^{0.00009} _{0.00012} $ & 0.00030 & $\pm 0.00027 $ & $ \pm ^{0.00004} _{0.00005} $ \\ 
$ -1.2$ &$ -0.6   $ & 0.00243 & $\pm 0.00035 $ & $ \pm ^{0.00018} _{0.00018} $ & 0.00196 & $\pm 0.00028 $ & $ \pm ^{0.00021} _{0.00016} $ \\ 
$-0.6 $ &$\ \ \ 0.0$ & 0.00653 & $\pm 0.00050 $ & $ \pm ^{0.00047} _{0.00050} $ & 0.00377 & $\pm 0.00040 $ & $ \pm ^{0.00036} _{0.00029} $ \\ 
$ 0.0 $ &$\ \ \ 0.6$ & 0.00761 & $\pm 0.00053 $ & $ \pm ^{0.00056} _{0.00057} $ & 0.00439 & $\pm 0.00047 $ & $ \pm ^{0.00051} _{0.00044} $ \\ 
$ 0.6 $ &$\ \ \ 1.2$ & 0.00724 & $\pm 0.00057 $ & $ \pm ^{0.00058} _{0.00064} $ & 0.00571 & $\pm 0.00054 $ & $ \pm ^{0.00052} _{0.00051} $ \\ 
$ 1.2 $ &$\ \ \ 1.8$ & 0.00462 & $\pm 0.00076 $ & $ \pm ^{0.00082} _{0.00082} $ & 0.00257 & $\pm 0.00079 $ & $ \pm ^{0.00020} _{0.00021} $ \\ 

\hline
\end{tabular}
}
\end{center}
\caption{Double differential \dstar\ cross sections as a function of 
$\eta(\dstar)$ and $\pt(\dstar)$ and as a function of $\eta(\dstar)$ 
and $\pt^*(\dstar)$ 
in the kinematic range of $5 < Q^2 < 100\ {\rm GeV}^2$,
$0.02 < y < 0.7$, $|\eta(\dstar)|<1.8$, 
$\pt(\dstar) > 1.25\ {\rm GeV}$.
The first quoted uncertainty is statistical and the second is systematic.
}
\label{Tab_xsec_2d}
\end{table}

\begin{table}[htbp]
\begin{center}
{\scriptsize
\begin{tabular}{|r r|d l l|d l l|}
\hline
\multicolumn{2}{| c |}{} & 
\multicolumn{3}{ c |}{${{\rm d}^2\sigma}/{{\rm d} y}{{\rm d}  Q^2} \
[{{\rm nb}}/{\rm GeV}^2]$}  &
\multicolumn{3}{ c |}{${{\rm d}^2\sigma}/{{\rm d} y}{{\rm d}  Q^2} \
[{{\rm nb}}/{\rm GeV}^2]$}  
\\ 
\multicolumn{2}{| c |}{\rb{$y\ {\rm range}$}} & 
\multicolumn{3}{ c |}{$ {\rm\ for\ } \pt(\dstar) > 1.25 ~{\rm GeV}, 
 |\eta(\dstar)| < 1.8$}  &
\multicolumn{3}{ c |}{$ {\rm\ for\ } \pt(\dstar) > 1.5 ~{\rm GeV}, 
 |\eta(\dstar)| < 1.5$}  
\\ \hline
& & \multicolumn{6}{ c |}{$ 5 < Q^2 < 9\ {\rm GeV}^2$} 
\\ \hline
$ 0.02$ & $0.05$ & 2.27  & $\pm 0.19  $ & $ \pm ^{0.22}  _{0.24}  $ & 1.23  & $\pm 0.14 $  & $ \pm ^{0.11}  _{0.11}  $\\
$ 0.05$ & $0.09$ & 1.95  & $\pm 0.16  $ & $ \pm ^{0.14}  _{0.14}  $ & 1.57  & $\pm 0.12 $  & $ \pm ^{0.12}  _{0.14}  $\\
$ 0.09$ & $0.16$ & 1.767 & $\pm 0.096 $ & $ \pm ^{0.127} _{0.127} $ & 1.378 & $\pm 0.077 $ & $ \pm ^{0.114} _{0.131} $\\
$ 0.16$ & $0.32$ & 0.954 & $\pm 0.052 $ & $ \pm ^{0.077} _{0.072} $ & 0.839 & $\pm 0.039 $ & $ \pm ^{0.077} _{0.080} $\\
$ 0.32$ & $0.70$ & 0.361 & $\pm 0.024 $ & $ \pm ^{0.030} _{0.027} $ & 0.243 & $\pm 0.018 $ & $ \pm ^{0.024} _{0.025} $\\
\hline\hline 
& & \multicolumn{6}{ c |}{$ 9 < Q^2 < 14\ {\rm GeV}^2$} 
\\ \hline
$ 0.02$ & $0.05$ & 0.845 & $\pm 0.101 $ & $ \pm ^{0.078} _{0.086} $ & 0.457 & $\pm 0.082 $ & $ \pm ^{0.045} _{0.042} $\\
$ 0.05$ & $0.09$ & 0.953 & $\pm 0.085 $ & $ \pm ^{0.072} _{0.075} $ & 0.745 & $\pm 0.068 $ & $ \pm ^{0.065} _{0.074} $\\
$ 0.09$ & $0.16$ & 0.687 & $\pm 0.054 $ & $ \pm ^{0.050} _{0.051} $ & 0.581 & $\pm 0.044 $ & $ \pm ^{0.048} _{0.052} $\\
$ 0.16$ & $0.32$ & 0.447 & $\pm 0.029 $ & $ \pm ^{0.032} _{0.031} $ & 0.414 & $\pm 0.028 $ & $ \pm ^{0.031} _{0.035} $\\
$ 0.32$ & $0.70$ & 0.193 & $\pm 0.015 $ & $ \pm ^{0.015} _{0.015} $ & 0.144 & $\pm 0.011 $ & $ \pm ^{0.013} _{0.013} $\\
\hline\hline 
& & \multicolumn{6}{ c |}{$ 14 < Q^2 < 23\ {\rm GeV}^2$} 
\\ \hline
$ 0.02$ & $0.05$ & 0.444 & $\pm 0.055 $ & $ \pm ^{0.052} _{0.049} $ & 0.249  & $\pm 0.032 $  & $ \pm ^{0.022} _{0.023} $\\
$ 0.05$ & $0.09$ & 0.434 & $\pm 0.040 $ & $ \pm ^{0.030} _{0.033} $ & 0.359  & $\pm 0.035 $  & $ \pm ^{0.030} _{0.033} $\\
$ 0.09$ & $0.16$ & 0.356 & $\pm 0.028 $ & $ \pm ^{0.030} _{0.029} $ & 0.303  & $\pm 0.021 $  & $ \pm ^{0.030} _{0.032} $\\
$ 0.16$ & $0.32$ & 0.249 & $\pm 0.015 $ & $ \pm ^{0.018} _{0.018} $ & 0.208  & $\pm 0.012 $  & $ \pm ^{0.016} _{0.019} $\\
$ 0.32$ & $0.70$ & 0.0887& $\pm 0.0078$ & $ \pm ^{0.0079}_{0.0077}$ & 0.0659 & $\pm 0.0055 $ & $ \pm ^{0.0065}_{0.0068}$\\
\hline\hline 
& &\multicolumn{6}{ c |}{$ 23 < Q^2 < 45\ {\rm GeV}^2$} 
\\ \hline
$ 0.02$ & $0.05$ & 0.105 & $\pm 0.016 $ & $ \pm ^{0.012} _{0.011} $ & 0.087  & $\pm 0.012 $  & $ \pm ^{0.014} _{0.014} $\\ 
$ 0.05$ & $0.09$ & 0.160 & $\pm 0.016 $ & $ \pm ^{0.013} _{0.014} $ & 0.120  & $\pm 0.013 $  & $ \pm ^{0.011} _{0.012} $\\ 
$ 0.09$ & $0.16$ & 0.125 & $\pm 0.011 $ & $ \pm ^{0.010} _{0.009} $ & 0.1211 & $\pm 0.0095 $ & $ \pm ^{0.0107}_{0.0120}$\\ 
$ 0.16$ & $0.32$ & 0.0885& $\pm 0.0059$ & $ \pm ^{0.0062}_{0.0064}$ & 0.0744 & $\pm 0.0046 $ & $ \pm ^{0.0056}_{0.0065}$\\ 
$ 0.32$ & $0.70$ & 0.0375& $\pm 0.0031$ & $ \pm ^{0.0031}_{0.0030}$ & 0.0304 & $\pm 0.0024 $ & $ \pm ^{0.0027}_{0.0029}$\\ 
\hline\hline 
& & \multicolumn{6}{ c |}{$ 45 < Q^2 < 100\ {\rm GeV}^2$} 
\\ \hline
$ 0.02$ & $0.05$ & 0.0150& $\pm 0.0085$ & $ \pm ^{0.0015}_{0.0020}$ & 0.0054 & $\pm 0.0026 $ & $ \pm ^{0.0007} _{0.0005} $\\ 
$ 0.05$ & $0.09$ & 0.0302& $\pm 0.0054$ & $ \pm ^{0.0024}_{0.0024}$ & 0.0249 & $\pm 0.0041 $ & $ \pm ^{0.0022} _{0.0025} $\\ 
$ 0.09$ & $0.16$ & 0.0258& $\pm 0.0034$ & $ \pm ^{0.0021}_{0.0020}$ & 0.0215 & $\pm 0.0028 $ & $ \pm ^{0.0023} _{0.0023} $\\ 
$ 0.16$ & $0.32$ & 0.0235& $\pm 0.0022$ & $ \pm ^{0.0019}_{0.0018}$ & 0.0236 & $\pm 0.0019 $ & $ \pm ^{0.0022} _{0.0024} $\\ 
$ 0.32$ & $0.70$ & 0.0097& $\pm 0.0011$ & $ \pm ^{0.0008}_{0.0008}$ & 0.00729& $\pm 0.00085$ & $ \pm ^{0.00065}_{0.00072}$\\ 

\hline
\end{tabular}
}
\end{center}
\caption{Double differential \dstar\ cross sections as a function of 
$y$ and $Q^2$ in two different kinematic ranges: $|\eta(\dstar)|<1.8$ and
$\pt(\dstar) > 1.25\ {\rm GeV}$ or $|\eta(\dstar)|<1.5$ and
$\pt(\dstar) > 1.5\ {\rm GeV}$.
The first quoted uncertainty is statistical and the second is systematic.
}
\label{Tab_xsec_yq2}
\end{table}

\begin{table}[htbp]
\begin{center}
{\scriptsize
\begin{tabular}{|r r|d l l|}
\hline
\multicolumn{2}{| c |}{$Q^2\ {\rm range} \  [{\rm GeV}^2]$} & 
\multicolumn{3}{ c |}{${{\rm d}\sigma}/{{\rm d}  Q^2} \
[{{\rm nb}}/{\rm GeV^2}]$} \\ \hline
$  5.0$ & $  6.0$ & 0.552  & $\pm 0.032  $ & $ \pm ^{0.046}  _{0.043}  $\\ 
$  6.0$ & $  8.0$ & 0.398  & $\pm 0.016  $ & $ \pm ^{0.031}  _{0.030}  $\\ 
$  8.0$ & $ 10.0$ & 0.278  & $\pm 0.013  $ & $ \pm ^{0.020}  _{0.020}  $\\ 
$ 10.0$ & $ 13.0$ & 0.1983 & $\pm 0.0088 $ & $ \pm ^{0.0143} _{0.0143} $\\ 
$ 13.0$ & $ 19.0$ & 0.1236 & $\pm 0.0042 $ & $ \pm ^{0.0088} _{0.0088} $\\ 
$ 19.0$ & $ 27.5$ & 0.0679 & $\pm 0.0028 $ & $ \pm ^{0.0048} _{0.0048} $\\ 
$ 27.5$ & $ 40.0$ & 0.0374 & $\pm 0.0017 $ & $ \pm ^{0.0027} _{0.0027} $\\ 
$ 40.0$ & $ 60.0$ & 0.01562& $\pm 0.00095$ & $ \pm ^{0.00113}_{0.00110}$\\ 
$ 60.0$ & $100.0$ & 0.00724& $\pm 0.00045$ & $ \pm ^{0.00053}_{0.00051}$\\ 

\hline\hline
\multicolumn{2}{| c |}{$y\ {\rm range}$} & 
\multicolumn{3}{ c |}{${{\rm d}\sigma}/{{\rm d}  y} \
[{{\rm nb}}]$} \\ \hline
 $0.02$ & $0.05$ & 12.21 & $\pm 0.64 $ & $ \pm ^{1.47} _{1.47} $\\ 
 $0.05$ & $0.09$ & 16.39 & $\pm 0.69 $ & $ \pm ^{1.21} _{1.20} $\\ 
 $0.09$ & $0.13$ & 15.89 & $\pm 0.72 $ & $ \pm ^{1.30} _{1.25} $\\ 
 $0.13$ & $0.18$ & 12.71 & $\pm 0.61 $ & $ \pm ^{0.92} _{0.92} $\\ 
 $0.18$ & $0.26$ & 10.90 & $\pm 0.42 $ & $ \pm ^{0.84} _{0.82} $\\ 
 $0.26$ & $0.36$ &  6.85 & $\pm 0.31 $ & $ \pm ^{0.53} _{0.51} $\\ 
 $0.36$ & $0.50$ &  4.24 & $\pm 0.22 $ & $ \pm ^{0.37} _{0.34} $\\ 
 $0.50$ & $0.70$ &  2.13 & $\pm 0.17 $ & $ \pm ^{0.20} _{0.18} $\\ 

\hline\hline
\multicolumn{2}{| c |}{$x\ {\rm range}$} & 
\multicolumn{3}{ c |}{${{\rm d}\sigma}/{{\rm d}  x} \
[{{\rm nb}}]$} \\ \hline
 $0.00007$ & $0.00020$ & 3320 & $\pm 200 $ & $ \pm ^{270} _{250} $\\ 
 $0.00020$ & $0.00035$ & 4780 & $\pm 220 $ & $ \pm ^{370} _{350} $\\ 
 $0.00035$ & $0.00060$ & 3430 & $\pm 130 $ & $ \pm ^{250} _{250} $\\ 
 $0.00060$ & $0.00100$ & 2034 & $\pm  81 $ & $ \pm ^{147} _{145} $\\ 
 $0.00100$ & $0.00170$ & 1225 & $\pm  46 $ & $ \pm  ^{90}  _{89} $\\ 
 $0.00170$ & $0.00330$ &  446 & $\pm  18 $ & $ \pm  ^{34}  _{34} $\\ 
 $0.00330$ & $0.05000$ & 10.15& $\pm 0.44$ & $ \pm^{0.84} _{0.82} $\\ 

\hline
\end{tabular}
}
\end{center}
\caption{Differential \dstar\ cross section as a function of  $Q^2$, 
$y$ and $x$   in the kinematic range of 
$5 < Q^2 < 100\ {\rm GeV}^2$,
$0.02 < y < 0.7$, $|\eta(\dstar)|<1.5$ and 
$\pt(\dstar) > 1.5\ {\rm GeV}$.
The first quoted uncertainty is statistical and the second is systematic.
}
\label{Tab_xsec_kin_redphasespace}
\end{table}
\begin{table}[htbp]
\begin{center}
{\scriptsize
\begin{tabular}{|r r|d l l|}
\hline
\multicolumn{2}{| c |}{$p_T\ {\rm range} \  [{\rm GeV}]$} & 
\multicolumn{3}{ c |}{${{\rm d}\sigma}/{{\rm d}  p_T} \
[{{\rm nb}}/{\rm GeV}]$} \\ \hline
$ 1.50$ & $\ \ 1.88$ & 2.34   & $\pm 0.15   $ & $ \pm ^{0.17} _{0.17} $\\ 
$ 1.88$ & $\ \ 2.28$ & 2.042  & $\pm 0.093  $ & $ \pm ^{0.144} _{0.144} $\\ 
$ 2.28$ & $\ \ 2.68$ & 1.959  & $\pm 0.070  $ & $ \pm ^{0.140} _{0.140} $\\ 
$ 2.68$ & $\ \ 3.08$ & 1.384  & $\pm 0.050  $ & $ \pm ^{0.096} _{0.096} $\\ 
$ 3.08$ & $\ \ 3.50$ & 1.152  & $\pm 0.043  $ & $ \pm ^{0.080} _{0.079} $\\ 
$ 3.50$ & $\ \ 4.00$ & 0.814  & $\pm 0.028  $ & $ \pm ^{0.056} _{0.056} $\\ 
$ 4.00$ & $\ \ 4.75$ & 0.575  & $\pm 0.018  $ & $ \pm ^{0.040} _{0.040} $\\ 
$ 4.75$ & $\ \ 6.00$ & 0.2714 & $\pm 0.0088 $ & $ \pm ^{0.0189} _{0.0187} $\\ 
$ 6.00$ & $\ \ 8.00$ & 0.0851 & $\pm 0.0037 $ & $ \pm ^{0.0058} _{0.0058} $\\ 
$ 8.00$ & $   11.00$ & 0.0211 & $\pm 0.0015 $ & $ \pm ^{0.0017} _{0.0016} $\\ 
$11.00$ & $   20.00$ & 0.00178& $\pm 0.00028$ & $ \pm ^{0.00013} _{0.00012} $\\ 

\hline\hline
\multicolumn{2}{| c |}{$\eta\ {\rm range}$} & 
\multicolumn{3}{ c |}{${{\rm d}\sigma}/{{\rm d}  \eta} \
[{{\rm nb}}]$} \\ \hline
 $-1.50$ & $-1.25$ & 1.229 & $\pm 0.077 $ & $ \pm ^{0.090} _{0.088} $\\ 
 $-1.25$ & $-1.00$ & 1.319 & $\pm 0.062 $ & $ \pm ^{0.098} _{0.094} $\\ 
 $-1.00$ & $-0.75$ & 1.501 & $\pm 0.061 $ & $ \pm ^{0.113} _{0.108} $\\ 
 $-0.75$ & $-0.50$ & 1.635 & $\pm 0.065 $ & $ \pm ^{0.118} _{0.116} $\\ 
 $-0.50$ & $-0.25$ & 1.569 & $\pm 0.063 $ & $ \pm ^{0.112} _{0.109} $\\ 
 $-0.25$ & $ 0.00$ & 1.629 & $\pm 0.066 $ & $ \pm ^{0.118} _{0.116} $\\ 
 $ 0.00$ & $ 0.25$ & 1.667 & $\pm 0.070 $ & $ \pm ^{0.117} _{0.117} $\\ 
 $ 0.25$ & $ 0.50$ & 1.677 & $\pm 0.074 $ & $ \pm ^{0.121} _{0.119} $\\ 
 $ 0.50$ & $ 0.75$ & 1.756 & $\pm 0.078 $ & $ \pm ^{0.126} _{0.124} $\\ 
 $ 0.75$ & $ 1.00$ & 1.746 & $\pm 0.080 $ & $ \pm ^{0.131} _{0.128} $\\ 
 $ 1.00$ & $ 1.25$ & 2.024 & $\pm 0.095 $ & $ \pm ^{0.150} _{0.146} $\\ 
 $ 1.25$ & $ 1.50$ & 1.73  & $\pm 0.12  $ & $ \pm ^{0.13} _{0.12} $\\ 

\hline\hline
\multicolumn{2}{| c |}{$p_T^*\ {\rm range} \  [{\rm GeV}]$} & 
\multicolumn{3}{ c |}{${{\rm d}\sigma}/{{\rm d}  p_T^*} \
[{{\rm nb}}/{\rm GeV}]$} \\ \hline
 $0.300 $& $\ \ 0.700$ & 0.75	& $\pm 0.13   $ & $ \pm ^{0.13} _{0.13} $\\ 
 $0.700 $& $\ \ 1.125$ & 1.34	& $\pm 0.12   $ & $ \pm ^{0.14} _{0.14} $\\ 
 $1.125 $& $\ \ 1.500$ & 1.48	& $\pm 0.13   $ & $ \pm ^{0.16} _{0.16} $\\ 
 $1.500 $& $\ \ 1.880$ & 1.62	& $\pm 0.12   $ & $ \pm ^{0.11} _{0.12} $\\ 
 $1.880 $& $\ \ 2.280$ & 1.511  & $\pm 0.093  $ & $ \pm ^{0.106} _{0.108} $\\ 
 $2.280 $& $\ \ 2.680$ & 1.163  & $\pm 0.073  $ & $ \pm ^{0.087} _{0.086} $\\ 
 $2.680 $& $\ \ 3.080$ & 0.884  & $\pm 0.055  $ & $ \pm ^{0.061} _{0.061} $\\ 
 $3.080 $& $\ \ 3.500$ & 0.570  & $\pm 0.039  $ & $ \pm ^{0.041} _{0.040} $\\ 
 $3.500 $& $\ \ 4.250$ & 0.403  & $\pm 0.020  $ & $ \pm ^{0.030} _{0.030} $\\ 
 $4.250 $& $\ \ 6.000$ & 0.1785 & $\pm 0.0069 $ & $ \pm ^{0.0166} _{0.0156} $\\ 
 $6.000 $& $   11.000$ & 0.0269 & $\pm 0.0015 $ & $ \pm ^{0.0026} _{0.0025} $\\ 
 $11.000$ &$   20.000$ & 0.00186& $\pm 0.00041$ & $ \pm ^{0.00020}_{0.00016} $\\ 

\hline\hline
\multicolumn{2}{| c |}{$z\ {\rm range}$} & 
\multicolumn{3}{ c |}{${{\rm d}\sigma}/{{\rm d}  z} \
[{{\rm nb}}]$} \\ \hline
$ 0.000$ & $0.100$ & 3.29 & $\pm 0.41 $ & $ \pm ^{0.29} _{0.29} $\\ 
$ 0.100$ & $0.200$ & 7.02 & $\pm 0.44 $ & $ \pm ^{0.58} _{0.58} $\\ 
$ 0.200$ & $0.325$ & 8.22 & $\pm 0.36 $ & $ \pm ^{0.70} _{0.70} $\\ 
$ 0.325$ & $0.450$ & 7.59 & $\pm 0.31 $ & $ \pm ^{0.65} _{0.63} $\\ 
$ 0.450$ & $0.575$ & 7.40 & $\pm 0.28 $ & $ \pm ^{0.59} _{0.59} $\\ 
$ 0.575$ & $0.800$ & 4.06 & $\pm 0.13 $ & $ \pm ^{0.34} _{0.35} $\\ 
$ 0.800$ & $1.000$ & 0.861& $\pm 0.064$ & $ \pm ^{0.233} _{0.204} $\\ 

\hline
\end{tabular}
}
\end{center}
\caption{Differential \dstar\ cross section as a function of 
$\pt(\dstar)$, $\eta(\dstar)$, $\pt^*(\dstar)$ and $z(\dstar)$ 
in the kinematic range of 
$5 < Q^2 < 100\ {\rm GeV}^2$,
$0.02 < y < 0.7$, $|\eta(\dstar)|<1.5$ and 
$\pt(\dstar) > 1.5\ {\rm GeV}$.
The first quoted uncertainty is statistical and the second is systematic.
}
\label{Tab_xsec_dstar1_redphasespace}
\end{table}

\begin{table}[htbp]
\begin{center}
{\scriptsize
\begin{tabular}{|d|c|d l|d|l||d l|}
\hline
\multicolumn{1}{| c |}{$$} & 
\multicolumn{1}{ c |}{$$} & 
\multicolumn{2}{ c |}{HVQDIS} &   
\multicolumn{1}{ c |}{$$} &   
\multicolumn{1}{ c ||}{$$} &
\multicolumn{2}{ c |}{CASCADE}    
\\   
\multicolumn{1}{| c |}{\rb{$Q^2\  [{\rm GeV}^2]$}} & 
\multicolumn{1}{ c |}{\rb{$x$}} & 
\multicolumn{1}{ c }{\ftwocc} &   
\multicolumn{1}{ c |}{$\delta_{ext}\ [\%]$} &   
\multicolumn{1}{ c |}{\rb{$\delta_{stat}\ [\%]$}} &   
\multicolumn{1}{ c ||}{\rb{$\delta_{syst}\ [\%]$}} &    
\multicolumn{1}{ c }{\ftwocc} &   
\multicolumn{1}{ c |}{$\delta_{ext}\ [\%]$}    
\\ \hline

 6.5 & $1.3\cdot 10^{-4}$ & 0.2160 & $ \pm _{8.7} ^{8.5} $ & \pm 6.7 & $ \pm ^{7.7} _{8.1}  $ & 0.2005 & $ \pm _{7.3}  ^{16.6} $  \\  
 6.5 & $3.2\cdot 10^{-4}$ & 0.1576 & $ \pm _{3.2} ^{4.3} $ & \pm 5.5 & $ \pm ^{7.7} _{8.0}  $ & 0.1634 & $ \pm _{10.7} ^{12.3} $ \\   
 6.5 & $5.0\cdot 10^{-4}$ & 0.1516 & $ \pm _{4.5} ^{4.2} $ & \pm 5.4 & $ \pm ^{7.2} _{7.3}  $ & 0.1597 & $ \pm _{11.9} ^{11.1} $ \\   
 6.5 & $8.0\cdot 10^{-4}$ & 0.1036 & $ \pm _{3.4} ^{5.7} $ & \pm 8.1 & $ \pm ^{7.2} _{7.2}  $ & 0.1153 & $ \pm _{12.6} ^{6.0} $  \\   
 6.5 & $2.0\cdot 10^{-3}$ & 0.0735 & $ \pm _{7.2} ^{10.8}$ & \pm 8.6 & $ \pm ^{9.9} _{10.4} $ & 0.1044 & $ \pm _{12.2} ^{7.2} $  \\[0.2cm]

12.0 & $3.2\cdot 10^{-4}$ & 0.2829 & $ \pm _{5.6} ^{8.7} $ & \pm 7.7 & $ \pm ^{7.6} _{7.9}  $ & 0.2727 & $ \pm _{8.6}  ^{15.5} $  \\  
12.0 & $5.0\cdot 10^{-4}$ & 0.2123 & $ \pm _{2.9} ^{3.1} $ & \pm 6.6 & $ \pm ^{7.1} _{7.1}  $ & 0.2169 & $ \pm _{10.7} ^{12.5} $ \\   
12.0 & $8.0\cdot 10^{-4}$ & 0.1689 & $ \pm _{2.3} ^{4.6} $ & \pm 7.8 & $ \pm ^{7.4} _{7.3}  $ & 0.1779 & $ \pm _{11.7} ^{10.0} $ \\   
12.0 & $2.0\cdot 10^{-3}$ & 0.1226 & $ \pm _{3.5} ^{6.1} $ & \pm 8.9 & $ \pm ^{7.7} _{7.7}  $ & 0.1353 & $ \pm _{14.4} ^{8.7} $  \\   
12.0 & $3.2\cdot 10^{-3}$ & 0.0773 & $ \pm _{7.4} ^{11.6}$ & \pm 12.0& $ \pm ^{9.6} _{9.8}  $ & 0.1125 & $ \pm _{16.5} ^{5.7} $  \\[0.2cm]   

18.0 & $5.0\cdot 10^{-4}$ & 0.3221 & $ \pm _{5.0} ^{4.6} $ & \pm 8.8 & $ \pm ^{8.5} _{9.1}  $ & 0.3045 & $ \pm _{9.2}  ^{15.2} $  \\  
18.0 & $8.0\cdot 10^{-4}$ & 0.2899 & $ \pm _{2.1} ^{3.8} $ & \pm 6.1 & $ \pm ^{7.3} _{7.2}  $ & 0.2964 & $ \pm _{10.6} ^{11.7} $ \\   
18.0 & $1.3\cdot 10^{-3}$ & 0.2167 & $ \pm _{2.9} ^{4.0} $ & \pm 8.0 & $ \pm ^{8.2} _{8.3}  $ & 0.2202 & $ \pm _{12.9} ^{11.0} $ \\   
18.0 & $3.2\cdot 10^{-3}$ & 0.1368 & $ \pm _{3.5} ^{5.3} $ & \pm 9.3 & $ \pm ^{7.4} _{7.2}  $ & 0.1471 & $ \pm _{16.0} ^{10.1} $ \\   
18.0 & $5.0\cdot 10^{-3}$ & 0.1033 & $ \pm _{6.0} ^{13.6}$ & \pm 12.5& $ \pm ^{11.3} _{11.5}$ & 0.1455 & $ \pm _{13.6} ^{8.7} $  \\[0.2cm]   

35.0 & $8.0\cdot 10^{-4}$ & 0.3958 & $ \pm _{3.0} ^{3.6} $ & \pm 8.3 & $ \pm ^{8.0} _{8.2}  $ & 0.3620 & $ \pm _{11.7} ^{14.0} $ \\   
35.0 & $1.3\cdot 10^{-3}$ & 0.3188 & $ \pm _{2.4} ^{2.8} $ & \pm 6.7 & $ \pm ^{7.2} _{7.1}  $ & 0.3092 & $ \pm _{13.5} ^{11.9} $ \\   
35.0 & $3.2\cdot 10^{-3}$ & 0.2015 & $ \pm _{2.4} ^{3.7} $ & \pm 8.5 & $ \pm ^{7.6} _{7.6}  $ & 0.2000 & $ \pm _{14.2} ^{7.5} $  \\   
35.0 & $5.0\cdot 10^{-3}$ & 0.1616 & $ \pm _{2.7} ^{4.2} $ & \pm 9.9 & $ \pm ^{8.3} _{8.7}  $ & 0.1684 & $ \pm _{12.5} ^{9.0} $  \\   
35.0 & $8.0\cdot 10^{-3}$ & 0.0854 & $ \pm _{6.5} ^{11.2}$ & \pm 14.9& $ \pm ^{9.9} _{12.4} $ & 0.1253 & $ \pm _{18.1} ^{7.6} $  \\[0.2cm]   

60.0 & $1.3\cdot 10^{-3}$ & 0.3952 & $ \pm _{1.5} ^{2.8} $ & \pm 11.3& $ \pm ^{8.2} _{8.3}  $ & 0.3606 & $ \pm _{15.1} ^{10.5} $ \\   
60.0 & $3.2\cdot 10^{-3}$ & 0.3040 & $ \pm _{1.3} ^{3.4} $ & \pm 9.5 & $ \pm ^{7.8} _{8.0}  $ & 0.2957 & $ \pm _{16.4} ^{9.5} $  \\   
60.0 & $5.0\cdot 10^{-3}$ & 0.1860 & $ \pm _{2.6} ^{3.5} $ & \pm 13.2& $ \pm ^{7.9} _{8.0}  $ & 0.1778 & $ \pm _{19.2} ^{9.3} $  \\   
60.0 & $8.0\cdot 10^{-3}$ & 0.1417 & $ \pm _{1.4} ^{5.5} $ & \pm 17.9& $ \pm ^{8.0} _{7.9}  $ & 0.1457 & $ \pm _{19.4} ^{6.9} $  \\   
60.0 & $2.0\cdot 10^{-2}$ & 0.0519 & $ \pm _{6.8} ^{10.9}$ & \pm 56.4& $ \pm ^{9.9} _{13.4} $ & 0.0834 & $ \pm _{19.9} ^{3.4} $  \\

\hline
\end{tabular}
}
\end{center}
\caption{\ftwocc\ in bins of $Q^2$ and $x$ extracted from measured 
\dstar\ cross sections
with two different programs, HVQDIS and CASCADE. The extrapolation uncertainty 
$\delta_{ext}$ is determined by varying model parameters within a program. The statistical 
($\delta_{stat}$) and systematic ($\delta_{syst}$) uncertainties arise from the
determination of the \dstar\ cross section and are the same for both programs.}
\label{Tab_f2cc}
\end{table}

\clearpage


\newpage


\end{document}